\documentclass[conference]{IEEEtran}
\usepackage{cite}
\usepackage{amsmath,amssymb,amsfonts}
\usepackage{algorithmic}
\usepackage{graphicx}
\usepackage{textcomp}
\usepackage{subfigure}
\usepackage{amssymb}
\usepackage[table]{xcolor}
\usepackage{pifont}
\usepackage{makecell}
\usepackage{epstopdf}
\def\BibTeX{{\rm B\kern-.05em{\sc i\kern-.025em b}\kern-.08em
    T\kern-.1667em\lower.7ex\hbox{E}\kern-.125emX}}
\AtBeginDocument{\definecolor{ojcolor}{cmyk}{0.93,0.59,0.15,0.02}}

\begin{document}

\title{Analysis and Prediction of Coverage and Channel Rank for UAV Networks in Rural Scenarios with Foliage}
\author{Donggu Lee$^1$, Ozgur Ozdemir$^1$, Asokan Ram$^2$, and Ismail Guvenc$^1$
\\ $^1$Department of Electrical and Computer Engineering, North Carolina State University, Raleigh, NC, USA 
\\ $^2$Wireless Research Center of North Carolina, Wake Forest, NC, USA
\\ E-mail: \{dlee42, oozdemi, iguvenc\}@ncsu.edu,
asokan.ram@wrc-nc.org
}






\maketitle

\begin{abstract} 
Unmanned aerial vehicles (UAVs) are expected to play a key role in 6G-enabled vehicular-to-everything (V2X) communications requiring high data rates, low latency, and reliable connectivity for mission-critical applications. Multi-input multi-output (MIMO) technology is essential for meeting these demands. However, UAV link performance is significantly affected by environmental factors such as signal attenuation, multipath propagation, and blockage from obstacles, particularly dense foliage in rural areas. In this paper, we investigate RF coverage and channel rank over UAV channels in foliage-dominated rural environments using ray tracing (RT) simulations. We conduct RT-based channel rank and RF coverage analysis over Lake Wheeler Field Labs at NC State University to examine the impact on UAV links. Custom-modeled trees are integrated into the RT simulations using NVIDIA Sionna, Blender, and Open Street Map (OSM) database to capture realistic blockage effects. Results indicate that tree-induced blockage impacts RF coverage and channel rank at lower UAV altitudes. We also propose a Kriging interpolation-based 3D channel rank interpolation scheme, leveraging the observed spatial correlation of channel rank in the given environments. The accuracy of the proposed scheme is evaluated using the mean absolute error (MAE) metric and compared against baseline interpolation methods. Finally, we compare the RT-based received signal strength (RSS) and channel rank results with real-world measurements from the NSF AERPAW testbed demonstrating reasonable consistency between simulation results and the measurements.
\end{abstract}



\section{Introduction}
\label{sec:introduction}
Unmanned aerial vehicles (UAVs) are expected to play a key role in 6G-enabled networks, especially in the context of vehicular-to-everything (V2X) communication. Using UAVs in 6G networks has many promising applications such as surveillance, infrastructure inspections, search-and-rescue missions, and among others \cite{access_UAV_application, proc_ieee_UAV_application}. High data rate, low latency, and reliable connectivity are critical requirements to support mission-oriented applications in V2X communications, which need to be specifically studied for UAV scenarios.

UAVs provide the advantages of forming flexible coverage, operating in challenging terrains, and handling dynamic network loads. To achieve high data rate and reliability requirements, multi-input multi-output (MIMO) technology is essential in UAV networks. Spatial multiplexing in MIMO systems enhances data rates by utilizing parallel data stream transmissions without allocating additional wireless communication resources. Various performance metrics such as singular value spread, condition number, and correlation matrix distance can be considered for evaluating MIMO systems \cite{MIMO_performance_metric}. Especially, channel rank is a key metric in MIMO system design as it quantifies how many parallel spatial streams can be supported in a given MIMO link \cite{MIMO_white_paper, MIMO_adaptation_1, MIMO_adaptation_2}. However, analyzing channel rank and coverage for UAV links involves unique challenges such as propagation modeling in unfavorable terrains, blockage from obstacles, and prediction of coverage and rank at locations where no past data is available. 

To our knowledge, ray tracing (RT)-based UAV coverage analysis and Kriging interpolation-based 3D channel rank interpolation for UAV channels in foliage-dominated rural scenarios are not available in the existing literature. In this paper, we substantially extend our earlier related work in \cite{previous_work} with a wider target area and multiple base station settings. The main contributions of this paper can be summarized as follows. 

\textbf{Foliage Modeling and Blockage Analysis.} Blockage effects in an environment can severely impact network reliability \cite{ray_tracing_foliage}. In rural areas, dense foliage and buildings can obstruct propagation, leading to signal degradation and deviations in UAV communication scenarios. To accurately model these effects, we introduce custom-modeled trees in the target area using NVIDIA’s RT tool Sionna~\cite{nvidia_sionna}, the 3D modeling tool Blender~\cite{blender}, and the Open Street Map (OSM) database~\cite{osm}. This approach allows for realistic simulation of signal blockage, attenuation, and reflection due to environmental obstacles.

\textbf{RF Coverage and Channel Rank Analysis.} Understanding RF coverage and channel rank is fundamental for UAV link analysis in rural environments. RF coverage defines the effective operational range of UAVs, ensuring consistent connectivity. Meanwhile, channel rank determines the number of parallel spatial streams available for spatial multiplexing directly influencing link capacity. Using extensive RT simulations, we analyze the impact of foliage and terrain on RF coverage and channel rank at varying UAV altitudes. The RT tool enables precise evaluation of propagation including beam patterns, angular resolution, and multipath components, providing a detailed representation of UAV communication in complex environments \cite{RT_tutorial}.

\textbf{Kriging-Based 3D Channel Rank Interpolation.} We introduce a Kriging interpolation-based 3D channel rank interpolation scheme. This method leverages the spatial correlation of channel rank incorporating both horizontal and vertical variations. The interpolation process consists of: 1) utilizing channel ranks across different UAV altitudes to compute spatial correlation and semi-variograms for Kriging interpolation; and 2) integrating channel rank data within a defined horizontal sampling radius from target UAV locations. We evaluate the accuracy of the proposed Kriging-based interpolation using the mean absolute error (MAE) metric and compare it against two baseline interpolation methods.

\textbf{Validation with Real-World Measurements.} To ensure the practical applicability of our RT-based simulations, we compare the RT simulation results of RF coverage and channel rank with real-world measurements collected in the NSF AERPAW testbed~\cite{aerpawWebsite}. Our goal is to assess how well RT simulations capture real-world conditions in the given environments. The results show that RT-based modeling provides reasonable approximations of UAV communication performance in foliage-dense rural environments.

The rest of this paper is organized as follows. Relevant works from the existing literature are discussed in Section~\ref{ch:related_works}. The description of the system model is given in Section~\ref{ch:system_model}. The RT and measurement scenarios are introduced in Section~\ref{ch:RT_setup} and Section~\ref{ch:measurement_setup}, respectively. The proposed Kriging-based 3D channel rank interpolation scheme and baseline interpolation schemes are described in Section~\ref{ch:Kriging_3D_Channel_interpolation}. 
Simulation results for RF coverage, channel rank, Kriging-based 3D channel rank interpolation, and comparison of measurements and RT simulations are provided in Section~\ref{ch:numerical_results}. Finally, the last section concludes the paper.

\begin{table*}[t!]
    \centering
    \caption{Literature review on RT-based RF coverage and channel rank analysis, and Kriging interpolation scheme.}
    \begin{tabular}{|c|c|c|c|c|c|c|c|}
    \hline
        \textbf{Ref.} & \textbf{Analysis Objectives} & \makecell{\textbf{UAV} \\ \textbf{Channels}}  & \makecell{\textbf{Ray} \\ \textbf{Tracing}}  & \makecell{\textbf{Channel} \\ \textbf{Rank} \\ \textbf{Analysis}} & \makecell{\textbf{Kriging} \\ \textbf{Interpolation}} & \textbf{Trees} & \textbf{Measurements} \\ \hline
        
        \cite{duke_sionna_paper} & RT-based RF signal mapping and ML integration & \cellcolor{red!25} \ding{55} & \cellcolor{green!25} \ding{51} & \cellcolor{red!25} \ding{55} & \cellcolor{red!25} \ding{55} & \cellcolor{red!25} \ding{55} & \cellcolor{green!25} \ding{51} \\ \hline  
        
        \cite{boston_sionna} & RT-based digital twin for urban scenarios & \cellcolor{red!25} \ding{55} & \cellcolor{green!25} \ding{51} & \cellcolor{red!25} \ding{55} & \cellcolor{red!25} \ding{55} & \cellcolor{red!25} \ding{55} & \cellcolor{red!25} \ding{55} \\ \hline  

        \cite{RT_mode_urban} & RT-based channel analysis in urban scenarios & \cellcolor{red!25} \ding{55} & \cellcolor{green!25} \ding{51} & \cellcolor{red!25} \ding{55} & \cellcolor{red!25} \ding{55} & \cellcolor{red!25} \ding{55} & \cellcolor{red!25} \ding{55} \\ \hline

        \cite{uav_channel_model_urban} & UAV channel modeling with measurement
        & \cellcolor{green!25} \ding{51} & \cellcolor{red!25} \ding{55} & \cellcolor{red!25} \ding{55} & \cellcolor{red!25} \ding{55} & \cellcolor{red!25} \ding{55} &  \cellcolor{green!25} \ding{51} \\ \hline

        \cite{uav_channel_model_arxiv} & Stochastic channel model for urban UAV scenarios & \cellcolor{green!25} \ding{51} & \cellcolor{red!25} \ding{55} & \cellcolor{red!25} \ding{55} & \cellcolor{red!25} \ding{55} & \cellcolor{red!25} \ding{55} & \cellcolor{red!25} \ding{55} \\ \hline
        
        \cite{wahab_paper} & UAV channel modeling with foliage & \cellcolor{green!25} \ding{51} & \cellcolor{red!25} \ding{55} & \cellcolor{red!25} \ding{55} & \cellcolor{red!25} \ding{55} & \cellcolor{green!25} \ding{51} & \cellcolor{green!25} \ding{51} \\ \hline      
        \cite{Channel_rank_massive_MIMO} & Channel rank analysis for massive MIMO & \cellcolor{red!25} \ding{55} & \cellcolor{green!25} \ding{51} & \cellcolor{green!25} \ding{51} & \cellcolor{red!25} \ding{55} & \cellcolor{red!25} \ding{55} & \cellcolor{green!25} \ding{51}  \\ \hline     
        \cite{RT_channel_modeling, RT_terahz_UAV} & RT-based UAV channel modeling & \cellcolor{green!25} \ding{51} & \cellcolor{green!25} \ding{51} & \cellcolor{red!25} \ding{55} & \cellcolor{red!25} \ding{55} & \cellcolor{red!25} \ding{55} & \cellcolor{green!25} \ding{51}  \\ \hline

        \cite{rank_irs} & Reflector-based channel rank improvement & \cellcolor{green!25} \ding{51} & \cellcolor{red!25} \ding{55} & \cellcolor{green!25} \ding{51} & \cellcolor{red!25} \ding{55} & \cellcolor{red!25} \ding{55} & \cellcolor{red!25} \ding{55} \\ \hline
        
        \cite{rank_MIMO_bounds} & Theoretical channel rank analysis & \cellcolor{red!25} \ding{55} & \cellcolor{red!25} \ding{55} & \cellcolor{green!25} \ding{51} & \cellcolor{red!25} \ding{55} & \cellcolor{red!25} \ding{55} & \cellcolor{red!25} \ding{55} \\ \hline
        \cite{vertical_MIMO_paper, vertical_MIMO_paper_2} & Channel rank analysis for antenna separation &  \cellcolor{red!25} \ding{55} &  \cellcolor{red!25} \ding{55} & \cellcolor{green!25} \ding{51} & \cellcolor{red!25} \ding{55} & \cellcolor{red!25} \ding{55} & \cellcolor{green!25} \ding{51} \\ \hline
        
        \cite{Kriging_coverage, fixed_rank_kriging} & Kriging-based coverage estimation & \cellcolor{red!25} \ding{55} & \cellcolor{red!25} \ding{55} & \cellcolor{red!25} \ding{55} & \cellcolor{green!25} \ding{51} & \cellcolor{red!25} \ding{55} & \cellcolor{green!25} \ding{51} \\ \hline
        \cite{Kriging_map_construction} & Kriging interpolation for radio map construction & \cellcolor{red!25} \ding{55} & \cellcolor{red!25} \ding{55} & \cellcolor{red!25} \ding{55} & \cellcolor{green!25} \ding{51} & \cellcolor{red!25} \ding{55} & \cellcolor{green!25} \ding{51} \\ \hline
        
        \cite{Kriging_mapping} & Coverage analysis using Kriging interpolation & \cellcolor{red!25} \ding{55} & \cellcolor{red!25} \ding{55} & \cellcolor{red!25} \ding{55} & \cellcolor{green!25} \ding{51} & \cellcolor{red!25} \ding{55} & \cellcolor{green!25} \ding{51} \\ \hline  
        
        \cite{previous_work} & RT-based UAV link analysis & \cellcolor{green!25} \ding{51} & \cellcolor{green!25} \ding{51} & \cellcolor{green!25} \ding{51} & \cellcolor{red!25} \ding{55} & \cellcolor{red!25} \ding{55} & \cellcolor{red!25} \ding{55} \\ \hline
        \cite{ray_tracing_foliage} & RT-based channel measurement on trees  & \cellcolor{red!25} \ding{55} & \cellcolor{green!25} \ding{51} & \cellcolor{red!25} \ding{55} & \cellcolor{red!25} \ding{55} & \cellcolor{green!25} \ding{51} & \cellcolor{green!25} \ding{51} \\ \hline 
        \cite{ray_tracing_tree_urban} & Impact of foilage in mmWave urban channels & \cellcolor{red!25} \ding{55} & \cellcolor{green!25} \ding{51} & \cellcolor{red!25} \ding{55} & \cellcolor{red!25} \ding{55} & \cellcolor{green!25} \ding{51} & \cellcolor{red!25} \ding{55} \\ \hline
        \cite{ray_tracing_vegetation} & RT-based propagation modeling for vegetation & \cellcolor{red!25} \ding{55} & \cellcolor{green!25} \ding{51} & \cellcolor{red!25} \ding{55} & \cellcolor{red!25} \ding{55} & \cellcolor{green!25} \ding{51} & \cellcolor{green!25} \ding{51} \\ \hline 
       
        \cite{openGERT_paper} & RT-based propagation modeling for digital twins & \cellcolor{red!25} \ding{55} & \cellcolor{green!25} \ding{51} &
        \cellcolor{red!25} \ding{55} & \cellcolor{red!25} \ding{55} & \cellcolor{red!25} \ding{55} & \cellcolor{red!25} \ding{55} \\ \hline
       
        \cite{maeng2023kriging} & 3D radio map generation using Kriging interpolation & \cellcolor{green!25} \ding{51} & \cellcolor{red!25} \ding{55} & \cellcolor{red!25} \ding{55} & \cellcolor{green!25} \ding{51} & \cellcolor{red!25} \ding{55} & \cellcolor{green!25} \ding{51} \\ \hline
        
        This work & RT-based link analysis and Kriging interpolation & \cellcolor{green!25} \ding{51} & \cellcolor{green!25} \ding{51} & \cellcolor{green!25} \ding{51} & \cellcolor{green!25} \ding{51} & \cellcolor{green!25} \ding{51} & \cellcolor{green!25} \ding{51} \\ \hline     
    \end{tabular}
    
    \label{tab:literature_review}
\end{table*}

\section{Related Works}\label{ch:related_works}
There are limited studies investigating channel rank and coverage for UAV links in the literature with RT studies or measurements. Our literature review with representative publications related to our work and the differences with the present work is summarized in Table \ref{tab:literature_review}.

\subsection{UAV Channels}
In \cite{RT_channel_modeling} and \cite{RT_terahz_UAV}, RT-based UAV channel modeling has been investigated. The altitude-dependent channel parameters, such as path loss, power delay profile, and angular information, are simulated to characterize the UAV channel. The developed channel model consisted of a deterministic line-of-sight (LoS) path and ground reflection, and stochastic non-LoS (NLoS) components. The developed model has been validated using RT simulations and measurements.  

A stochastic geometry model is developed in \cite{uav_channel_model_arxiv} for investigating coverage and outage probability in an urban environment. The study investigates UAV backhaul links considering building density, antenna beamwidth, and interference effects. The results highlight the existence of an optimal UAV altitude that maximizes coverage while ensuring reliable backhaul connectivity. Similarly, in \cite{uav_channel_model_urban}, measurement-based UAV channel modeling is conducted in an urban scenario. The study extracts key parameters for path loss models and develops a simplified Saleh-Valenzuela channel model for UAV-ground links.

In our previous work~\cite{previous_work}, UAV channel characteristics are analyzed in rural and urban scenarios. Specifically, RF coverage, channel rank, and condition number distribution have been analyzed for UAV networks using Matlab-based RT simulations. Centennial Campus and Lake Wheeler Field Labs at NC State University are considered as urban and rural scenarios, respectively. Due to the blockage from the buildings, an outage can be observed at the lower altitude of the UAV. Moreover, the probability of having a channel rank of 3 or 4 is getting lower because of the LoS dominant path to the UAV as the altitude of the UAV increases.

\subsection{Ray Tracing}
In \cite{duke_sionna_paper}, an RF signal mapping scheme for cellular networks has been investigated by integrating RT simulation using Sionna with geographic databases and machine learning (ML) techniques. A cascaded neural network refines signal strength predictions by leveraging geographical information from RT and sparse real-world measurements. The proposed method improves computational efficiency while maintaining high accuracy, outperforming conventional RT-based approaches in real-world evaluations.

In \cite{boston_sionna} and \cite{RT_mode_urban}, RT-based RF coverage and channel capacity analysis have been studied for urban scenarios. Nodetably in \cite{boston_sionna}, a large-scale digital twin framework for downtown Boston has been developed. The proposed BostonTwin framework integrates a high-fidelity 3D model of the downtown Boston area with geographical data, enabling RT simulations. By leveraging Sionna, the framework facilitates large-scale coverage mapping and signal-to-noise ratio (SNR) evaluations. Moreover, the authors highlight the areas where BostonTwin meets the requirements for 6G use cases. On the other hand, ~\cite{RT_mode_urban} uses site maps and a RT software for computing MIMO gain matrices for a given mobile station position in urban areas of Boston and Manhattan. Subsequently, authors use the MIMO gain matrices to determine achievable rates for MIMO transmission modes including spatial multiplexing, beamforming, and diversity. The results suggest that the use of site-specific RT for data rate prediction works more effectively than the use of stochastic models.

An automated geometry extraction framework for precise RT simulation is developed in \cite{openGERT_paper}. With the capability of open-source resources from OSM, Microsoft Global ML Building Footprints, and the US Geological Survey (USGS), high-fidelity RT simulation environments can be extracted. The authors conducted sensitivity analyses to investigate the impact on the accuracy of RT simulation with the environmental factors including building heights, locations, and material settings. Channel statistics of path gain, delay spread, and link outage are shown with the developed framework.

\subsection{Channel Rank Analysis}
In \cite{Channel_rank_massive_MIMO}, the channel rank for an outdoor-to-indoor massive MIMO system has been analyzed. The angular characteristics in terms of the angle of arrival and departure have been investigated for the given wireless communication environment. Then, singular values of each channel rank have been simulated and measured. It is observed that the singular values decrease rapidly as the range of eigenvalue to be captured increases. 

Theoretical lower and upper bounds of channel capacity for high-rank MIMO systems are studied in \cite{rank_MIMO_bounds}. The study investigates how LoS propagation affects channel rank and capacity. It shows that strong LoS components increase channel correlation, reducing the efficiency of spatial multiplexing. However, it also demonstrates that optimizing antenna placement and spacing helps maintain orthogonality, preserve channel rank, and improve system performance.

In \cite{rank_irs}, the use of passive intelligent reflecting surfaces (IRSs) has been investigated to improve channel rank and spatial multiplexing in urban UAV scenarios. This study proposes an IRS placement optimization framework to maximize the average channel capacity along predefined UAV trajectories. Numerical results at various carrier frequencies demonstrate that IRS-assisted channels significantly improve MIMO capacity compared to LoS-dominant UAV channels. The authors introduce the potential of IRSs as a cost-effective solution for enhancing UAV communications in dense urban environments. Other work on studying channel rank includes \cite{vertical_MIMO_paper} and \cite{vertical_MIMO_paper_2}, which use measurements for an outdoor LTE MIMO network to explore how the channel rank varies for various scenarios. 


\subsection{Kriging Interpolation}
Coverage analysis and map construction schemes using spatial interpolation for mobile systems have been investigated in \cite{Kriging_coverage, fixed_rank_kriging, Kriging_mapping, Kriging_map_construction}. Especially in \cite{Kriging_coverage}, a radio map has been developed by interpolating geo-located measurements. A fixed-rank Kriging interpolation scheme has been used to generate the radio map to reduce computational complexity. The simulation results show the trade-off between computational complexity and accuracy in coverage estimation over rural scenarios. 
In \cite{maeng2023kriging}, a Kriging interpolation-based 3D radio map generation scheme for radio dynamic zones is proposed. A realistic propagation model is developed by using measurements over the air-to-ground link in terms of path loss, shadowing, and spatial correlation. Horizontal and vertical correlations of received signal strength (RSS) are used for accurate radio map generation. The generated radio map is more accurate than the model with perfect knowledge of the path loss due to the benefits of spatial correlation. 

\subsection{Trees}
In \cite{ray_tracing_foliage}, an RT-based case study of channel measurements to model propagation effects due to trees is provided. Statistics of angular properties and penetration loss over multiple types of trees have been analyzed to characterize blockage and propagation for mmWave channels. The authors highlighted that the tree-specific propagation model can be used for realistic RT simulation in non-controlled environments. 

An RT-based 3D mmWave propagation model through vegetation is investigated in \cite{ray_tracing_vegetation}. To develop the propagation model, measurements of radiation, angular, and 3D scattering profiles are used in the scenario of groups of trees and various dimensions of the trees. The proposed 3D propagation model is assessed by comparing 3D directional measurements within different positions and angles.  

In \cite{ray_tracing_tree_urban}, the impact of foliage in an urban environment over the mmWave channel is investigated. Here, a simplified urban scenario consists of four buildings in a grid-wise position and four trees at the end of each road between the buildings. A hybrid channel model with the knowledge of a conventional correlation-based channel model is integrated with the RT results to characterize the foliage impact.

The impact of foliage on UAV air-to-ground propagation channels has been investigated using channel-sounding measurements in \cite{wahab_paper} for ultra-wideband signals. The authors examine how tree obstructions affect signal propagation analyzing path loss, multipath fading, and coherence bandwidth. Results show that foliage significantly increases path loss and reduces coherence bandwidth, leading to stronger multipath fading and degraded reliability.

\subsection{Measurements}
In \cite{wahab_paper}, channel-sounding measurement-based UAV air-to-ground propagation channel modeling has been investigated. The study conducts comprehensive measurements for various UAV communication scenarios and derives statistical models for path loss, multipath propagation, and small-scale fading. The measurements capture high-resolution time and frequency domain characteristics. The collected data is analyzed to extract power delay profiles and coherence bandwidth, which provides insights into UAV channel behavior under different propagation conditions.

Studies in \cite{vertical_MIMO_paper} and \cite{vertical_MIMO_paper_2}, considering terrestrial cellular network scenarios, analyze the impact of antenna configurations on channel rank and system performance using LTE MIMO measurements. In \cite{vertical_MIMO_paper}, increased antenna spacing and optimized polarization, achieved by comparing vertically and horizontally spaced configurations, are shown to improve channel rank and throughput. Similarly, \cite{vertical_MIMO_paper_2} demonstrates that reducing mutual correlation enhances throughput by enabling spatial multiplexing with a higher channel rank. Both studies highlight the critical role of precise measurement campaigns in evaluating channel rank under various conditions.

\subsection{Contributions of This Work}

 Based on Table~\ref{tab:literature_review} and to our knowledge, there are no works that study the effects of foliage on the channel conditions of UAV networks using RT simulations or measurements. Other than \cite{RT_channel_modeling,RT_terahz_UAV}, there are also no works that compare RT simulations with measurements for analyzing coverage for UAV networks.  
In this paper, we extend the scope of our previous work in \cite{previous_work} to address the gaps in the literature. In particular, the target area of the rural scenario has been extended with multiple base stations to understand long-distance propagation effects. Trees in the target area have been included to investigate realistic propagation in the rural area. Moreover, a constant channel rank threshold has been used to investigate channel rank using the ratio of the strongest singular value. Lastly, the spatial correlation, which can be observed in the channel rank distribution over the rural area scenario, has been adopted for the Kriging interpolation-based 3D channel rank interpolation scheme.

     \begin{figure*}[t]
     \centering
     \includegraphics[width=1.6\columnwidth]{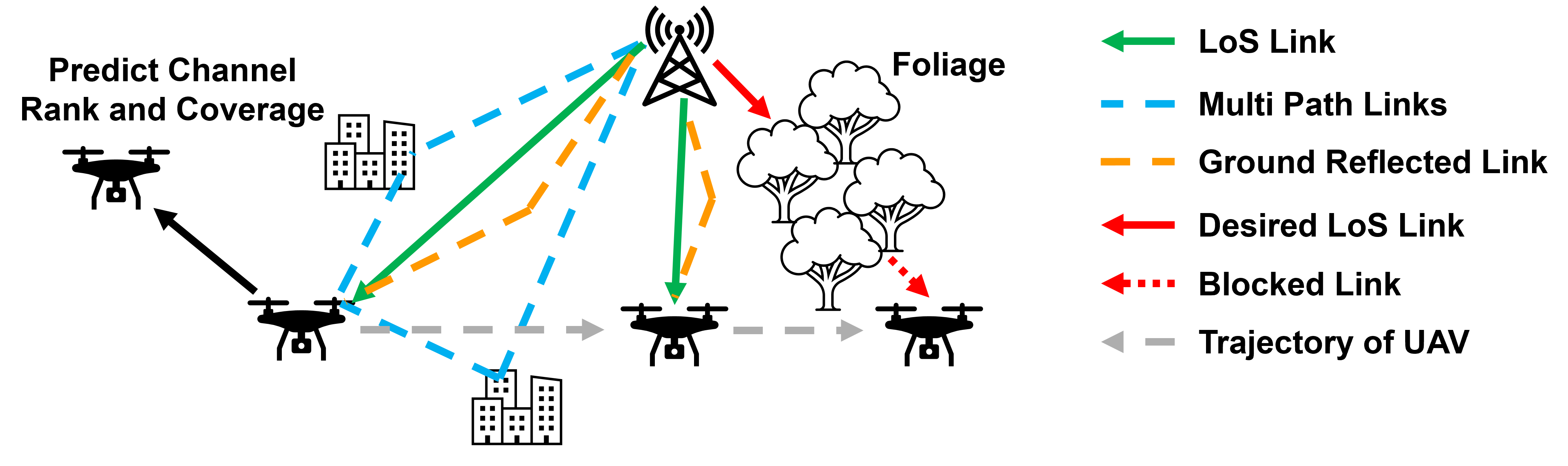}
     \caption{UAV connectivity scenario in a rural area. UAV coverage can be blocked due to buildings and foliage. It is of interest to predict channel characteristics, such as channel rank and coverage, at a location where no measurements have been collected before, based on measurements at other locations.}     \label{fig:introduction_figure}
 \end{figure*}

 \section{System Model}\label{ch:system_model}
In this section, we provide the system model for the UAV MIMO network and discuss our assumptions for calculating channel rank. The $N_{\rm r}\times 1$ received signal vector $\boldsymbol{y}$ in a MIMO link as in Figure~\ref{fig:introduction_figure} can be expressed as 
\begin{equation}
    \boldsymbol{y} = \boldsymbol{Hx + n},    \label{Eq:MIMO}
\end{equation}
where $\boldsymbol{H}$ is the $N_{\rm r} \times N_{\rm t}$ channel matrix of the MIMO link, $\boldsymbol{x}$ denotes the $N_{\rm t} \times 1$ transmit signal vector, and $\boldsymbol{n}$ is the noise vector having the same dimensions as $\boldsymbol{y}$. 

The rank of the channel matrix in~\eqref{Eq:MIMO} affects the number of parallel data streams that can be transmitted with spatial multiplexing over a given link~\cite{lte_book, MIMO_white_paper}. The channel rank can be obtained from the number of non-zero singular values after singular value decomposition of the channel matrix $\boldsymbol{H}$. This can be expressed as 
\begin{equation}
    \boldsymbol{H} = \boldsymbol{U \Sigma V^{*}},    
\end{equation}
where $\boldsymbol{U}$ represents the $\textit{m} \times \textit{m}$ complex unitary matrix, $\boldsymbol{\Sigma}$ is the $\textit{m} \times \textit{n}$ rectangular diagonal matrix, with diagonal elements $\sigma_i$, with $1<i<\min\{m, n\}$, and $\boldsymbol{V}$ is the $\textit{n} \times \textit{n}$ complex unitary matrix, respectively. The diagonal elements, $\sigma_i$, are sorted in descending order as $i$ increases. The channel rank, $R$, is bounded into the range of $1\leq R\leq \min\{m, n\}$. 

To develop fundamental insights on the channel rank behavior of UAV links for different RT scenarios, in this paper, we consider that the channel rank is determined by the number of eigenvalues of the channel matrix that are higher than a predetermined threshold. In particular, we consider that the channel rank for a given link at a UAV location $\boldsymbol{p}=(x, y, h)$, where $h \in (h_{1}, h_{2}, ..., h_{N_{h}})$ for $N_{h}$ different altitude settings, is given by  
\begin{align}    
\label{eq:channel_rank}
   R_{\delta_{K_j}}(\boldsymbol{p})&= \sum_{i=1}^{\min(N_r,N_t)}{\mathrm I}\{\sigma_{i}(\boldsymbol{p})> \delta_{K_j} \} \nonumber \\ &=\sum_{i=1}^{\min(N_r,N_t)}{\mathrm I}\bigg\{\sigma_{i}(\boldsymbol{p})> \frac{\sigma_{1}(\boldsymbol{p})}{K_{j}} \bigg\},    
\end{align}
where $I\{ \cdot \}$ is an indicator function that returns $1$ if its input is satisfied and $0$ otherwise, and $\delta_{K_j}$ is a threshold for the singular value with the $j$-th threshold ratio constant $K$. In this paper, we set the threshold for the singular value at a UAV location as $\delta_{K_j} = \sigma_{1}(\boldsymbol{p}) / K_{j}$ to focus on the ratio compared to the first strongest singular value and analyze the performance for different values of $K$. Here, the threshold is extended with a different threshold ratio constant $K$, i.e., $\delta_{K_{j}} = \sigma_{1}(\boldsymbol{p}) / K_{j}$ for $j=1,2, ..., N_{K}$, where $N_K$ is the total number of threshold ratio constant $K$. From this, the channel rank is determined as the number of non-zero singular values that exceed the threshold.  
 
\section{Ray Tracing Simulation Setup}\label{ch:RT_setup}
\begin{figure}[t!]
    \centering    
    \subfigure[Satellite view]{
    \includegraphics[width=0.45\columnwidth]{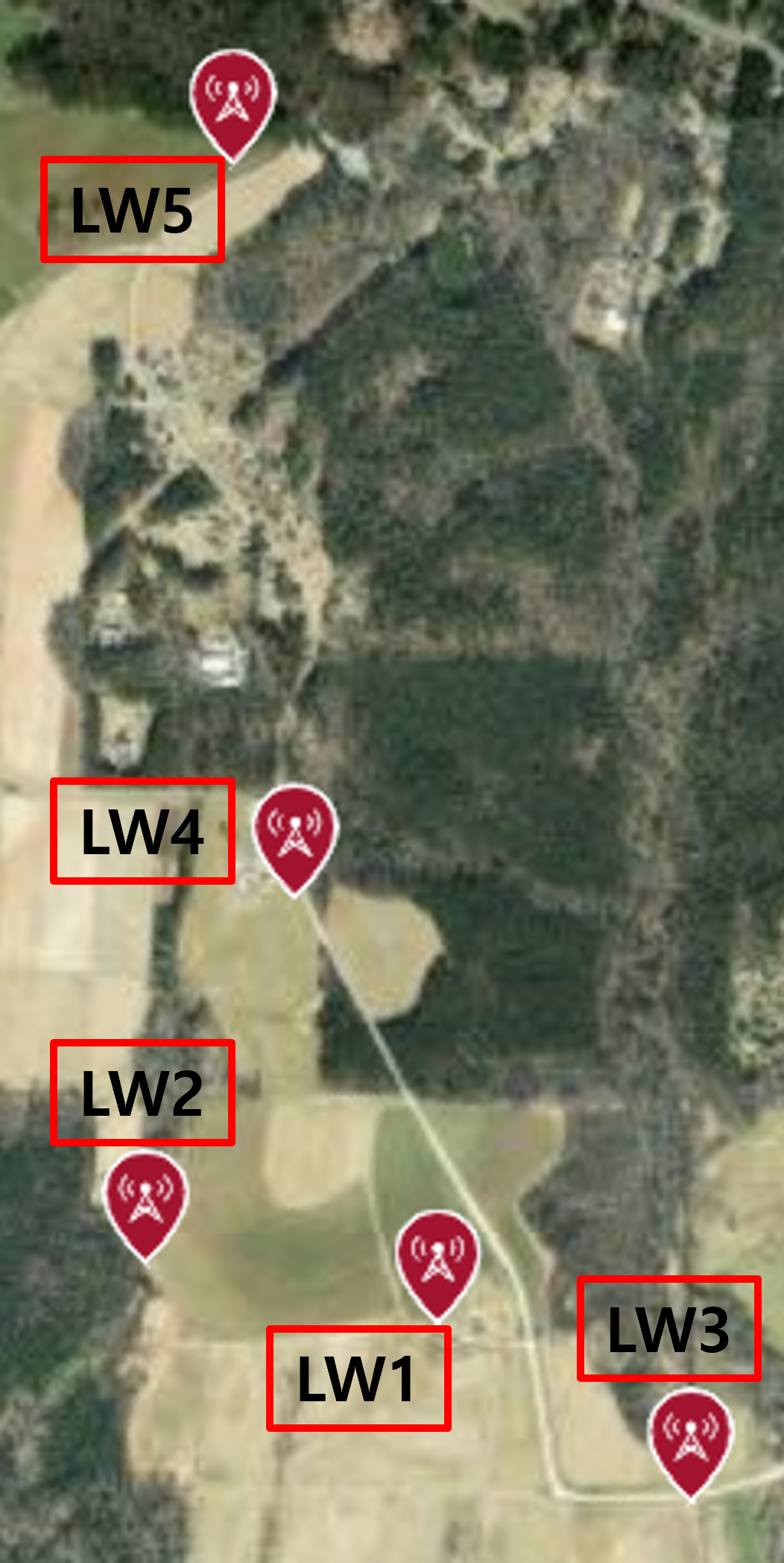}
    \label{fig:map_LW}
    }
    \subfigure[Blender scene]{
    \includegraphics[width=0.44\columnwidth]{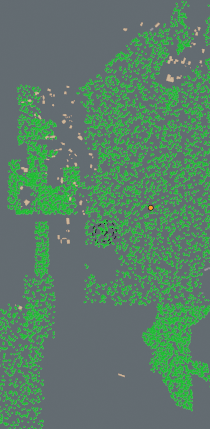}
    \label{fig:map_blender}
    }
    \subfigure[Tree model in Blender]{    \includegraphics[width=0.35\columnwidth]{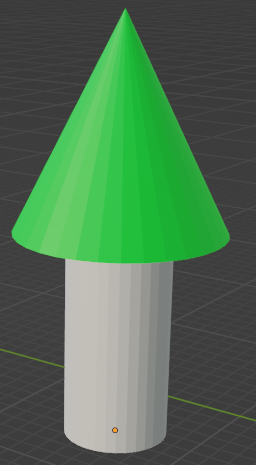}
    \label{fig:tree}
    }
   \caption{Map view of the location of towers used in RT simulations: (a) Satellite view, (b) Blender scene view, and (c) tree model.}
    \label{fig:map}
\end{figure}

We consider NSF AERPAW's fixed node locations at NC State University for the target area of RT simulations~\cite{aerpawWebsite}. NVIDIA Sionna is used as the RT simulation tool \cite{nvidia_sionna}. It is known that Sionna employs the Fibonacci lattice unit sphere on the transmitter's side to calculate the candidates for the possible trajectory of rays. A configurable parameter of the number of samples for the sphere is adopted to align the balance between computational complexity and accuracy in calculating the trajectories. The numbers of samples for the sphere for RF coverage and channel rank simulations are set to $10^6$ and $10^3$, respectively, to reduce the computational load in channel rank simulation.  
Figure \ref{fig:map_LW} shows a satellite view of the target area and the locations of the towers with red markers in the Lake Wheeler Field Labs. Geographical information including buildings is obtained from the OSM database \cite{osm, osm_buildings}. Moreover, Figure \ref{fig:map_blender} shows the Blender scene view of the target area, where green dots are the trees and orange rectangles indicate the buildings in the area.

The shooting and bounce ray (SBR) method-based RT model \cite{ray_tracing_access_paper, sbr_paper} has been applied in this work, where the ray tracer traces all rays at the receiver side after calculating reflection or diffraction in the trajectory. The procedure for the SBR can be categorized as 1) ray launching, 2) ray tracing, and 3) ray reception. After the launching of rays from the source, the trajectory of each ray is calculated in the 3D space \cite{sbr_paper}, which can be expressed as 
 \begin{equation}
    (x_1, y_1, z_1) = (x_0, y_0, z_0) + (s_x, s_y, s_z)t,     
 \end{equation}
where $(x_0, y_0, z_0)$ is the reference point, $(s_x, s_y, s_z)$ is the direction vector, and $t$ is the time duration of the trajectory.

The reflection losses of the ray with a surface can be expressed as follows~\cite{ITU-R-P.2040-3}: 
\begin{equation}
    \epsilon_{r} = \epsilon_{r}' + j\epsilon_{r}'',    
\end{equation}
 where $\epsilon_{r}' = af^{b}$ and $\epsilon_{r}'' = cf^{d}/2\pi\epsilon_{0}f$. Here, $a, b, c,$ and $d$ are the constant determined by the surface material, $\epsilon_{0}$ is the permittivity of the free space, and $f$ is the frequency in Hz, respectively. Those constants are decided under ITU recommendation  \cite{itu_R_vegetation, itu_R_buildings} for building, ground, and vegetation material settings in this work. Moreover, the electrical field after reflection and diffraction can be expressed as \cite{sbr_paper}
\begin{equation}
    \boldsymbol{E}(x_{i+1}, y_{i+1}, z_{i+1}) = D_i \Gamma_i \boldsymbol{E}(x_i, y_i, z_i) e^{j\theta},  
\end{equation}
where $D_i$ is the divergence factor related to the spreading of the ray right after the \textit{i}-th reflection, $\Gamma_i$ is the reflection coefficient, $\boldsymbol{E}$$(x_i, y_i, z_i)$ is the incident electric field, and $\theta$ is the phase shift of the electrical field. After the calculation, the receiving rays are derived by the overlapping area of the receiving field and trajectories of the rays.

 \begin{table}[t!]
    \centering
    \caption{Simulation parameters for ray tracing.}\vspace{-1mm}
    \begin{tabular}{|c|c|c|}
    \hline
      \textbf{Parameters}   & \textbf{Description} & \textbf{Value} \\
      \hline
        $f_{\mathrm{c}}$ & Carrier frequency & 3.4 GHz \\
        \hline
        $a_{\mathrm{BS}}$ & Height of base stations & 10 m \\        
        \hline
        $T_{\mathrm{LW}}$ & Target area of Lake Wheeler & 1080 m $\times$ 2130 m \\
        \hline
        $N_{\mathrm{ref}}$ & Maximum number of reflections & 2\\
        \hline
        $N_{\mathrm {ele}}$ & Number of antenna elements & 4 (TX and RX) \\
        \hline
        $\Delta_{\mathrm{tx}}$ & Element spacing for TX antennas & $0.5\lambda$ \\
        \hline
        $\Delta_{\mathrm{rx}}$ & Element spacing for RX antennas & $0.5\lambda$ 
        \\
        \hline
        $\mathrm{P_{TX}}$ & Transmit power of TX antennas & 10 W
        \\
        \hline
        $d_{\mathrm{rx}}$ & Horizontal interval of UAV & $30$~m
        \\
        \hline
    \end{tabular}    
    \label{tab:sim_param}
\end{table}

 The following assumptions are made for the RT simulation scenario of Figure \ref{fig:map_blender}: 1) the carrier frequency is set to $3.4$ GHz; 2) a 4-element linear array antenna is used for the transmitter and receiver in the MIMO cases with horizontal element spacing of $0.5\lambda$ for both transmitter and receiver, where $\lambda$ is the wavelength; 3) the height of the base station is set to $10~$m; 4) the maximum number of reflection is set to be 2; 5) surface materials for the buildings are set to concrete and the ground material for the area is medium dry ground, which is predefined as ``itu\_concrete" and ``itu\_medium\_dry\_ground" in Sionna, respectively; 6) the receivers are located on a uniform mesh grid with $30$~m spacing in both x-axis and y-axis in the horizontal directions covering the given area; 7) noise is neglected in RT simulations; and 8) channel rank is determined between the nearest base station and receiver in Lake Wheeler area. The simulation parameters are summarized in Table~\ref{tab:sim_param}. Lastly, the tree consists of a cylinder with ``itu\_wood" material and a cone on top of the cylinder with custom vegetation material setting calculated in \cite{itu_R_vegetation}. The tree model for the simulation is demonstrated in Figure~\ref{fig:tree}, which has a height of $20$~m and a maximum diameter of $10$~m.

 The information about rays such as the angle of arrival and departure, locations of the transmitter and receiver, and path types indicating LoS or reflective NLoS are extracted using Sionna for every UAV location. Then, an RT channel object is created with Matlab's RT tool \cite{matlab_ray_tracing} using ray information. The channel matrix is derived to determine the channel rank by using random input bits and output bits through the ray-tracing channel object. Finally, channel rank is decided as in (\ref{eq:channel_rank}) by singular value decomposition and the indicator function with a threshold compared to the strongest singular value aforementioned in the previous section.

\section{Real-World Measurement Scenario}\label{ch:measurement_setup}

\begin{figure}
     \centering
     \subfigure[Trajectory]{\includegraphics[trim={0.4cm 0.1cm 1.1cm 0.6cm},clip,width=0.98\columnwidth]{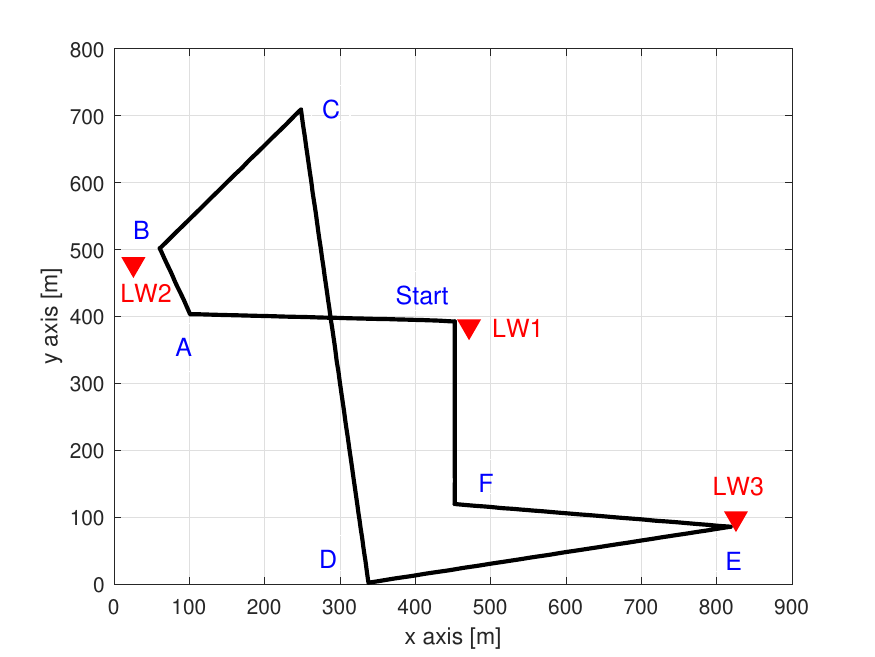}
     \label{fig:trajectory_measurement_RSS}
     }
     \subfigure[Altitude]{\includegraphics[trim={0.4cm 0.1cm 1.3cm 0.6cm},clip,width=0.98\columnwidth]{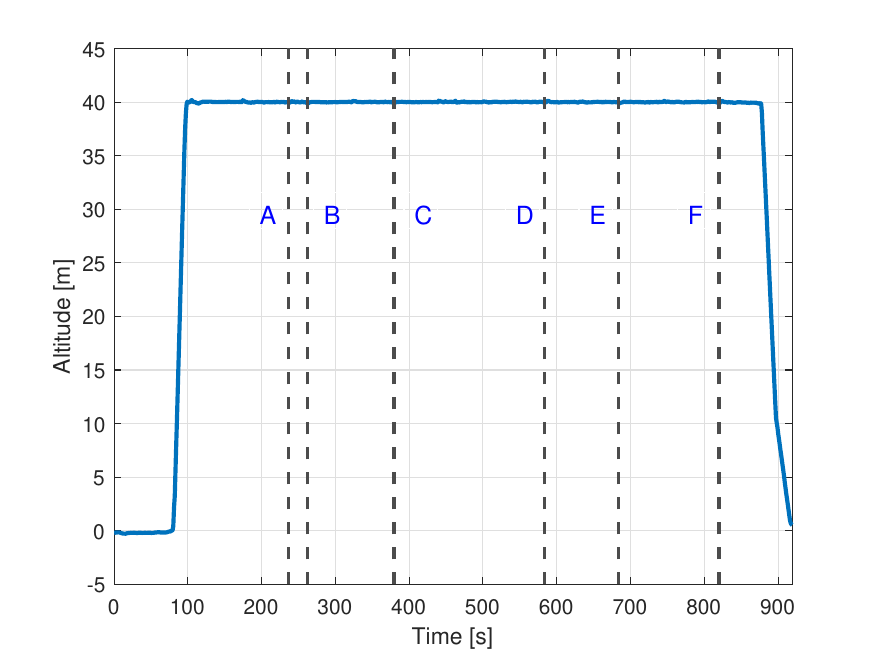}
     \label{fig:altitude_measurement_RSS}
     }     
     \caption{Trajectory and altitude of the signal coverage measurements and RT simulation.}
     \label{fig:trajectory_altitude_measurement_RSS}
 \end{figure}

 \begin{figure}
     \centering
     \subfigure[Trajectory]{\includegraphics[trim={0.4cm 0.1cm 1.1cm 0.6cm},clip,width=0.98\columnwidth]{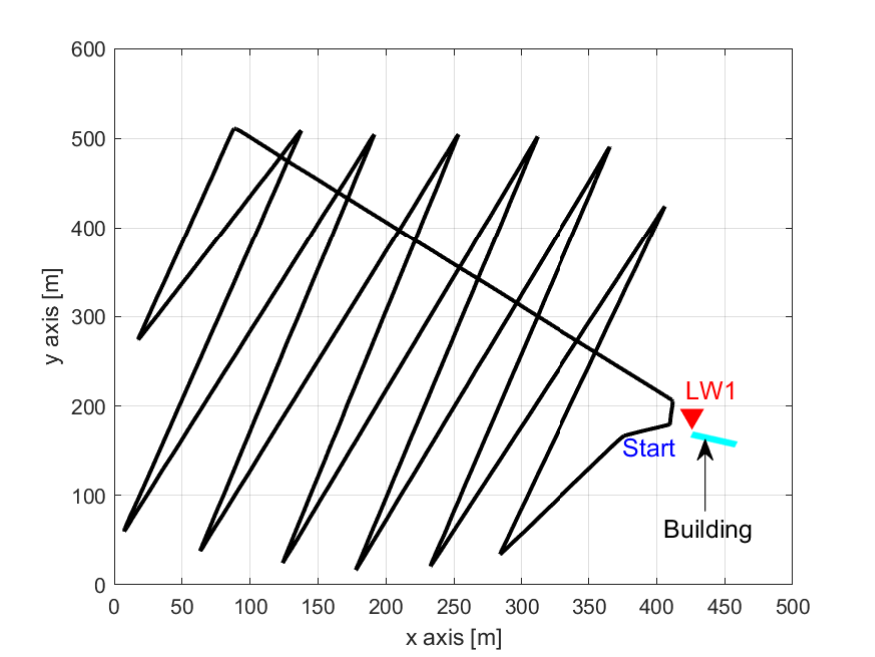}
     \label{fig:trajectory_measurement_rank}
     }
     \subfigure[Altitude]{\includegraphics[trim={0.4cm 0.1cm 1.3cm 0.6cm},clip,width=0.98\columnwidth]{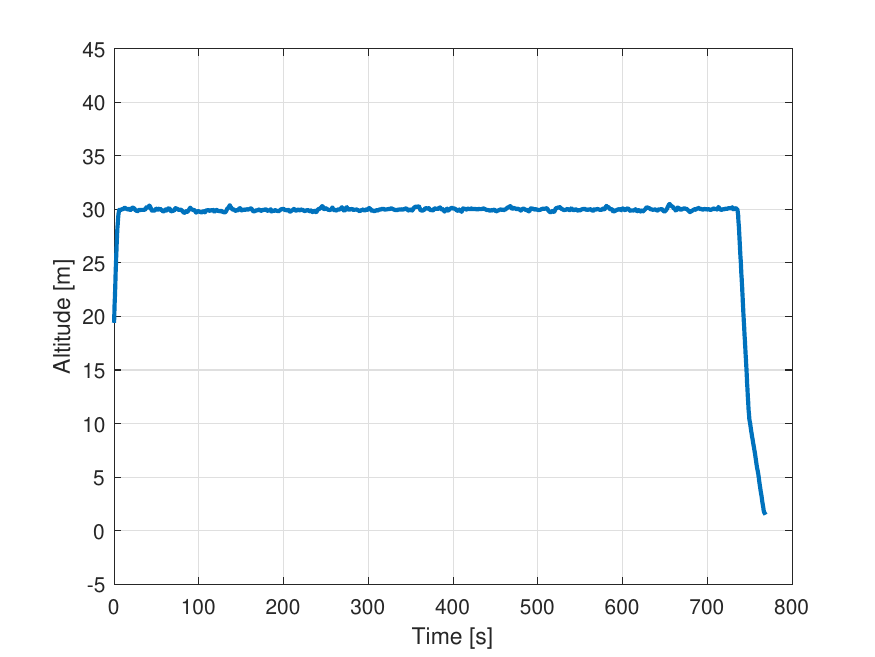}
     \label{fig:altitude_measurement_rank}
     }     
     \caption{Trajectory and altitude of the channel rank measurements and RT simulation.}
     \label{fig:trajectory_altitude_measurement_rank}
 \end{figure}

In addition to RT simulations, we also use real-world coverage and channel rank data from multiple measurement campaigns conducted at the NSF AERPAW platform in the same environments in Figure \ref{fig:map_LW}. While channel rank measurements are available only for one of the base stations, coverage results are available for all five of the base stations in the experiment area. 

\textbf{Coverage Measurements.} The predefined trajectory and altitude of the UAV over time for coverage measurements are shown in Figure~ \ref{fig:trajectory_altitude_measurement_RSS}. The UAV takes off near the LW1 fixed node. Then, the UAV sweeps LW2 and LW3 areas, and returns to the LW1. Each waypoint of the trajectory is marked with a letter and with vertical lines in the following figures. 

The following assumptions and setups are adopted for this scenario: 1) the antenna is a single-element antenna (SISO) with a carrier frequency of $3.3$~GHz; 2) the measurements are recorded by dual channel USRP B205 and GNU Radio; 3) the UAV collected data for $20 $~ms intervals every $100$~ms; 4) the offsets calculated by minimum root mean squared error (RMSE) between measurements and RT simulation results are adopted to each measurement for calibration purpose; 5) the proper offsets that have minimum RMSE is searched by a unit of $0.1$~dB in $[-50, 50]$ range in dB scale; 6) the altitude within the trajectory below $0.5$~m is rounded to $0.5$~m for RT simulation to obtain margin from the ground surface; 7) the UAV is the transmitter and the fixed node LW1 to LW5 act as the receivers; and 8) other parameters are the same for the RT simulation parameters. 

\textbf{Channel Rank Measurements.} The predefined trajectory and altitude over time of the UAV for channel rank measurements are shown in Figure \ref{fig:trajectory_altitude_measurement_rank}, where the building in the target area is marked in the cyan polygon. The UAV takes off near the LW1 tower and sweeps the target area with a zigzag pattern.

The following assumptions and configurations are used for this scenario: 1) $4 \times 4$ MIMO channel is used with a carrier frequency of $3.4$~GHz; 2) two AW3232 dual port sector antennas are used for fixed node LW1 as a transmitter; 3) two AW3232 antenna modules are departed by $1.68$~m; 4) Quectel RM502Q-AE 5G mobile modem with antenna module is mounted at the bottom side of the UAV as a receiver; 5) the direction of antennas of LW1 and UAV are placed to face each other, which leads the yaw angle of $45^\circ$ for LW1 and $315^\circ$ for UAV, respectively; 6) 5G new radio (NR) rank indicator (RI) is recorded over the trajectory; and 7) other parameters are the same for the RT simulation parameters.

 \section{Kriging-Based 3D Channel Rank Interpolation}\label{ch:Kriging_3D_Channel_interpolation}
 MIMO channel rank at a given location $\boldsymbol{p}$ can be interpolated based on the knowledge of channel rank at other locations in the 3D space, where the rank information has been logged before. In this work, we introduce the Kriging interpolation method for interpolating the channel rank. 

 \subsection{Spatial Correlation Model}\label{ch:spatial_correlation}
 In this section, we describe the spatial correlation of channel ranks at different UAV locations. It is well-known that the channel rank depends on the angular separation of alignments of the transmitter and receiver antennas within a given environment \cite{tse2005fundamentals}. 

 \begin{figure*}[t!]
    \centering
    \includegraphics[width=1.95\columnwidth]{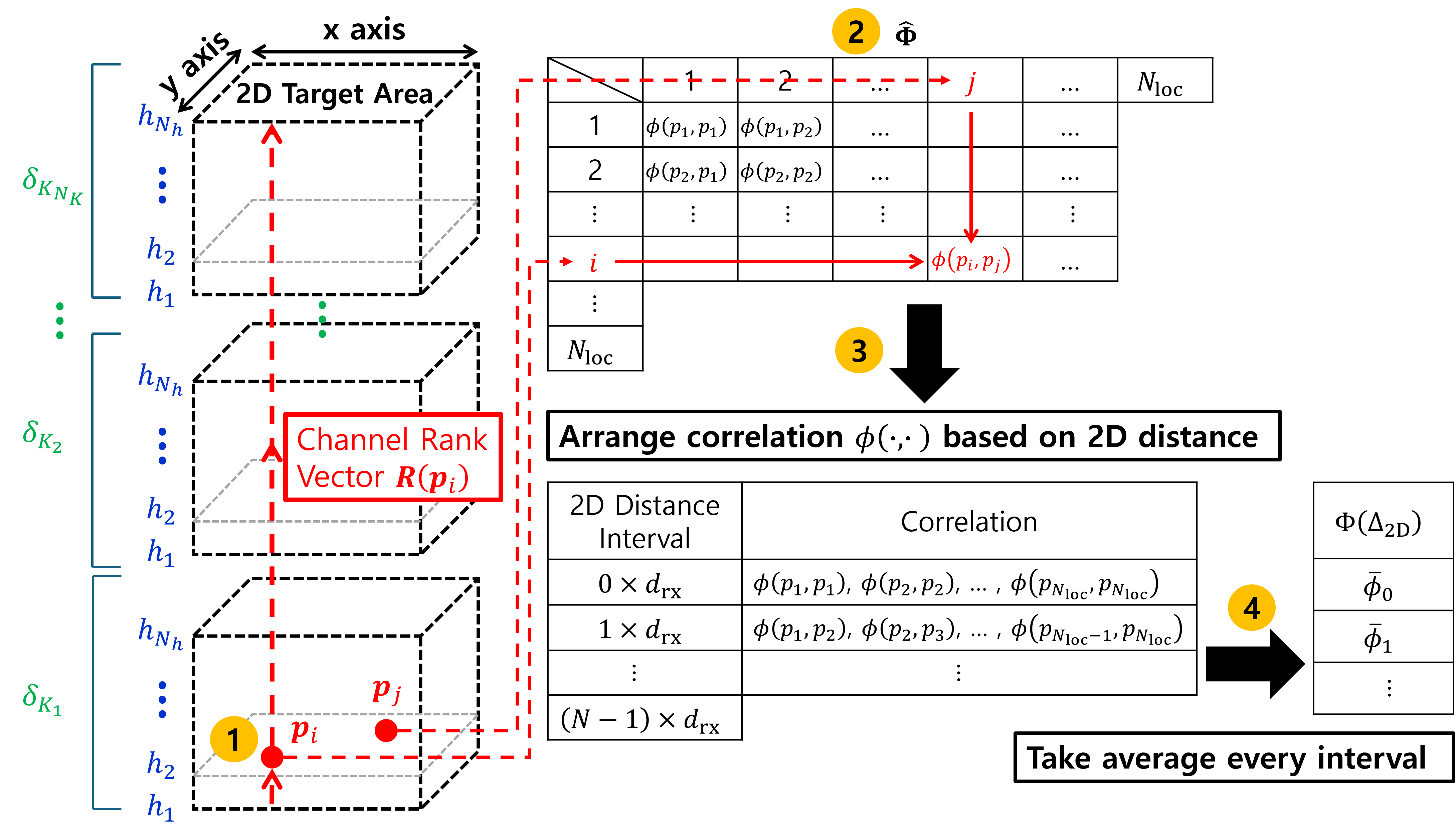}
    \caption{Spatial correlation framework for Kriging interpolation.}
    \label{fig:spatial_correlation_description}
\end{figure*}
 
 To take into account both horizontal and vertical correlation of the channel rank, we first define the channel rank vector at a location of UAV as 
\begin{align}\label{eq:channel_rank_vector}
    \boldsymbol{R}(\boldsymbol{p_i}) = \nonumber \\ \big\{ R_{\delta_{K_j}}(x_i &, y_i, h) \mid j = 1, \dots, N_K, \; h = h_1, \dots, h_{N_h} \big\}.
\end{align}
Here, we define an index for each UAV position using a subscript $i=1, ..., N_{\mathrm{loc}}$ at $\boldsymbol{p}$, where $N_{\mathrm{loc}}$ represents the total number of possible UAV locations within the 2D target area. These positions are distributed over a uniformly discretized 2D target area with the horizontal interval of $d_{\mathrm{rx}}$. The index follows a structured ordering: the index starts with the southwest corner in the target area. Then, the index increases along the x-axis from west to east. When the index reaches the end of the x-axis row, it moves one step along the y-axis before the index increases along the x-axis from west to east again. With this, the $N_{\mathrm{loc}}$-th UAV position is at the northeast corner on the 2D target area. This structure enables each UAV position to be assigned a unique index over the structured grid.

Moreover, the reason for using channel rank vector with all thresholds in (\ref{eq:channel_rank_vector}) rather than using channel rank with a single threshold is based on the fact that the correlation value that will be described is non-valid when the channel rank has the same value at all altitudes. In other word, we may have $R_{\delta_{K_j}}(x, y, h_1) = ... = R_{\delta_{K_j}}(x, y, h_{N_h})$. To avoid this, we set the channel rank vector to have channel ranks with all altitudes and thresholds. The defined channel rank vector $\boldsymbol{R}(\boldsymbol{p}_i)$ represents a collection of channel ranks computed at a position of $\boldsymbol{p}_i$ hierarchical ordering by spanning multiple altitudes first and expanded to the multiple thresholds, which can be expressed as
\begin{align}\label{eq:channel_rank_vector_structure}
    \boldsymbol{R}(\boldsymbol{p_i}) =& \big[ R_{\delta_{K_1}}(x_i, y_i, h_1), \dots, R_{\delta_{K_1}}(x_i, y_i, h_{N_h}),\nonumber \\
    &~R_{\delta_{K_2}}(x_i, y_i, h_1), \dots, R_{\delta_{K_2}}(x_i, y_i, h_{N_h}), \dots, \nonumber\\
    &~R_{\delta_{K_{N_K}}}(x_i, y_i, h_{1}), \dots R_{\delta_{K_{N_K}}}(x_i, y_i, h_{N_h}) \big].
\end{align}
In Figure \ref{fig:spatial_correlation_description}, the channel rank vector at $\boldsymbol{p}_i$ is highlighted in red-dashed arrows demonstrating channel rank through all altitudes and thresholds in the multiple 3D target area structure. 

Then, the correlation between two locations of the UAV can be calculated using the Pearson linear correlation function using the channel rank vector defined in (\ref{eq:channel_rank_vector}), which can be expressed as 
\begin{align}
        &\phi(\boldsymbol{p}_{i}, \boldsymbol{p}_{j}) = \nonumber \\
        &\frac{\sum_{u=1}^{N} \big(\boldsymbol{R}_u(\boldsymbol{p}_{i})-\bar{\boldsymbol{R}}(\boldsymbol{p}_{i}) \big) \big(\boldsymbol{R}_u(\boldsymbol{p}_{j}) - \bar{\boldsymbol{R}}(\boldsymbol{p}_{j}) \big)}{[ \sum_{u=1}^{N}\{ \boldsymbol{R}_u(\boldsymbol{p}_{i}) - \bar{\boldsymbol{R}}(\boldsymbol{p}_{i})\}^2  \sum_{o=1}^{N}\{\boldsymbol{R}_o(\boldsymbol{p}_{j}) -\bar{\boldsymbol{R}}(\boldsymbol{p}_{j}) \}^2 ]^{1/2}} ,
    \label{eq:correlation}
\end{align}
where $\boldsymbol{R}_{u}(\boldsymbol{p}_{i})$ denotes the $u$-th element in the channel rank vector $\boldsymbol{R}(\boldsymbol{p}_{i})$ at $\boldsymbol{p}_{i}$, and $\bar{\boldsymbol{R}}(\boldsymbol{p}_{i})$ indicates the mean of the channel rank vector at $\boldsymbol{p}_{i}$, respectively. Since channel rank vector in (\ref{eq:channel_rank_vector}) contains channel rank over all altitudes, the correlation between two locations in (\ref{eq:correlation}) represents correlation in both vertical and horizontal dimensions.

To derive generalized spatial correlation at  all locations of the UAV, the following procedure is considered in this work, which is also illustrated in Figure \ref{fig:spatial_correlation_description}: 
\begin{itemize}
    \item \textbf{Step-1:} Construct the channel rank vector for all UAV position as defined in (\ref{eq:channel_rank_vector}) and (\ref{eq:channel_rank_vector_structure}). 
    
    \item \textbf{Step-2:} Calculate correlation for all UAV positions using equation (\ref{eq:correlation}), i.e., $\phi(\boldsymbol{p}_{i},\boldsymbol{p}_{j})$ for $i, j = 1, \cdots , N_{\mathrm{loc}}$, which has the dimension of $N_{\mathrm{loc}} \times N_{\mathrm{loc}}$, where $N_{\mathrm{loc}}$ is the total number of possible UAV positions in the target area. It results in a symmetric correlation matrix as
    \begin{multline}        
       \hat{\boldsymbol{\Phi}} =  \\  \begin{bmatrix}
\phi(\boldsymbol{p}_1, \boldsymbol{p}_1) & \phi(\boldsymbol{p}_1, \boldsymbol{p}_2) & \dots & \phi(\boldsymbol{p}_1, \boldsymbol{p}_{N_{\mathrm{loc}}}) \\
\phi(\boldsymbol{p}_2, \boldsymbol{p}_1) & \phi(\boldsymbol{p}_2, \boldsymbol{p}_2) & \dots & \phi(\boldsymbol{p}_2, \boldsymbol{p}_{N_{\mathrm{loc}}}) \\
\vdots & \vdots & \ddots & \vdots \\
\phi(\boldsymbol{p}_{N_{\mathrm{loc}}}, \boldsymbol{p}_1) & \phi(\boldsymbol{p}_{N_{\mathrm{loc}}}, \boldsymbol{p}_2) & \dots & \phi(\boldsymbol{p}_{N_{\mathrm{loc}}}, \boldsymbol{p}_{N_{\mathrm{loc}}}) \\
\end{bmatrix}.
    \end{multline}
    \item \textbf{Step-3:} Arrange the correlation from the previous step according to the horizontal (2D) distance between two UAV locations $\boldsymbol{p}_i$ and $ \boldsymbol{p}_j$, which can be expressed as
    \begin{equation}\label{eq:horizontal_distance}
        \Delta_{\mathrm{2D}}(\boldsymbol{p}_i, \boldsymbol{p}_j) = \sqrt{(x_i - x_j)^2 + (y_i - y_j)^2},
    \end{equation} 
    to ensure the correlations have ascending order based on the horizontal distance between corresponding positions. This process results in arranged pairs of 
    \begin{equation}
        \big\{ \Delta_{\mathrm{2D}}(\cdot, \cdot), (\phi_n \mid \Delta_{\mathrm{2D}}^{(n)} = n \cdot d_{\mathrm{rx}} ) \big\}, \mathrm{for} ~ n=0,1,..., N-1,
    \end{equation} where $d_{\mathrm{rx}}$ is the interval of the neighboring UAV positions, e.g., $\boldsymbol{p}_i$ and $\boldsymbol{p}_{i+1}$, and $N$ is the total number of intervals for covering all pairs. Moreover, horizontal distance $\Delta_{\mathrm{2D}}(\cdot, \cdot)$ is discretized by the unit of interval $d_{\mathrm{rx}}$, which leads to $\Delta_{\mathrm{2D}}(\cdot, \cdot)=\Delta_{\mathrm{2D}}^{(n)} = n \cdot d_{\mathrm{rx}}$, where $n$ indicates $n$-th interval. 
    \item \textbf{Step-4:} Take average every $d_{\mathrm{rx}}$ from the smallest distance-correlation pairs until covering all pairs within the interval. It can be expressed as
    \begin{equation}
        \bar{\phi}_n = \frac{1}{M_n} \sum_{m \in \mathcal{I}_n} \phi(\boldsymbol{p}_{i_m}, \boldsymbol{p}_{j_m}),
    \end{equation}
    
    \begin{align}
        \Phi(\Delta_{\mathrm{2D}}(\boldsymbol{p}_i, \boldsymbol{p}_j)) = \nonumber \\
        & \begin{cases} 
        \bar{\phi}_0 & \text{if } \Delta_{\mathrm{2D}}(\boldsymbol{p}_i, \boldsymbol{p}_j) = 0, \\
        \bar{\phi}_1 & \text{if } \Delta_{\mathrm{2D}}(\boldsymbol{p}_i, \boldsymbol{p}_j) = d_{\mathrm{rx}}, \\
        \bar{\phi}_2 & \text{if } \Delta_{\mathrm{2D}}(\boldsymbol{p}_i, \boldsymbol{p}_j) = 2 \cdot d_{\mathrm{rx}}, \\
        \vdots & \vdots \\
        \bar{\phi}_{N-1} & \text{if } \Delta_{\mathrm{2D}}(\boldsymbol{p}_i, \boldsymbol{p}_j) = \\
        & ~~~~~~~~~~\quad (N-1) \cdot d_{\mathrm{rx}},
        \end{cases}\label{eq:correlation_cases}
    \end{align}
    where $M_n$ is the number of pairs with $\Delta_{\mathrm{2D}}^{(n)} = n \cdot d_{\mathrm{rx}}$ and $\mathcal{I}_n$ is the set of indices of UAV position pairs satisfying $\Delta_{\mathrm{2D}}^{(n)} = n \cdot d_{\mathrm{rx}}$.
\end{itemize}

As a result of the previous steps, the generalized correlation of channel ranks as a function of horizontal distance is derived, quantifying the correlation between channel rank vectors at different UAV positions. To further simplify the computation of the correlation function, $\Phi(\Delta_{\mathrm{2D}}(\cdot, \cdot))$, we approximate it using a bi-exponential curve-fitting model for two UAV positions $\boldsymbol{p}_i$ and $\boldsymbol{p}_j$:
\begin{equation}
     \Phi(\Delta_{\mathrm{2D}}(\boldsymbol{p}_i, \boldsymbol{p}_j)) \approx c_{1} e^{c_{2} \Delta_{\mathrm{2D}}(\boldsymbol{p}_i, \boldsymbol{p}_j)} + c_{3} e^{c_{4} \Delta_{\mathrm{2D}}(\boldsymbol{p}_i, \boldsymbol{p}_j)},
     \label{eq:bi_exponential}
 \end{equation}
 where coefficients $c_1, c_2, c_3$, and $c_4$ are selected based on curve-fitting.

 \subsection{Kriging Interpolation-based Channel Rank Interpolation}
 
The ordinary Kriging interpolation method provides optimal estimation when the spatial correlation is well-defined as a function of distance. The Kriging interpolation uses squared-error loss between the observation in a known spatial location and unknown locations \cite{Cressie_2015}. In particular, channel rank at a location $\boldsymbol{p_0}$ with a threshold $\delta_{K_j}$ is interpolated from the linear combination of the channel rank samples of the nearby locations, which can be expressed as \cite{maeng2023kriging}
 \begin{align}
     \min_{l_{1}, ..., l_{M}} \mathop{\mathbb{E}}\Big[\big(\hat{R}_{\delta_{K_j}}&(\boldsymbol{p}_{0})-{R}_{\delta_{K_j}}(\boldsymbol{p}_{0})\big)^2\Big],\\
     \mathrm{\mathbf{ s. t.}} ~~~~ \hat{R}_{\delta_{K_j}}(\boldsymbol{p}_{0}) = & \sum_{i=1}^{M} l_{i} R_{\delta_{K_j}}(\boldsymbol{p}_{i}),
     \label{eq:Kriging_estimation}\\
     \sum_{i=1}^{M} & l_{i} = 1,
 \end{align}
where $\boldsymbol{p}_0$ is the location to be interpolated, $l_i$ denotes the coefficient for the linear combination, $R_{\delta_{K_j}}(\boldsymbol{p}_{i})$ is $i$-th element of the channel rank within the sampled locations for the interpolation, $\hat{R}_{\delta_{K_j}}(\boldsymbol{p}_{0})$ indicates the estimated channel rank with threshold $\delta_{K_j}$, and $M$ indicates the number of samples used for the interpolation, respectively. The optimization problem above can be solved by converting the problem into an equivalent Lagrange equation as \cite{maeng2023kriging, Kriging_solution_paper}
\begin{multline} 
    \min_{l_{1}, ..., l_{M}} \mathop{\mathbb{E}}\left[\left(R_{\delta_{K_j}}(\boldsymbol{p}_{0}) - \sum_{i=1}^{M} l_{i} R_{\delta_{K_j}}(\boldsymbol{p}_{i}) \right)^2\right]   \\ - L\left(\sum_{i=1}^{M} l_{i} - 1\right),
\end{multline}
where $L$ denotes the Lagrange multiplier. After a few steps, the optimal solution can be derived by the linear matrix equation as
\begin{align}
        \begin{bmatrix}
    \gamma(\boldsymbol{p}_{1}, \boldsymbol{p}_{1}) & ... & \gamma(\boldsymbol{p}_{1}, \boldsymbol{p}_{M}) & 1 \\
    \gamma(\boldsymbol{p}_{2}, \boldsymbol{p}_{1}) & ... & \gamma(\boldsymbol{p}_{2}, \boldsymbol{p}_{M}) & 1 \\
    \vdots & \vdots & \vdots & \vdots \\
    \gamma(\boldsymbol{p}_{M}, \boldsymbol{p}_{1}) & ... & \gamma(\boldsymbol{p}_{M}, \boldsymbol{p}_{M}) & 1 \\
    1 & ... & 1 & 0
    \end{bmatrix}
    \begin{bmatrix}
        l_1 \\
        l_2 \\ 
        \vdots \\
        l_M \\
        L'
    \end{bmatrix}
    = \begin{bmatrix}
        \gamma(\boldsymbol{p}_{0}, \boldsymbol{p}_{1}) \\
        \gamma(\boldsymbol{p}_{0}, \boldsymbol{p}_{2}) \\
        \vdots \\
        \gamma(\boldsymbol{p}_{0}, \boldsymbol{p}_{M}) \\
        1
    \end{bmatrix},
    \label{eq:linear_matrix}
\end{align}
where $\gamma(\cdot, \cdot)$ indicates the semi-variogram of the Kriging interpolation for two locations of the UAV. The semi-variogram of the Kriging interpolation scheme can be expressed as
\begin{equation}
    \gamma(\boldsymbol{p}_{i},   \boldsymbol{p}_{j}) = v_{\delta_{K_j}}^{2} \Big(1-\Phi \big(\Delta_{\mathrm{2D}}(\boldsymbol{p}_i, \boldsymbol{p}_j)\big)\Big),
    \label{eq:semi_variogram}
\end{equation}
where $\Phi(\Delta_{\mathrm{2D}}(\boldsymbol{p}_i, \boldsymbol{p}_j))$ is as defined in (\ref{eq:correlation_cases}) and (\ref{eq:bi_exponential}), and $v_{\delta_{K_j}}^{2}$ is the variance of the channel rank calculated with a threshold at a location with all altitudes. This can be expressed as 
\begin{equation}
    v_{\delta_{K_j}}^{2} = \frac{1}{N_{h}-1}\sum_{s=1}^{N_{h}}|R_{\delta_{K_j}}(x, y, h_s) - \mu|^2,
\end{equation}
where $\mu$ is the mean of $R_{\delta_{K_j}}(x, y, h_{s})$ over all $h_{s}$ for $s=1, ..., N_{h}$. The semi-variogram indicates the spatial correlation in different locations. The expression above is derived with the assumption of the existence of the covariance function of a stationary process. 

\begin{figure}
     \centering
     \includegraphics[trim={0.15cm 0.15cm 0.8cm 0.4cm},clip,width=0.95\columnwidth]{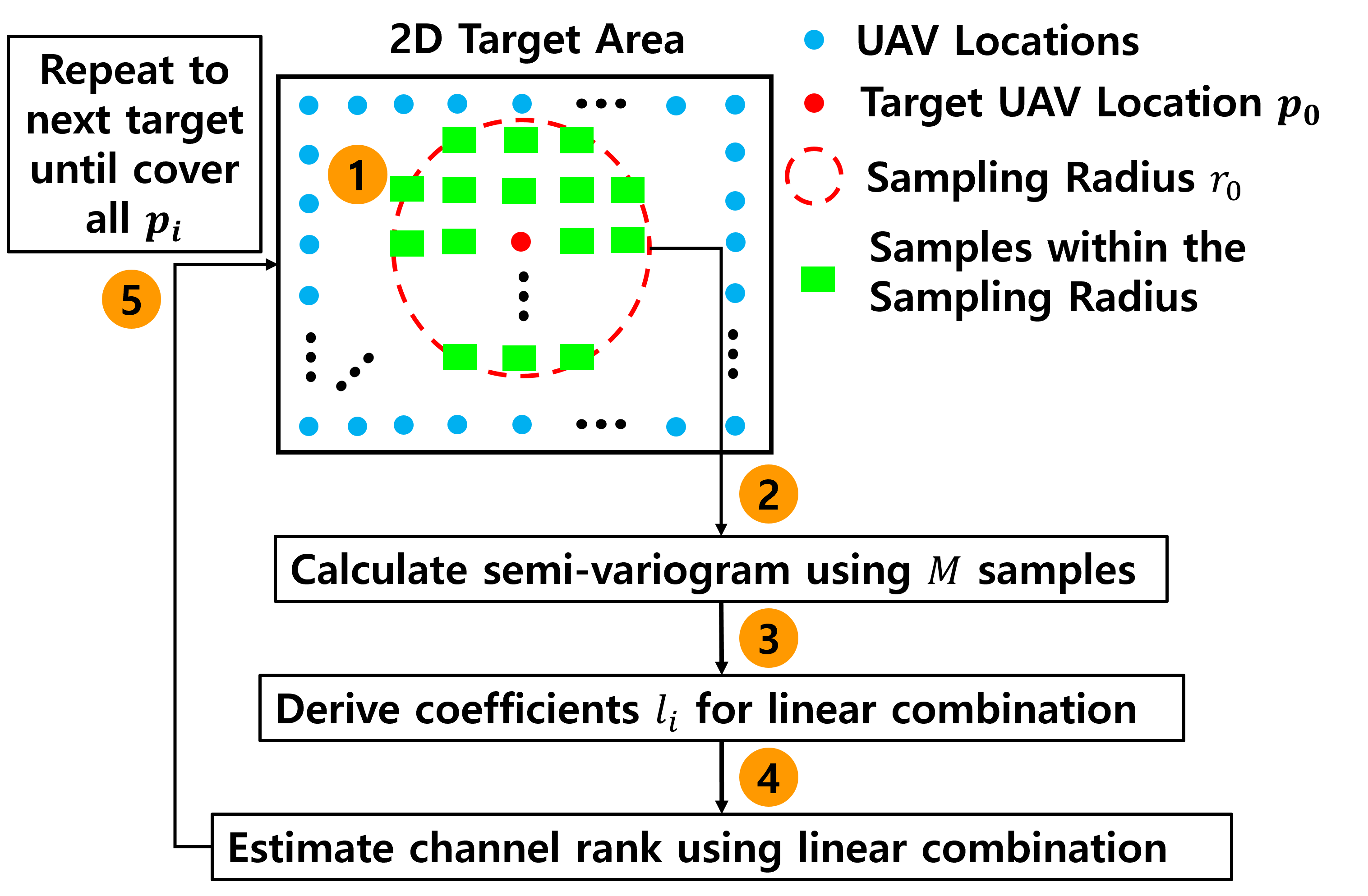}
     \caption{Procedure of the Kriging interpolation-based 3D channel rank interpolation scheme.}
     \label{fig:Kriging_description}
 \end{figure}

The procedure for the Kriging interpolation-based 3D channel rank interpolation scheme, which is also highlighted in Figure \ref{fig:Kriging_description}, can be summarized as follows:
\begin{itemize} 
 \item \textbf{Step-1:} Sample $M$ channel rank data within a radius of $r_{0}$ from the target UAV position, $\boldsymbol{p}_{0}$.
 
 \item \textbf{Step-2:} Calculate semi-variogram defined in equation (\ref{eq:semi_variogram}).

 \item \textbf{Step-3:} Derive coefficient for the linear combination, $l_i$, by substituting semi-variogram into linear matrix equation in (\ref{eq:linear_matrix}).

 \item \textbf{Step-4:} Process interpolation in equation (\ref{eq:Kriging_estimation}) with derived $l_i$ and channel rank for the target UAV position.

 \item \textbf{Step-5:} Repeat all the previous steps for all possible UAV locations.
\end{itemize} 

 \subsection{Other Baseline Interpolation Techniques}\label{ch:other_baseline}
We consider spline and makima (modified Akima cubic Hermite) interpolation methods as baseline interpolation approaches for comparison purposes. In the baseline interpolation approaches, the UAV position index as described right after (\ref{eq:channel_rank_vector}), $i=1, ..., N_{\mathrm{loc}}$, is used to identify UAV positions within a 2D target area that is discretized into a uniform grid with spacing $d_{\mathrm{rx}}$. 

The procedure for the baseline interpolation-based 3D channel rank interpolation can be summarized as follows: 1) Sample $M$ UAV positions within a radius of $r_{0}$ from the target UAV position $\boldsymbol{p}_0$, 2) sort the UAV position index and channel rank with respect to the index, 3) perform interpolation using sorted index and channel rank, i.e., $\hat{R}_{\delta_{K_j}}(\boldsymbol{p}_{0})=\mathsf{B}(\boldsymbol{X}, R_{\delta_{K_j}}(\boldsymbol{p}_{\boldsymbol{X}}), \boldsymbol{p}_{0})$, where $\mathsf{B}()$ indicates the baseline interpolation function with spline or makima approaches, $\boldsymbol{X}$ represents all the sampled UAV positions index within $r_0$ radius from the target point, and $R_{\delta_{K_j}}(\boldsymbol{p}_{\boldsymbol{X}})$ is the channel rank data of the corresponding locations $\boldsymbol{X}$, respectively.  

\section{Numerical Results}\label{ch:numerical_results}
In this section, we present numerical results on RSS, channel rank, and Kriging-based channel rank interpolation schemes for different UAV scenarios in a rural environment, in Section~\ref{ch:RF_coverage}, Section~\ref{ch:channel_rank}, and Section~\ref{ch:Kriging_interpolation}, respectively. The RT results are also compared with the real measurements in the NSF AERPAW testbed in Section~VII.\ref{ch:measurement}. In addition to the simulation parameters listed in Section~\ref{ch:RT_setup}, the following assumptions are used for the corresponding analysis: 1) the number of threshold ratio constant $N_{K} = 3$, where $K_1=10^1, K_2=10^2$, and $K_3=10^3$ leading $\delta_{K_1}=\sigma_1(\boldsymbol{p})/10^1$, $\delta_{K_2}=\sigma_1(\boldsymbol{p})/10^2$, and $\delta_{K_3}=\sigma_1(\boldsymbol{p})/10^3$ to capture singular values in $10$~dB, $20$~dB and $30$~dB range from the strongest singular value, respectively; 2) the horizontal interval between UAV positions $d_{\mathrm{rx}}$ is set to $30$~m, which leads to the total number of possible UAV locations $N_{\mathrm{loc}}= 36 \times 71 = 2556$ for the given 2D target area $T_{\mathrm{LW}}$ in Table~\ref{tab:sim_param}; 3) the number of altitudes of UAV $N_{h} = 9$ with $10$~m interval from $30$~m to $110$~m; 4) the altitudes of interest in RSS analysis is $30$~m, $70$~m, and $110$~m; and 5) in addition to the altitudes of interest in RSS simulation, $3$~m altitude configuration is also applied to the channel rank RT simulation to highlight the blockage effects.

 \subsection{Received Signal Strength and Coverage Analysis}\label{ch:RF_coverage}

 \begin{figure*}[t!]
    \centering    
    \subfigure[LW1 $30$~m]{
    \includegraphics[trim={0.1cm 1.0cm 0.15cm 1.7cm},clip,width=0.37\columnwidth]{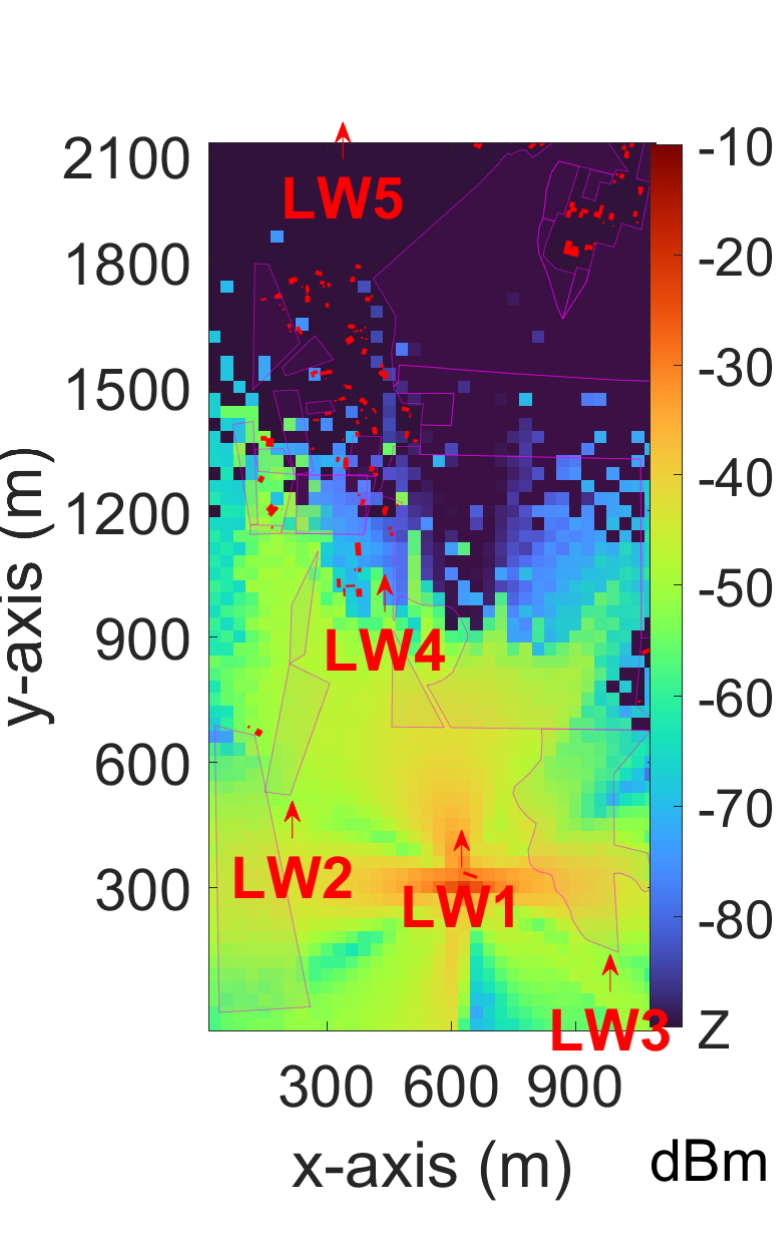}
    \label{fig:RSSI_MIMO_30_LW1}
    }
    \subfigure[LW2 $30$~m]{
    \includegraphics[trim={0.1cm 1.0cm 0.15cm 1.7cm},clip,width=0.37\columnwidth]{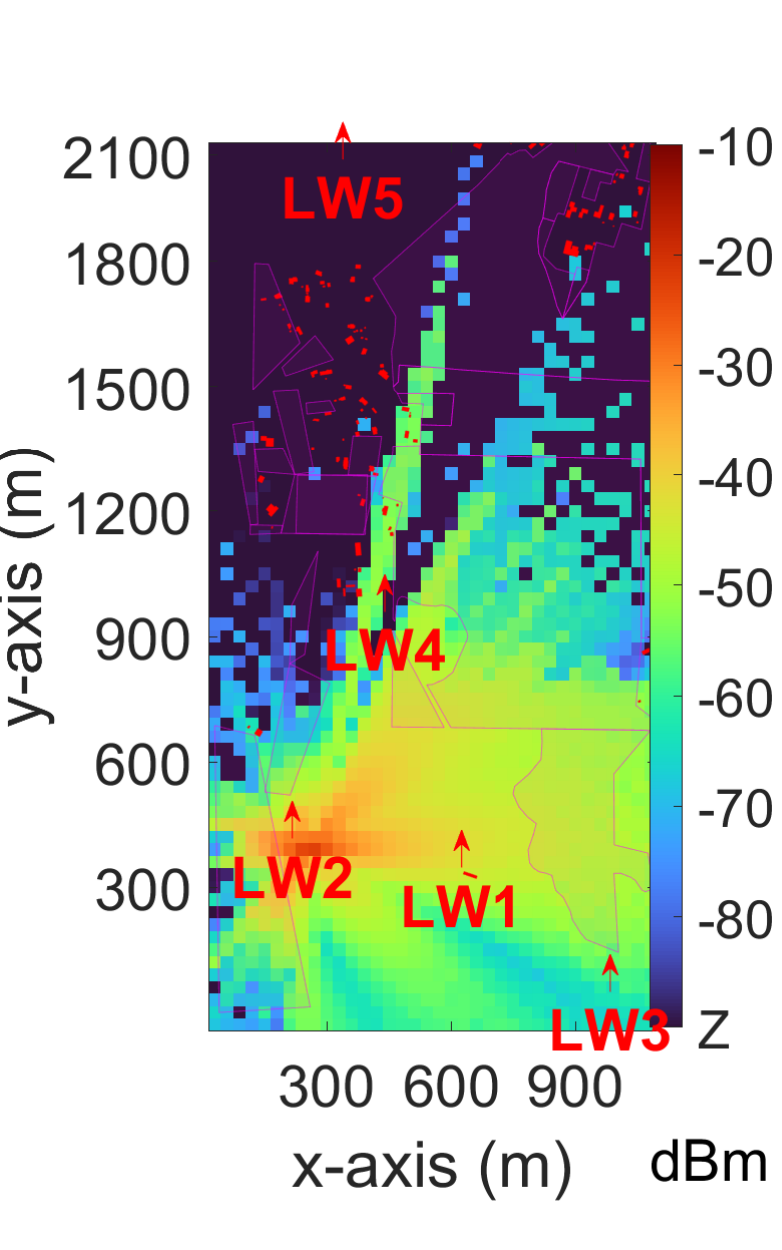}
    \label{fig:RSSI_MIMO_30_LW2}
    }
    \subfigure[LW3 $30$~m]{
    \includegraphics[trim={0.1cm 1.0cm 0.15cm 1.7cm},clip,width=0.37\columnwidth]{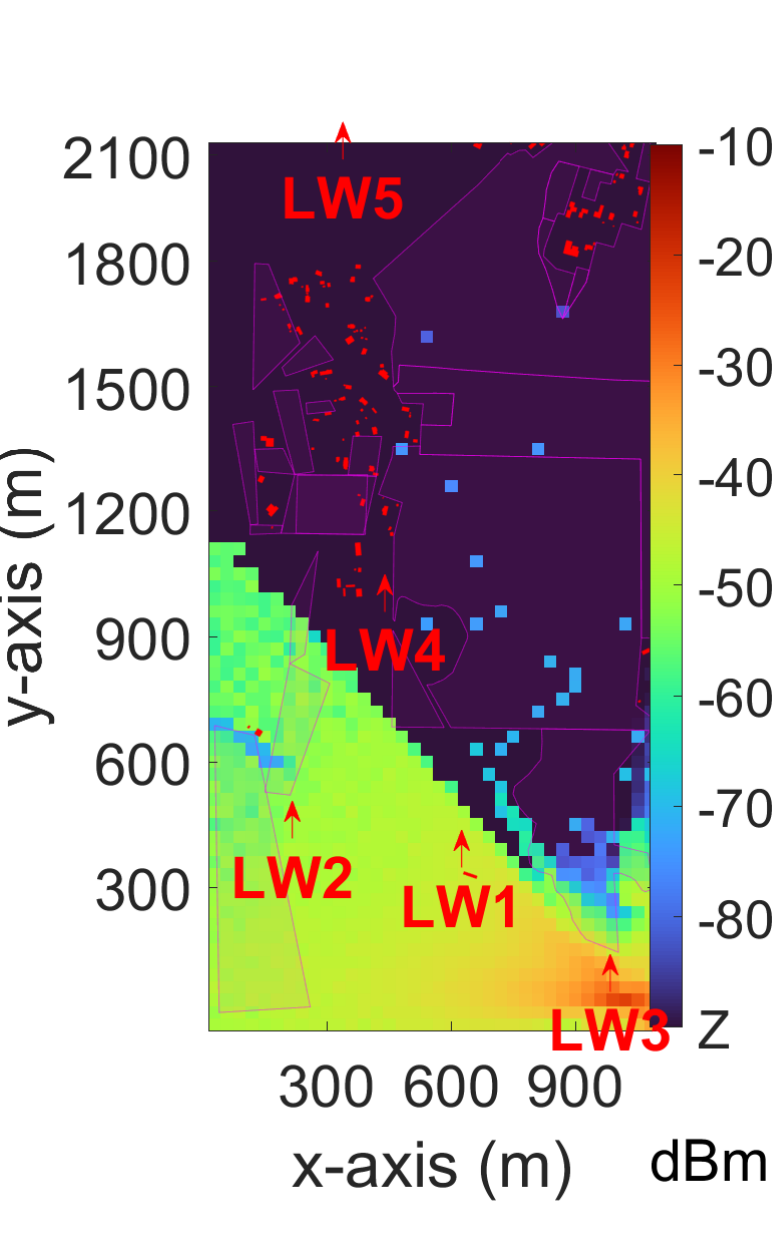}
    \label{fig:RSSI_MIMO_30_LW3}
    }
    \subfigure[LW4 $30$~m]{
    \includegraphics[trim={0.1cm 1.0cm 0.15cm 1.7cm},clip,width=0.37\columnwidth]{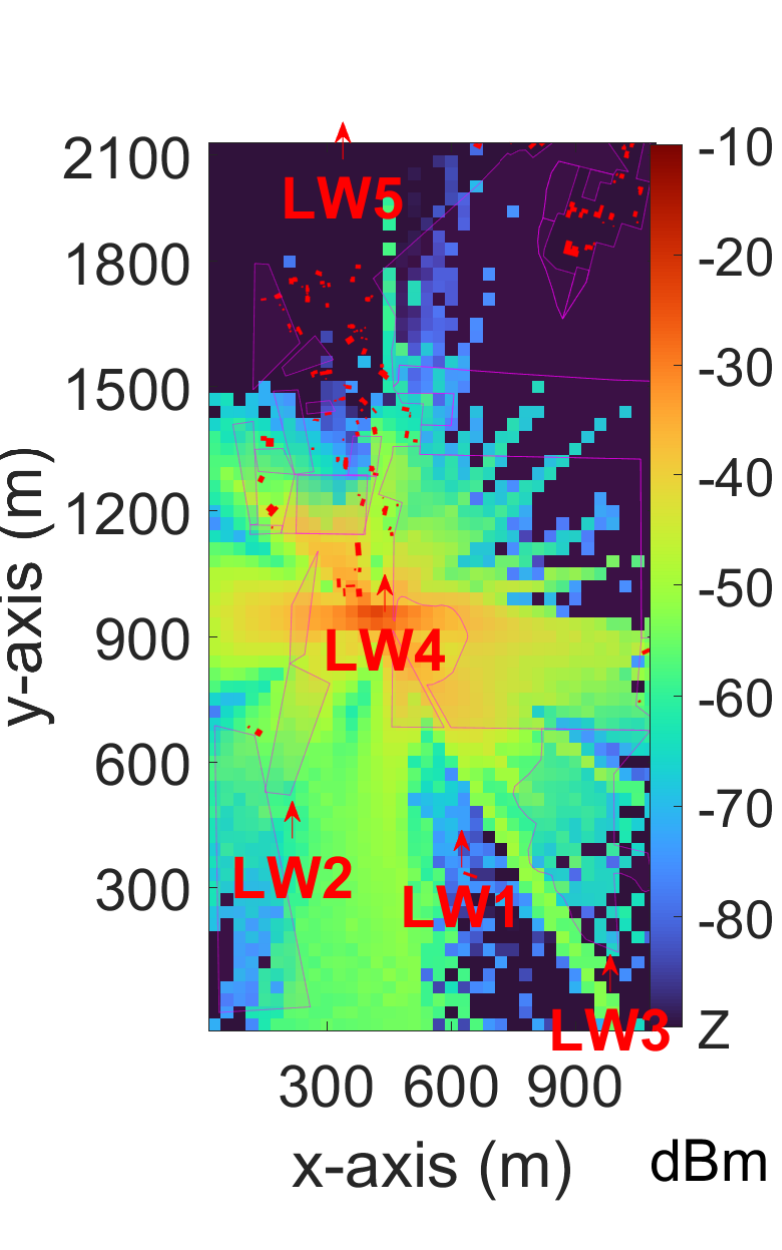}
    \label{fig:RSSI_MIMO_30_LW4}
    }
    \subfigure[LW5 $30$~m]{
    \includegraphics[trim={0.1cm 1.0cm 0.15cm 1.7cm},clip,width=0.37\columnwidth]{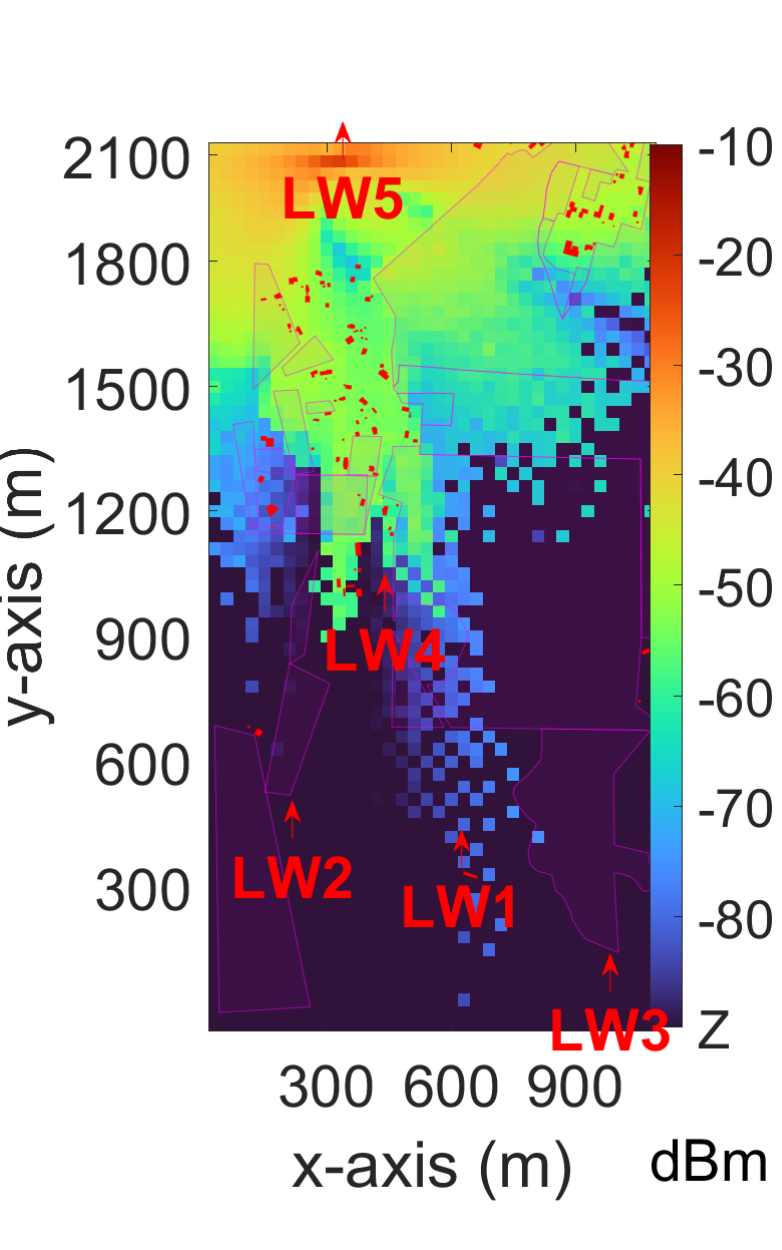}
    \label{fig:RSSI_MIMO_30_LW5}
    }    
    \subfigure[LW1 $70$~m]{
    \includegraphics[trim={0.1cm 1.0cm 0.15cm 1.7cm},clip,width=0.37\columnwidth]{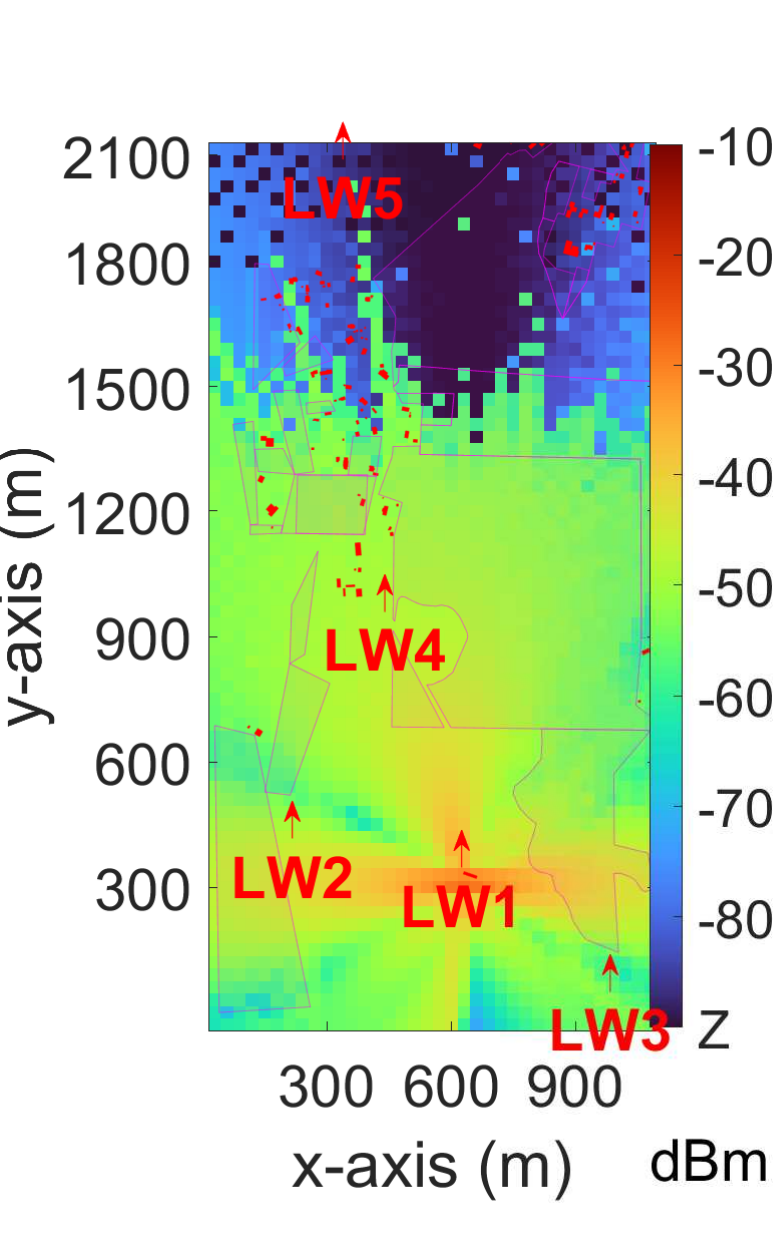}
    \label{fig:RSSI_MIMO_70_LW1}
    }
    \subfigure[LW2 $70$~m]{
    \includegraphics[trim={0.1cm 1.0cm 0.15cm 1.7cm},clip,width=0.37\columnwidth]{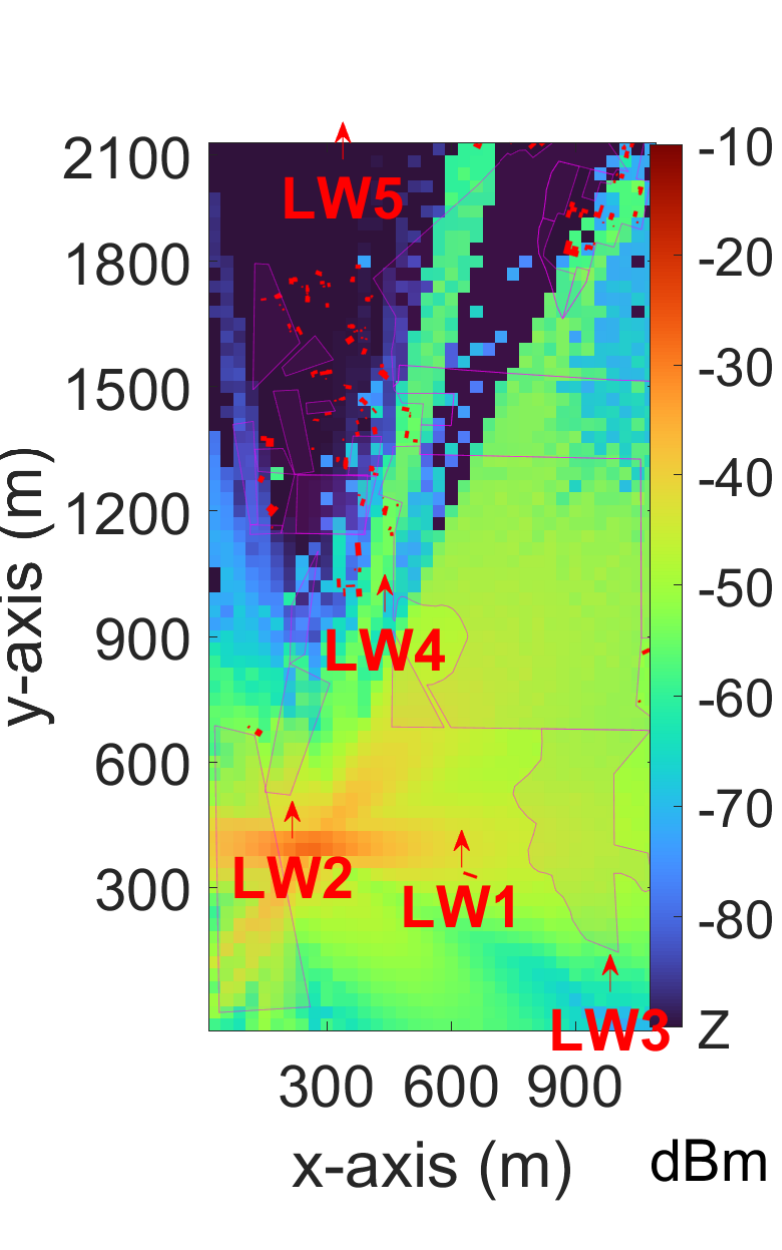}
    \label{fig:RSSI_MIMO_70_LW2}
    }
    \subfigure[LW3 $70$~m]{
    \includegraphics[trim={0.1cm 1.0cm 0.15cm 1.7cm},clip,width=0.37\columnwidth]{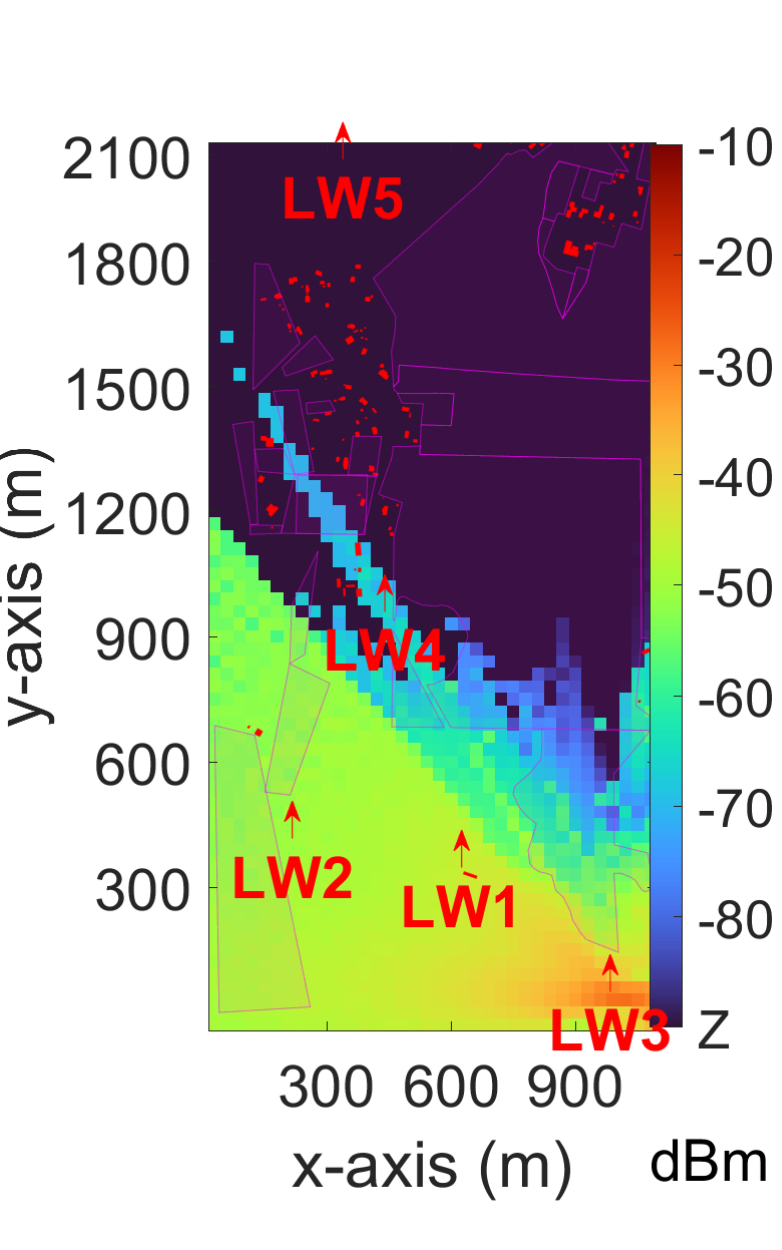}
    \label{fig:RSSI_MIMO_70_LW3}
    }
    \subfigure[LW4 $70$~m]{
    \includegraphics[trim={0.1cm 1.0cm 0.15cm 1.7cm},clip,width=0.37\columnwidth]{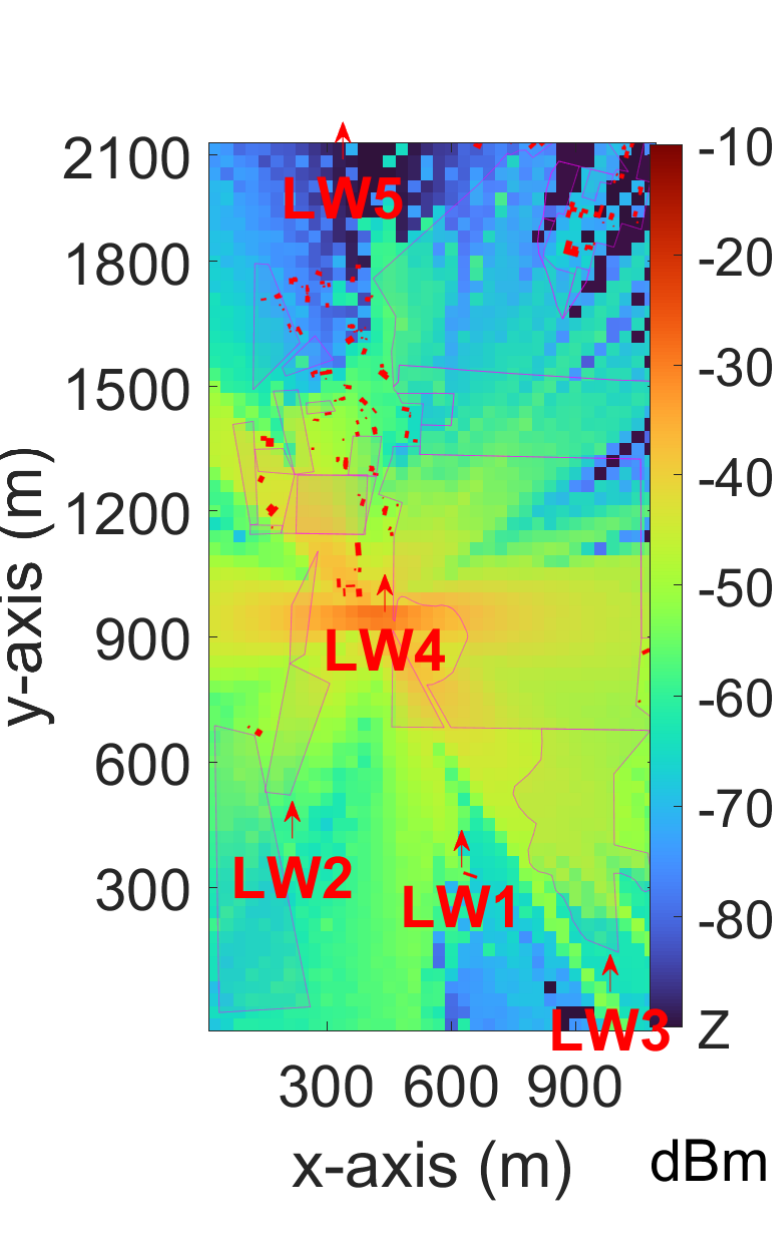}
    \label{fig:RSSI_MIMO_70_LW4}
    }
    \subfigure[LW5 $70$~m]{
    \includegraphics[trim={0.1cm 1.0cm 0.15cm 1.7cm},clip,width=0.37\columnwidth]{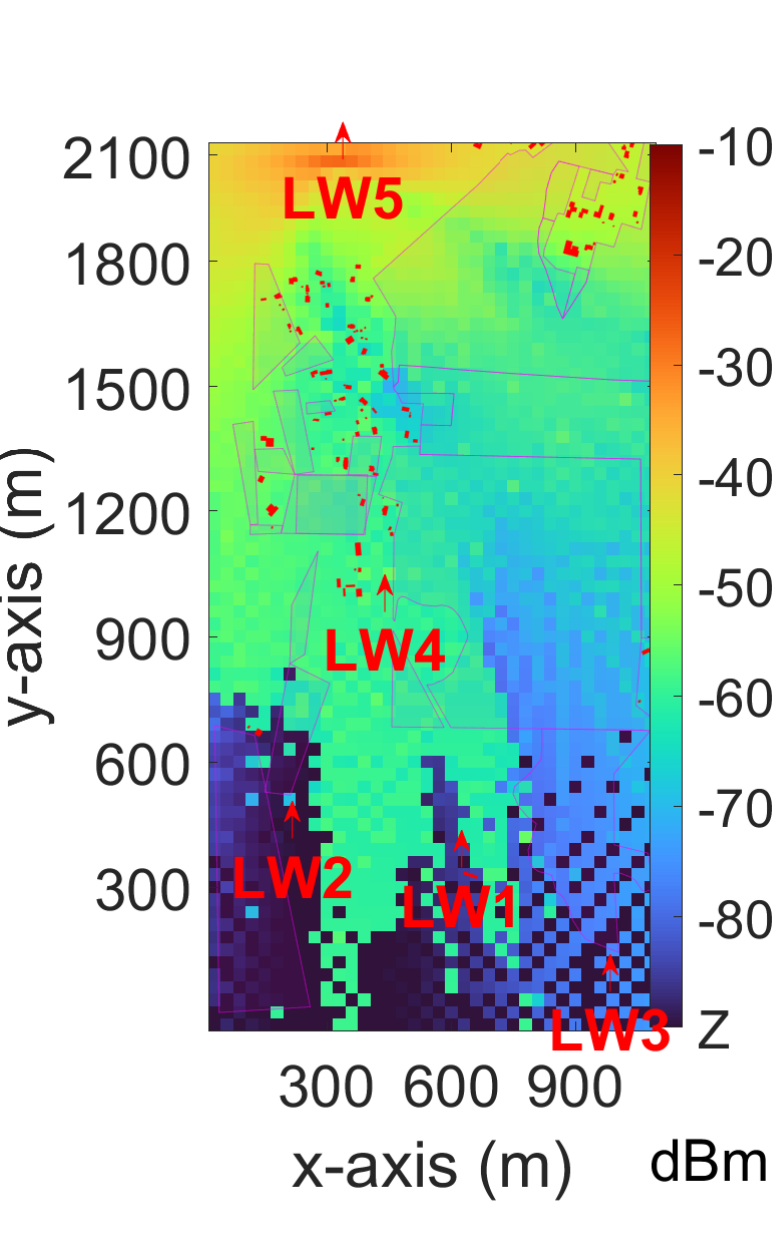}
    \label{fig:RSSI_MIMO_70_LW5}
    }    
    \subfigure[LW1 $110$~m]{
    \includegraphics[trim={0.1cm 1.0cm 0.15cm 1.7cm},clip,width=0.37\columnwidth]{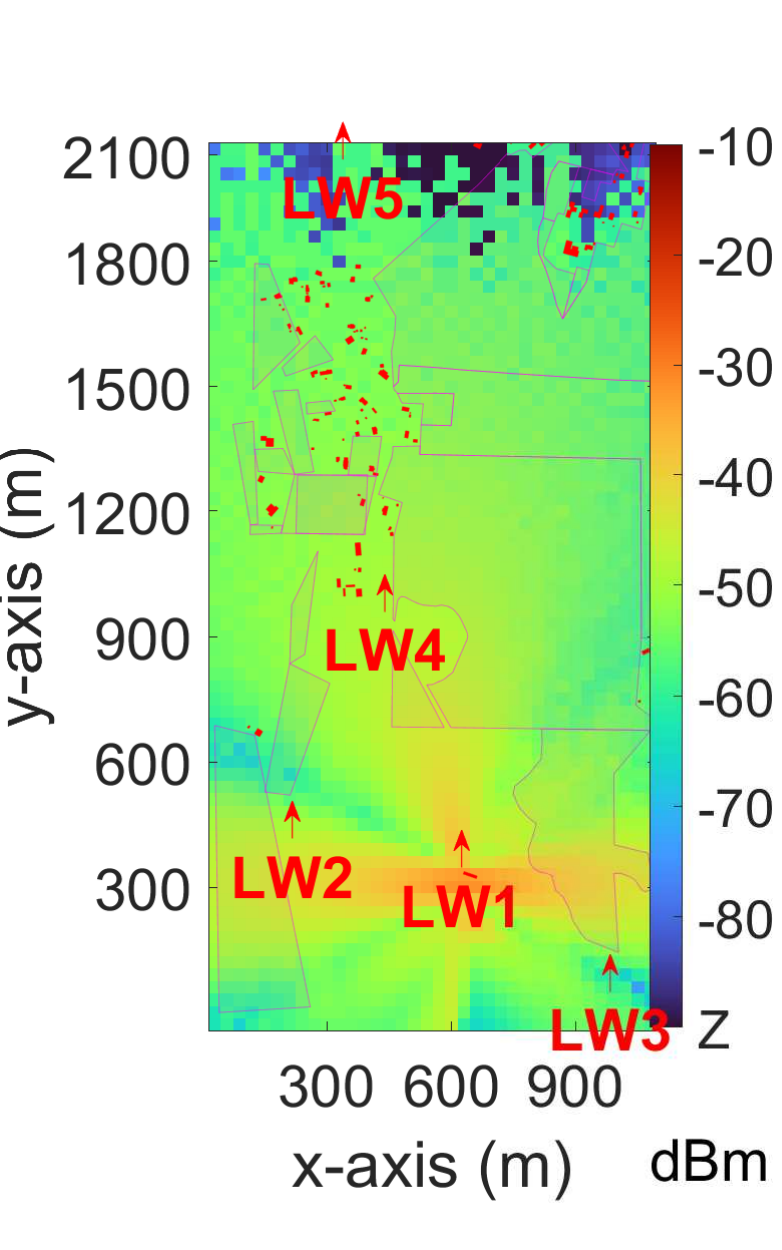}
    \label{fig:RSSI_MIMO_110_LW1}
    }
    \subfigure[LW2 $110$~m]{
    \includegraphics[trim={0.1cm 1.0cm 0.15cm 1.7cm},clip,width=0.37\columnwidth]{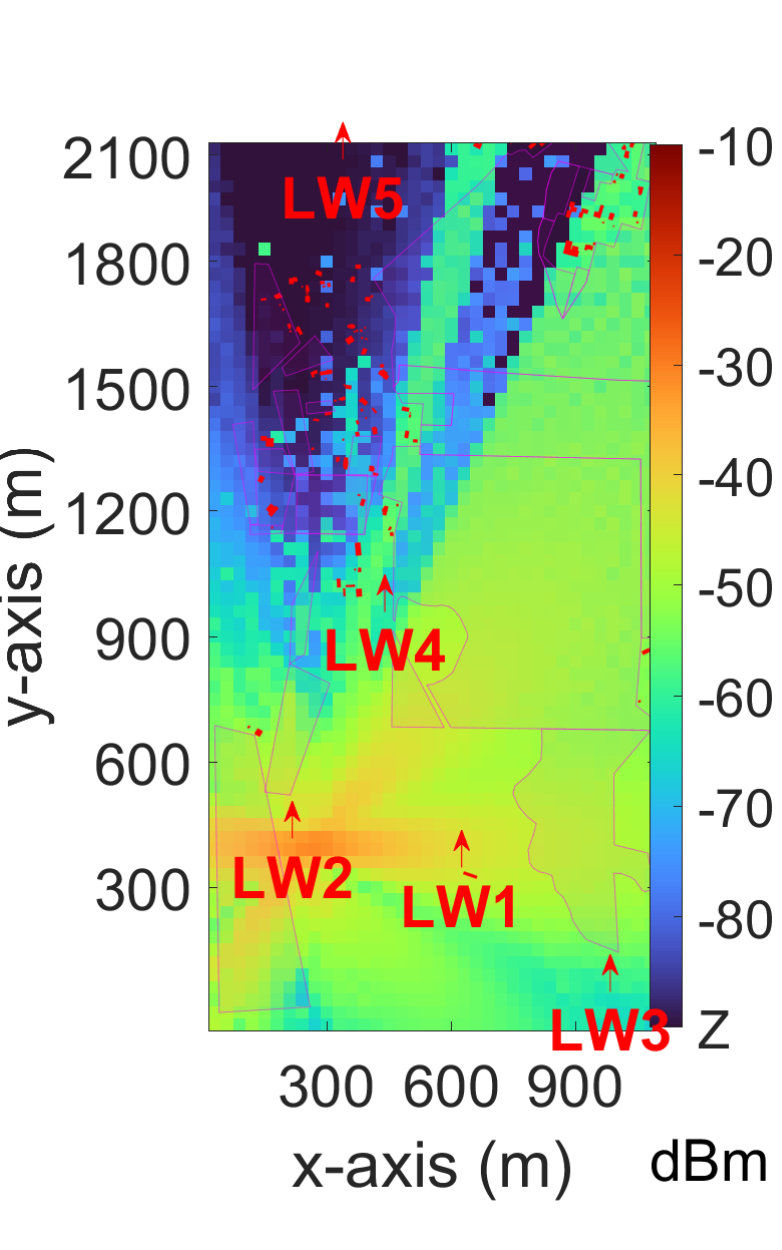}
    \label{fig:RSSI_MIMO_110_LW2}
    }
    \subfigure[LW3 $110$~m]{
    \includegraphics[trim={0.1cm 1.0cm 0.15cm 1.7cm},clip,width=0.37\columnwidth]{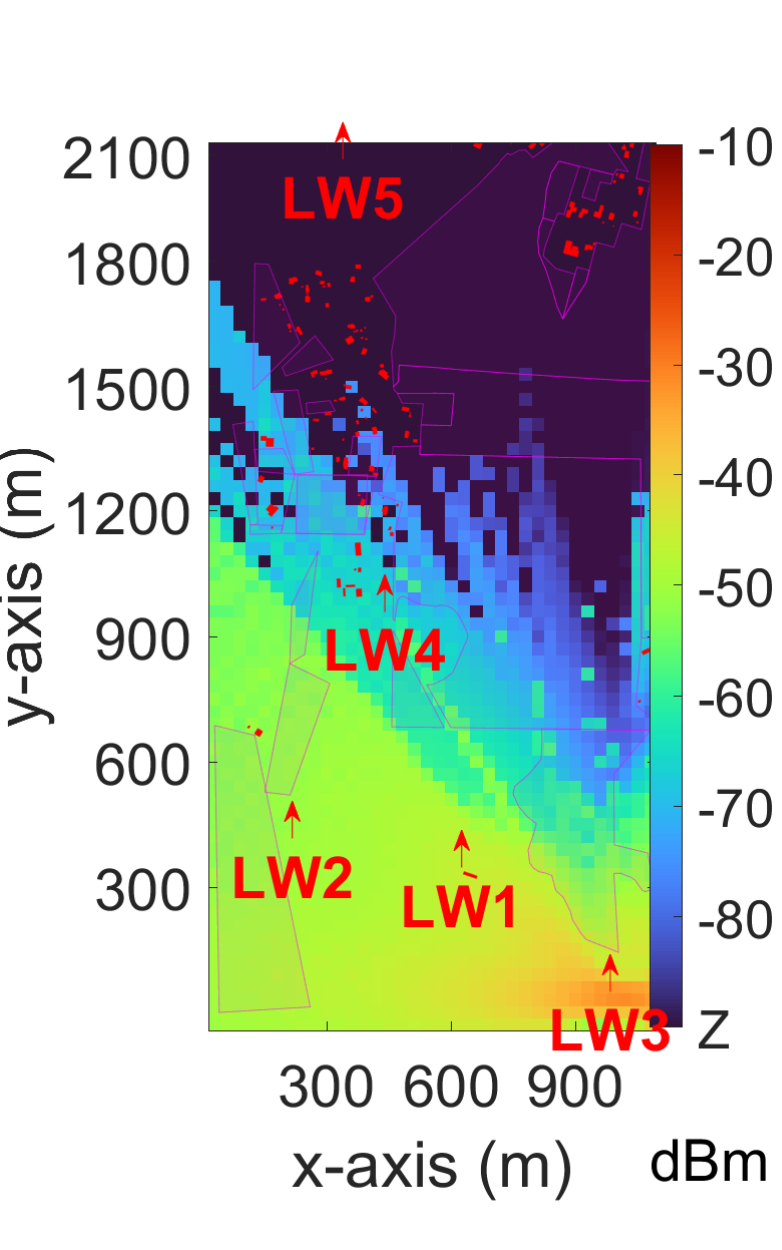}
    \label{fig:RSSI_MIMO_110_LW3}
    }
    \subfigure[LW4 $110$~m]{
    \includegraphics[trim={0.1cm 1.0cm 0.15cm 1.7cm},clip,width=0.37\columnwidth]{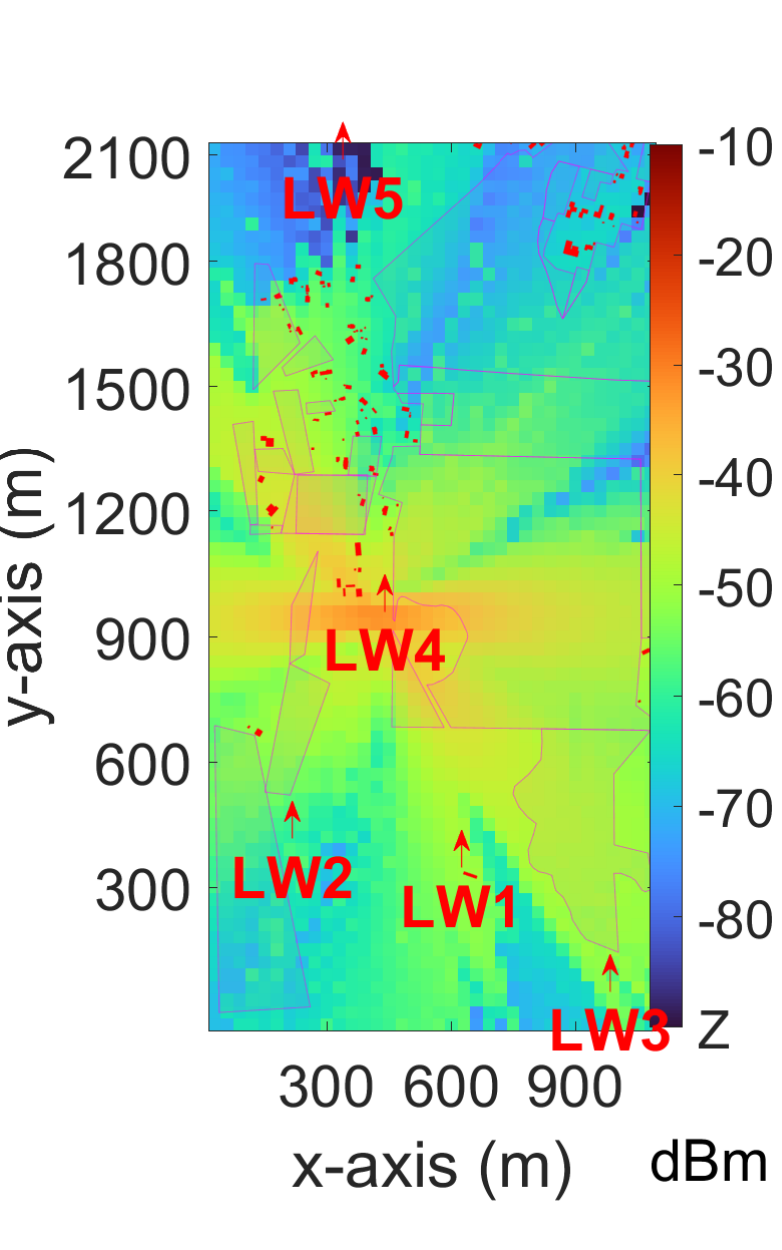}
    \label{fig:RSSI_MIMO_110_LW4}
    }
    \subfigure[LW5 $110$~m]{
    \includegraphics[trim={0.1cm 1.0cm 0.15cm 1.7cm},clip,width=0.37\columnwidth]{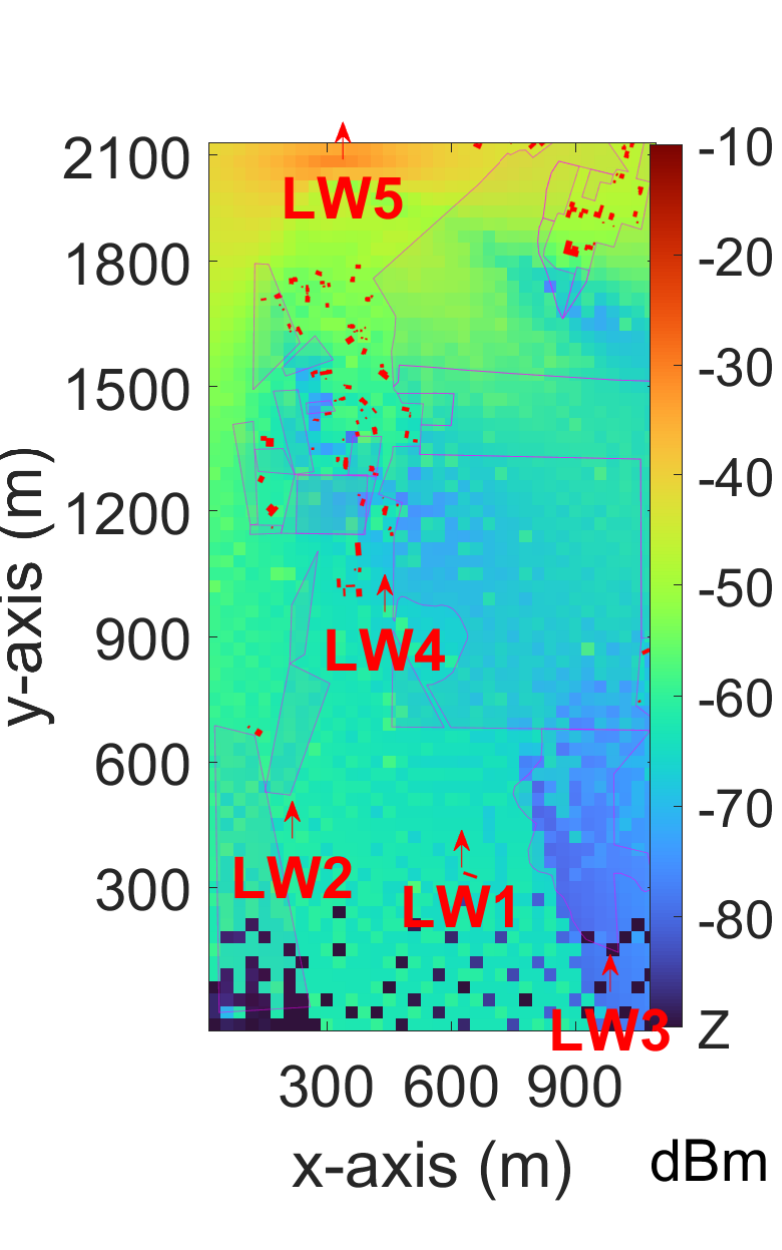}
    \label{fig:RSSI_MIMO_110_LW5}
    }    
   \caption{RSS in the Lake Wheeler Field Labs area with MIMO and various altitude configurations.}
    \label{fig:RSSI_MIMO_all_altitudes}
\end{figure*}

 \begin{figure*}[t!]
    \centering    
    \subfigure[LW1 $30$~m]{
    \includegraphics[trim={0.1cm 1.0cm 0.15cm 1.7cm},clip,width=0.37\columnwidth]{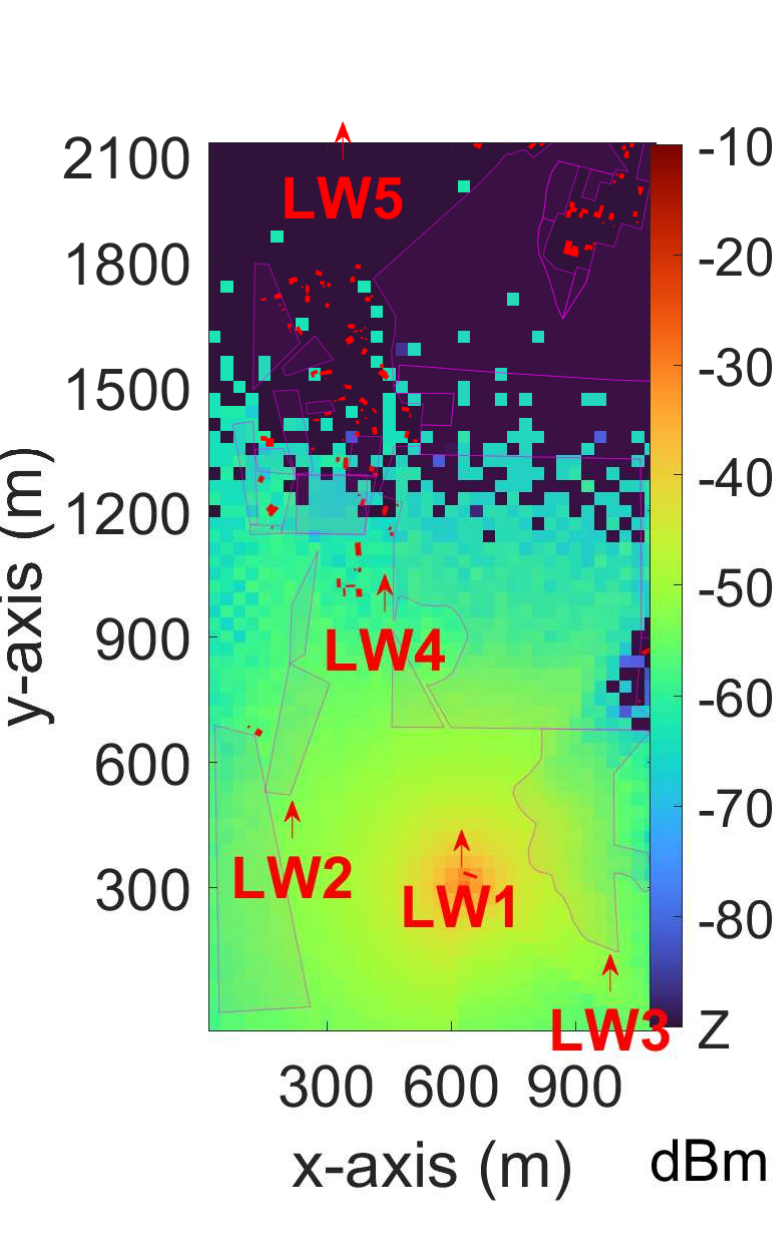}
    \label{fig:RSSI_SISO_30_LW1}
    }
    \subfigure[LW2 $30$~m]{
    \includegraphics[trim={0.1cm 1.0cm 0.15cm 1.7cm},clip,width=0.37\columnwidth]{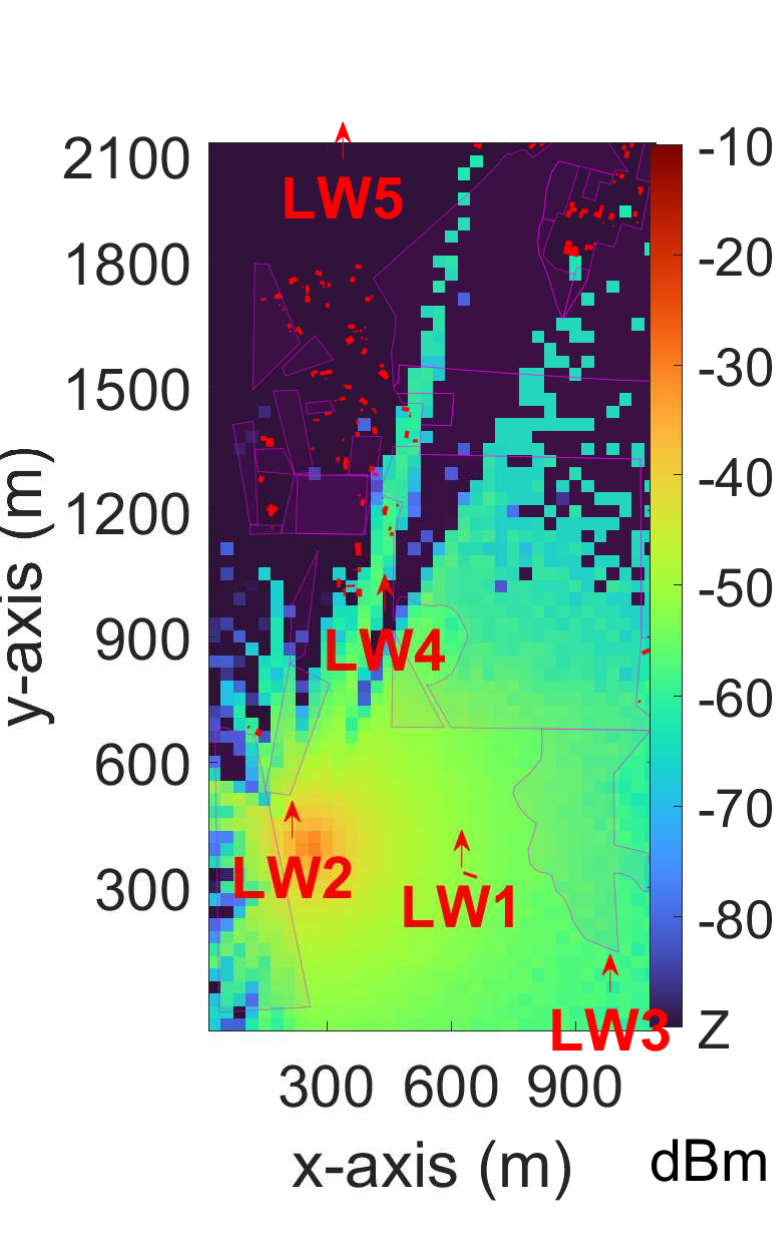}
    \label{fig:RSSI_SISO_30_LW2}
    }
    \subfigure[LW3 $30$~m]{
    \includegraphics[trim={0.1cm 1.0cm 0.15cm 1.7cm},clip,width=0.37\columnwidth]{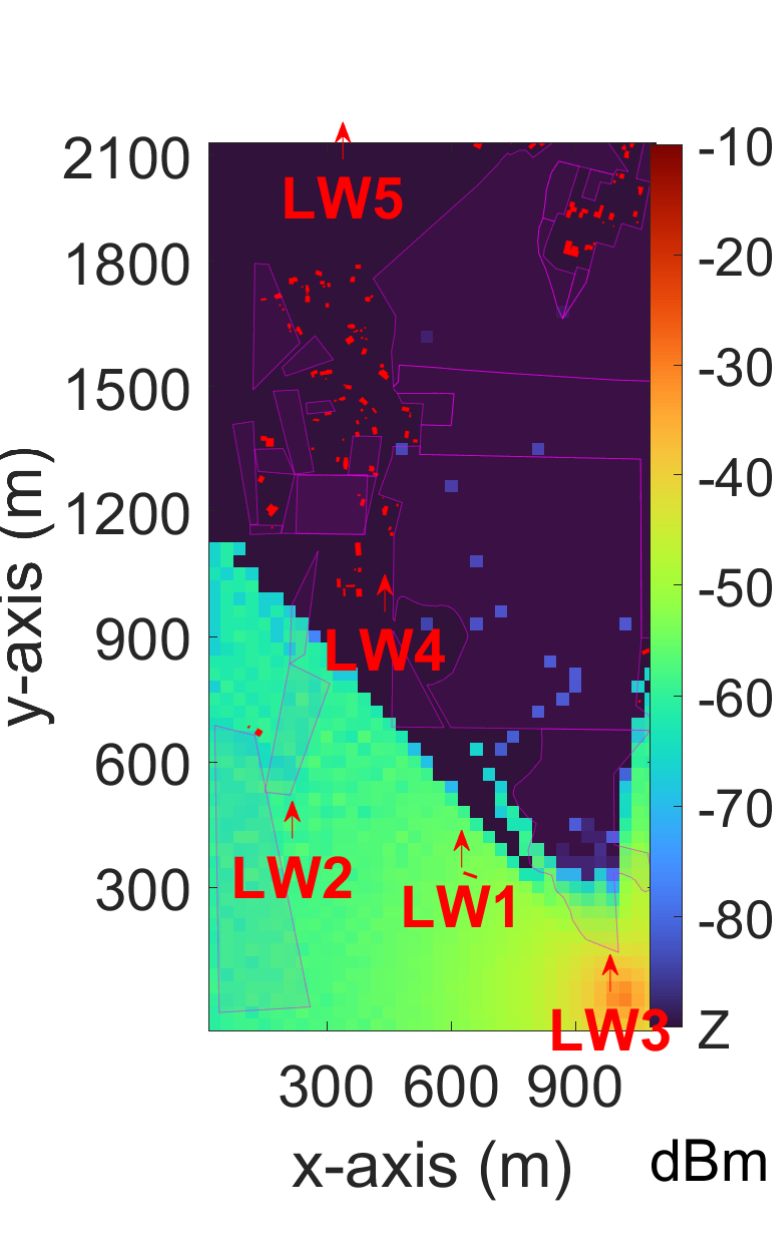}
    \label{fig:RSSI_SISO_30_LW3}
    }
    \subfigure[LW4 $30$~m]{
    \includegraphics[trim={0.1cm 1.0cm 0.15cm 1.7cm},clip,width=0.37\columnwidth]{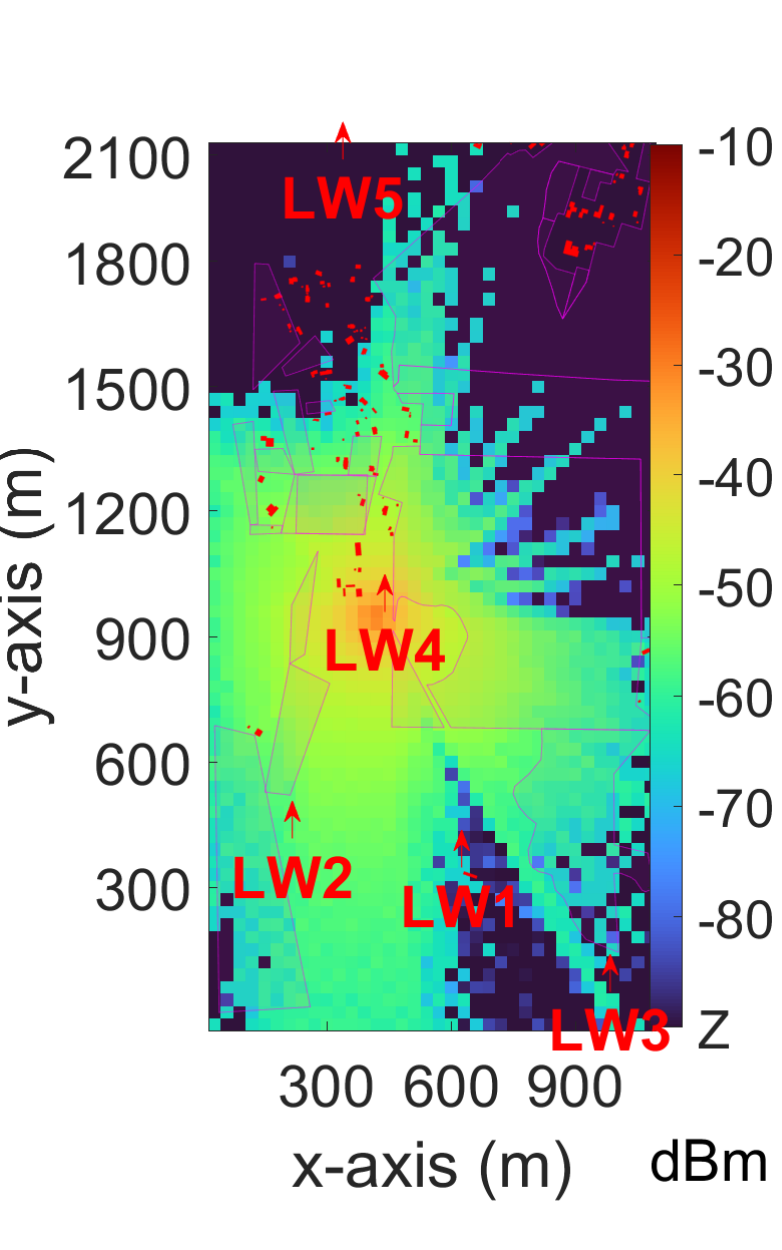}
    \label{fig:RSSI_SISO_30_LW4}
    }
    \subfigure[LW5 $30$~m]{
    \includegraphics[trim={0.1cm 1.0cm 0.15cm 1.7cm},clip,width=0.37\columnwidth]{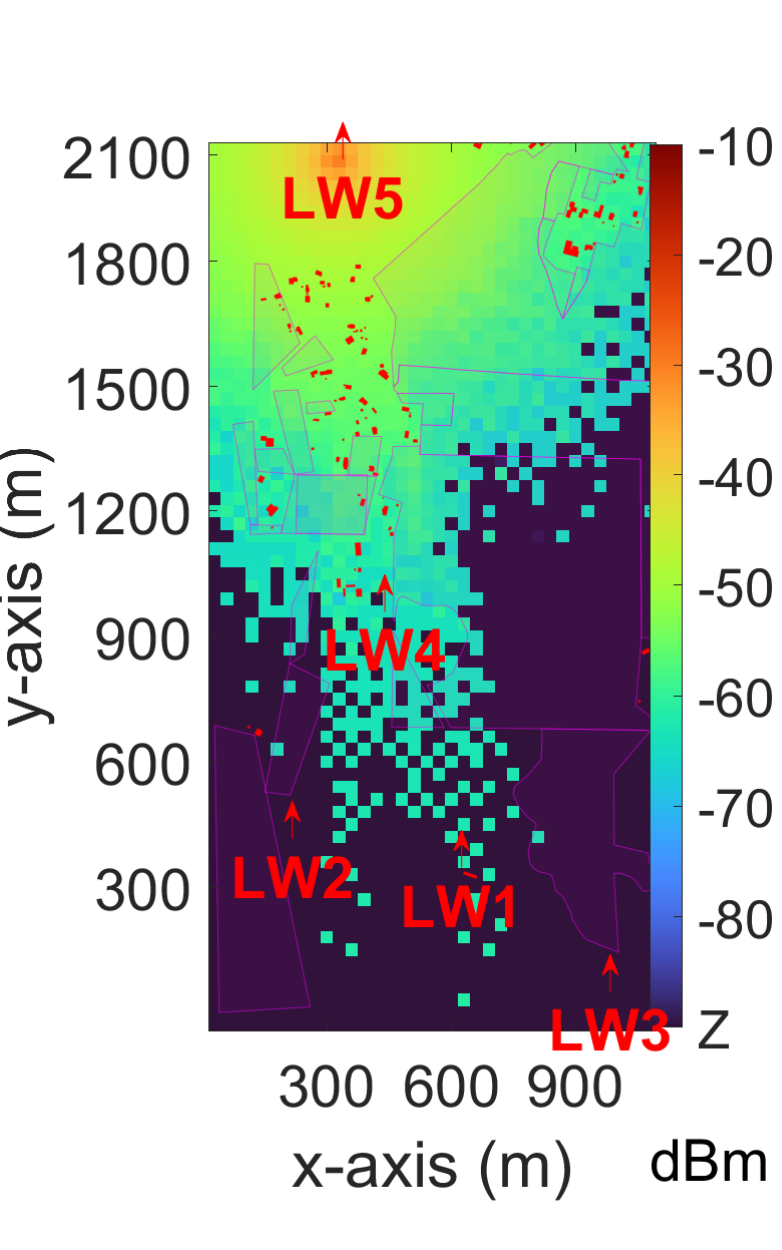}
    \label{fig:RSSI_SISO_30_LW5}
    }    
    \subfigure[LW1 $70$~m]{
    \includegraphics[trim={0.1cm 1.0cm 0.15cm 1.7cm},clip,width=0.37\columnwidth]{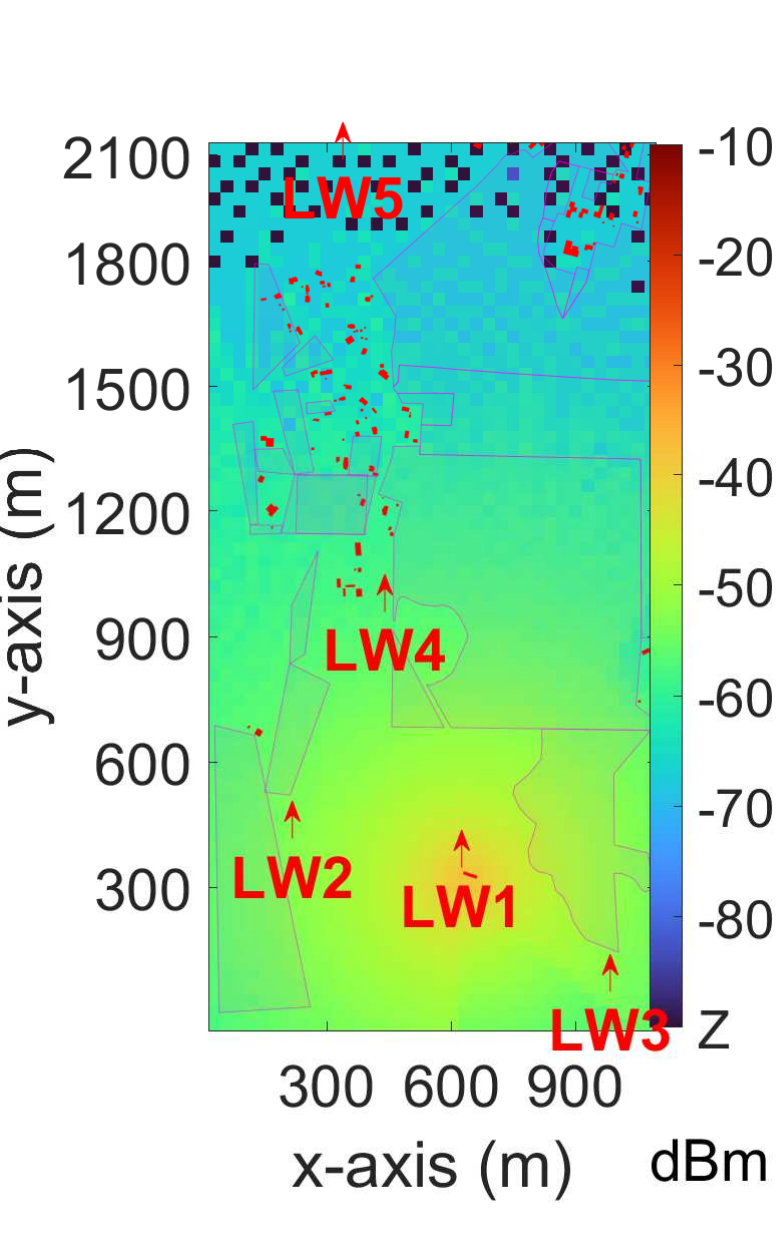}
    \label{fig:RSSI_SISO_70_LW1}
    }
    \subfigure[LW2 $70$~m]{
    \includegraphics[trim={0.1cm 1.0cm 0.15cm 1.7cm},clip,width=0.37\columnwidth]{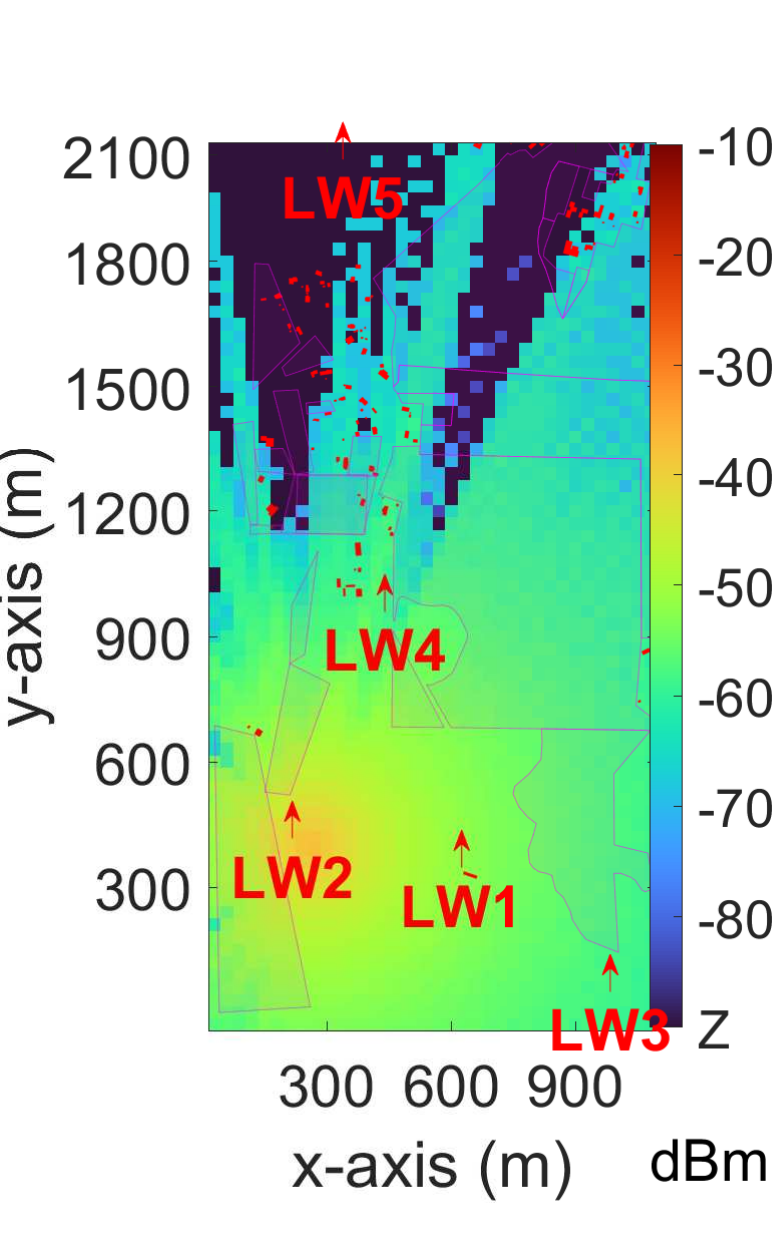}
    \label{fig:RSSI_SISO_70_LW2}
    }
    \subfigure[LW3 $70$~m]{
    \includegraphics[trim={0.1cm 1.0cm 0.15cm 1.7cm},clip,width=0.37\columnwidth]{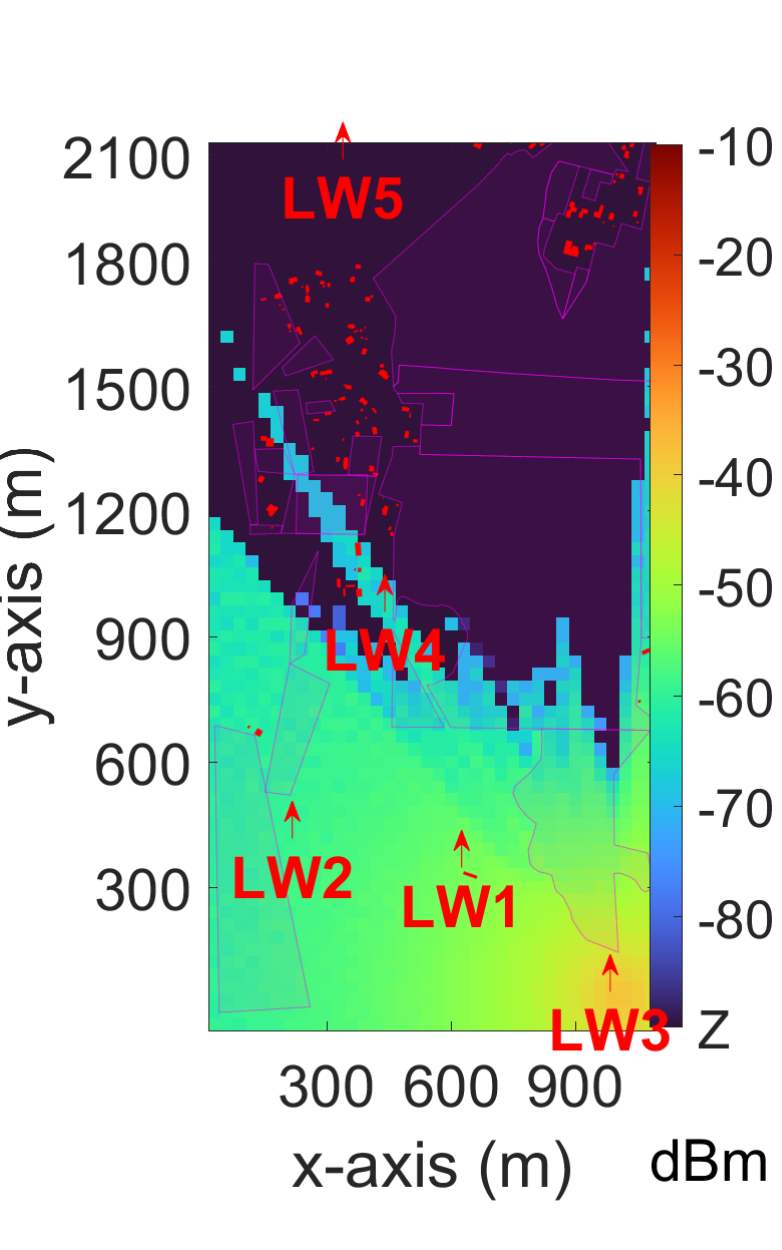}
    \label{fig:RSSI_SISO_70_LW3}
    }
    \subfigure[LW4 $70$~m]{
    \includegraphics[trim={0.1cm 1.0cm 0.15cm 1.7cm},clip,width=0.37\columnwidth]{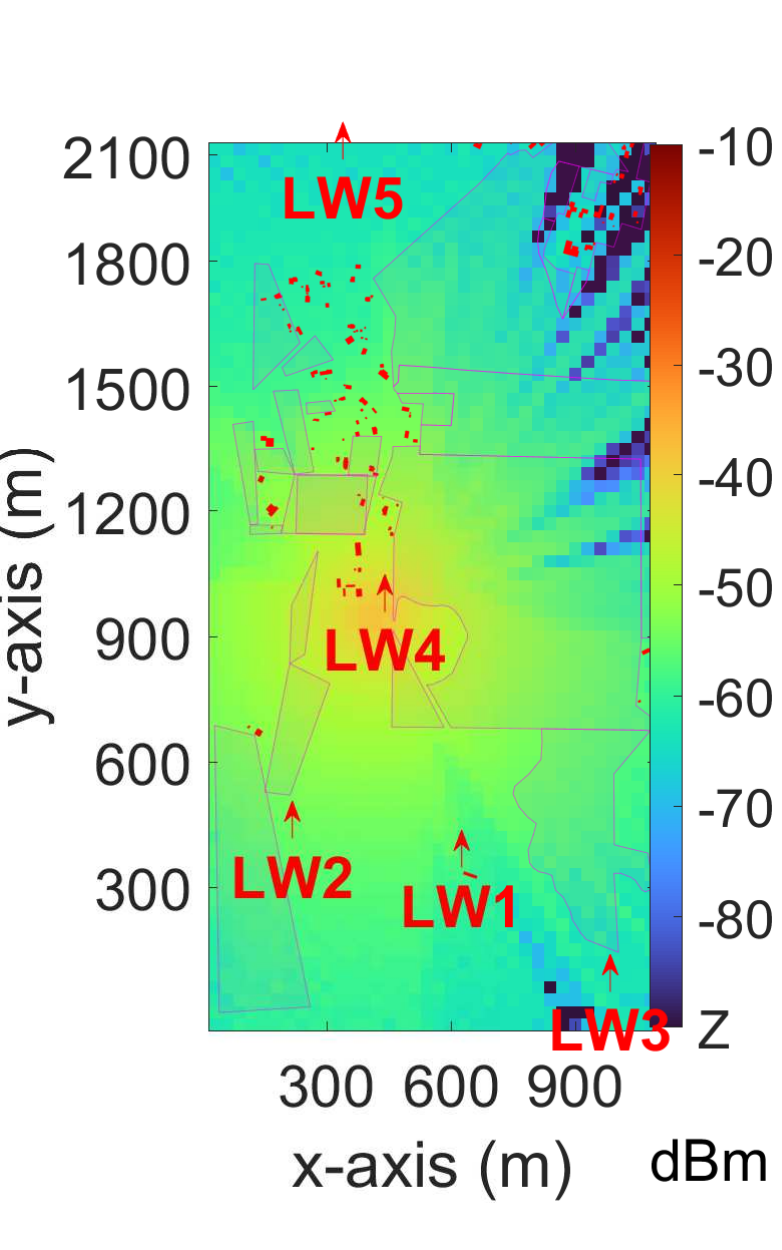}
    \label{fig:RSSI_SISO_70_LW4}
    }
    \subfigure[LW5 $70$~m]{
    \includegraphics[trim={0.1cm 1.0cm 0.15cm 1.7cm},clip,width=0.37\columnwidth]{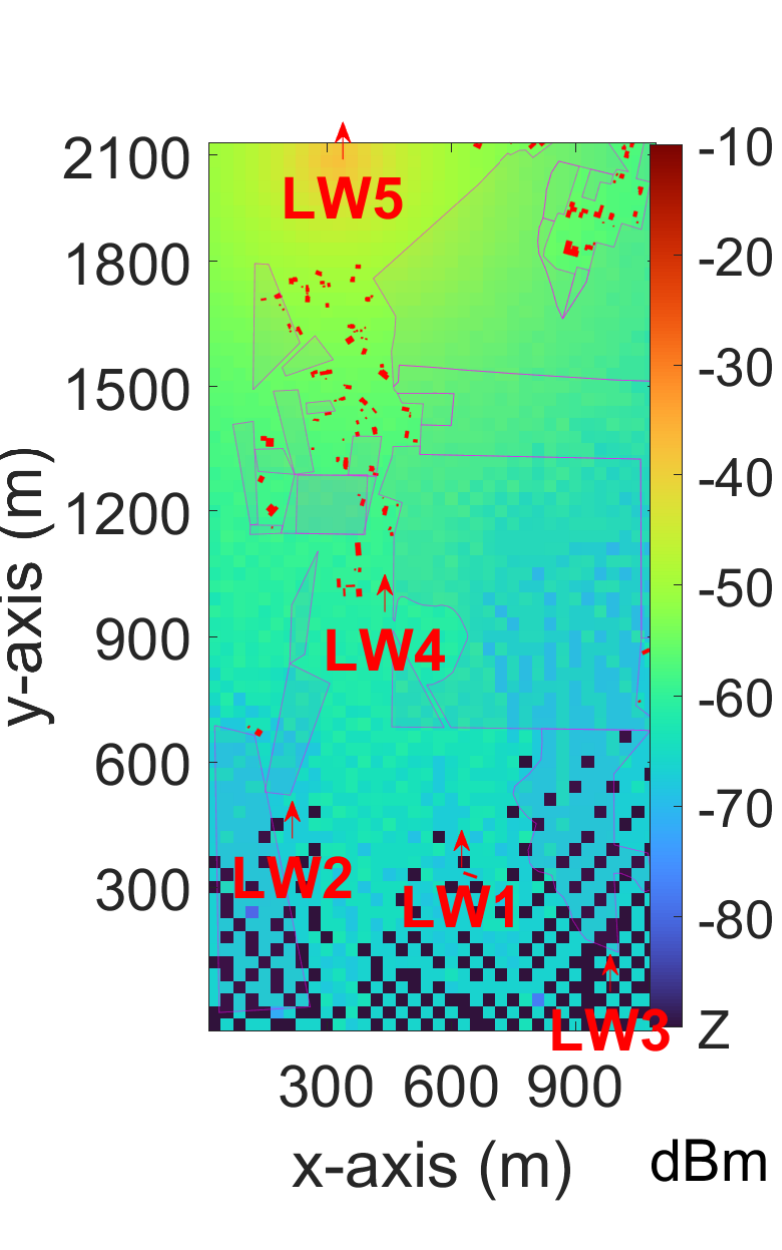}
    \label{fig:RSSI_SISO_70_LW5}
    }    
    \subfigure[LW1 $110$~m]{
    \includegraphics[trim={0.1cm 1.0cm 0.15cm 1.7cm},clip,width=0.37\columnwidth]{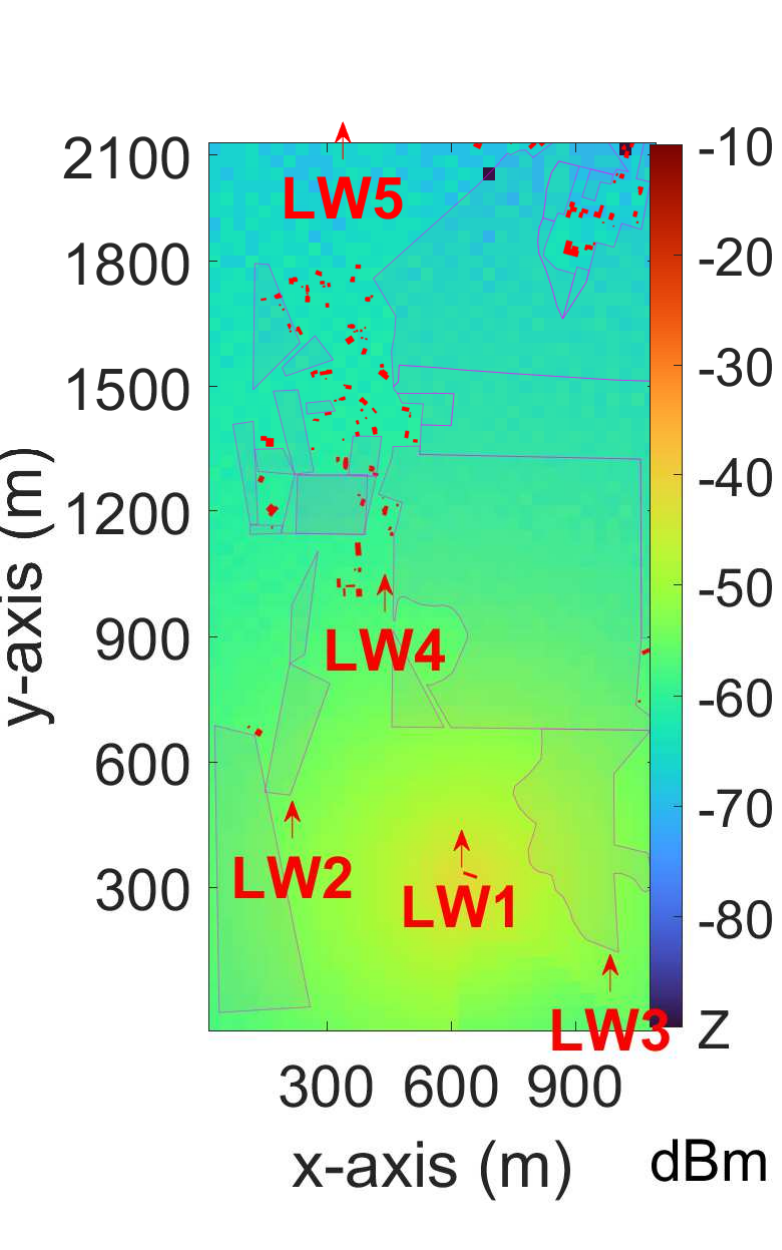}
    \label{fig:RSSI_SISO_110_LW1}
    }
    \subfigure[LW2 $110$~m]{
    \includegraphics[trim={0.1cm 1.0cm 0.15cm 1.7cm},clip,width=0.37\columnwidth]{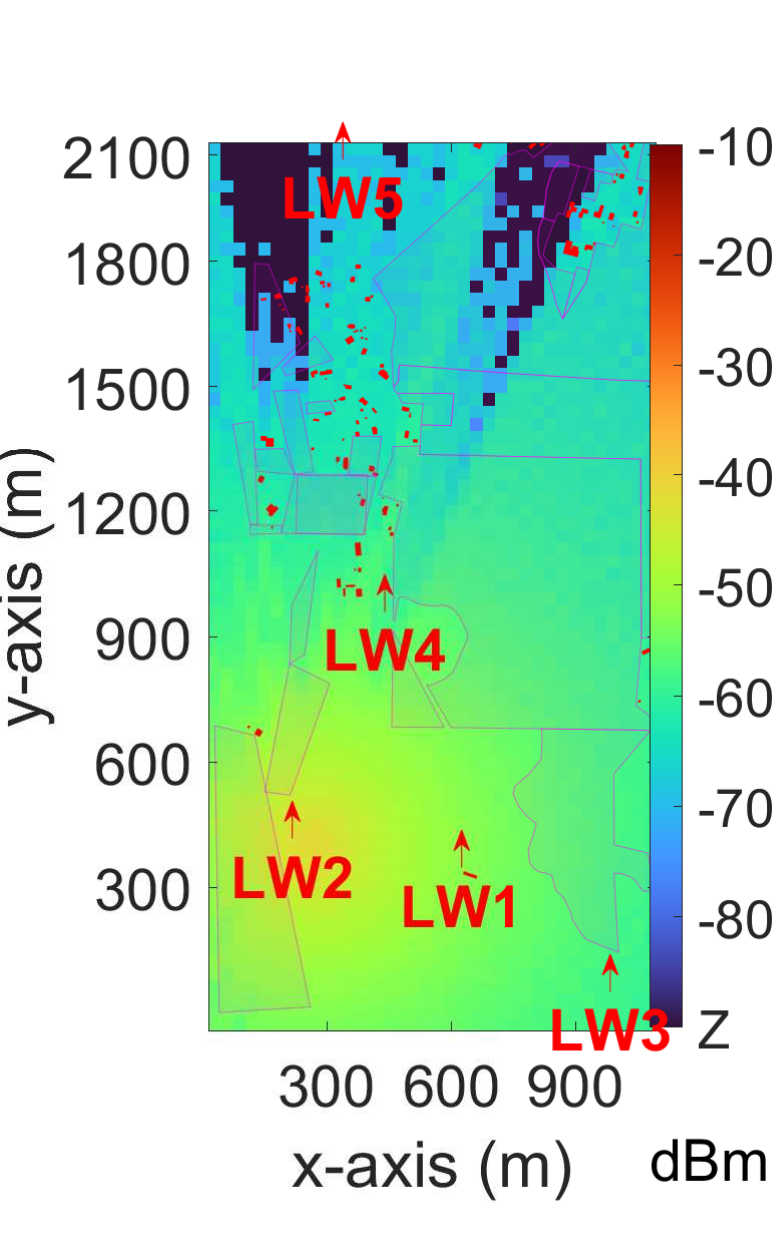}
    \label{fig:RSSI_SISO_110_LW2}
    }
    \subfigure[LW3 $110$~m]{
    \includegraphics[trim={0.1cm 1.0cm 0.15cm 1.7cm},clip,width=0.37\columnwidth]{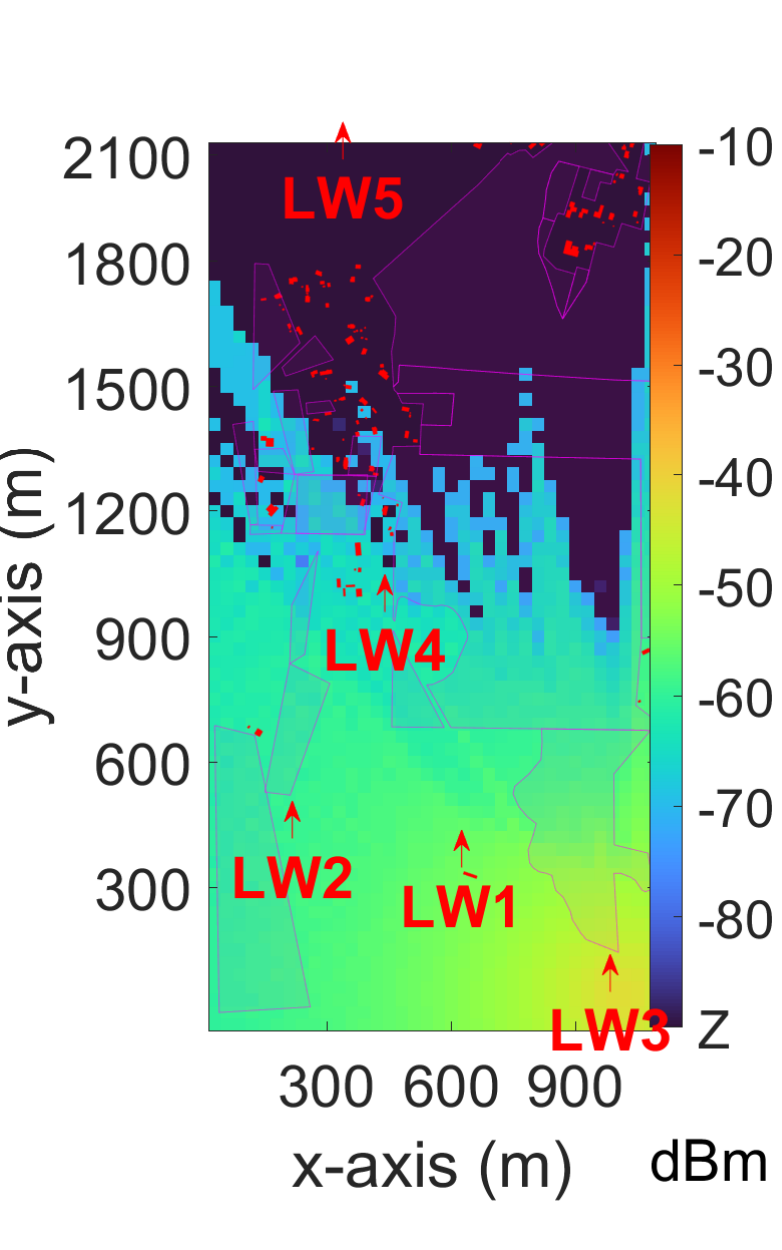}
    \label{fig:RSSI_SISO_110_LW3}
    }
    \subfigure[LW4 $110$~m]{
    \includegraphics[trim={0.1cm 1.0cm 0.15cm 1.7cm},clip,width=0.37\columnwidth]{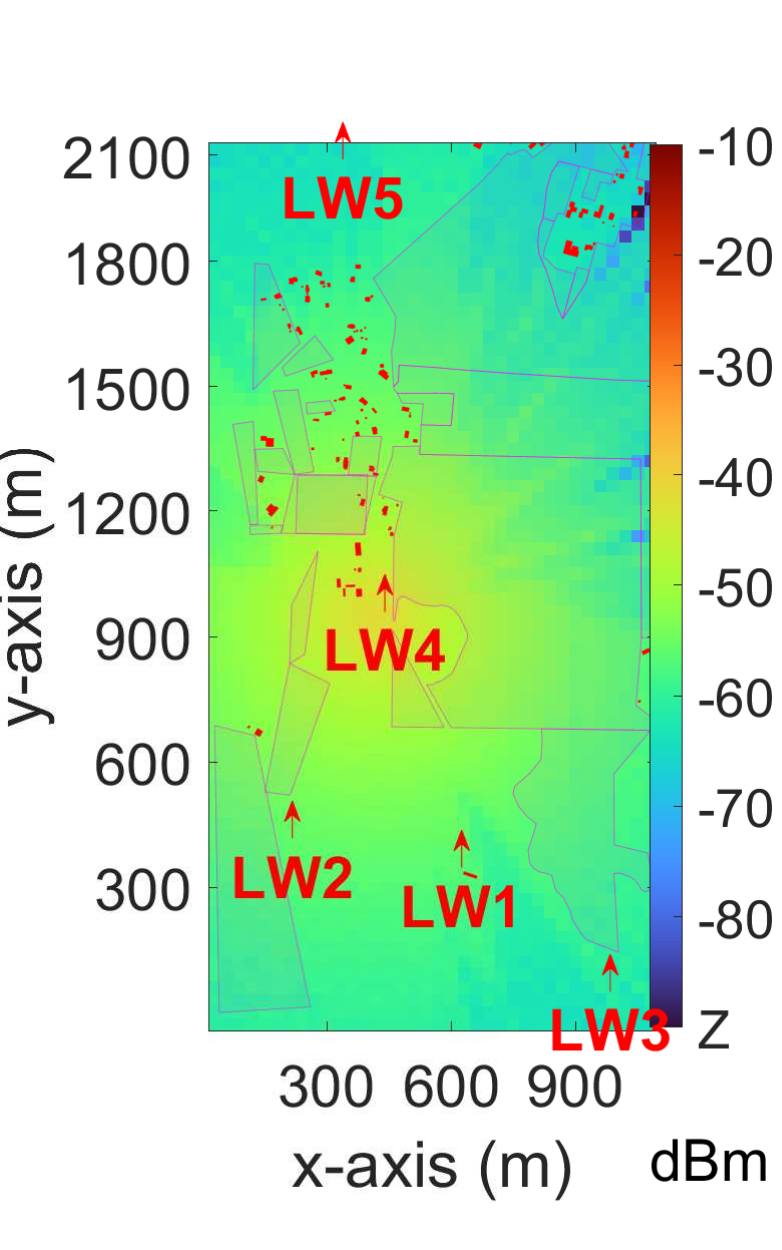}
    \label{fig:RSSI_SISO_110_LW4}
    }
    \subfigure[LW5 $110$~m]{
    \includegraphics[trim={0.1cm 1.0cm 0.15cm 1.7cm},clip,width=0.37\columnwidth]{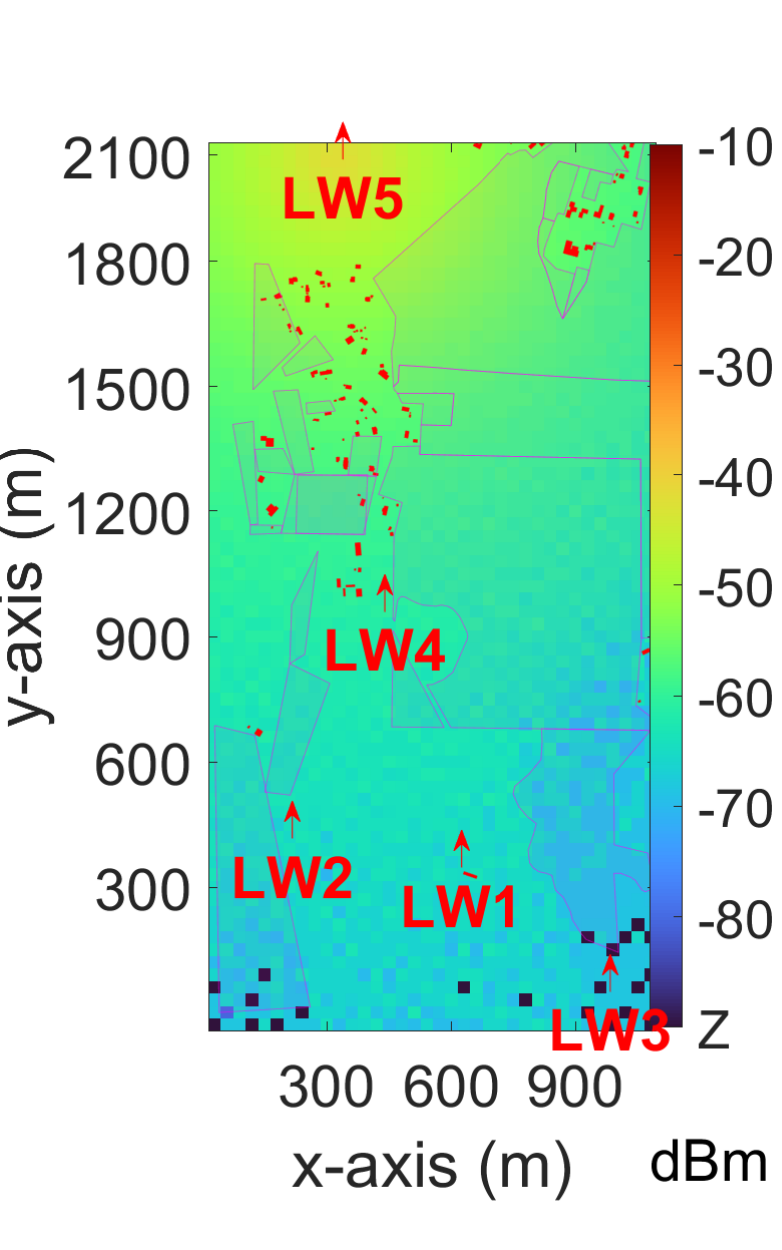}
    \label{fig:RSSI_SISO_110_LW5}
    }    
   \caption{RSS in the Lake Wheeler Field Labs area with SISO and $30, 70,$ and $110$~m altitude configurations.}
    \label{fig:RSSI_SISO_all_altitudes}
\end{figure*} 

\begin{figure}[t!]
    \centering
    \subfigure[MIMO $30$~m]{
    \includegraphics[trim={0.1cm 0.8cm 0.5cm 1.7cm},clip,width=0.45\columnwidth]{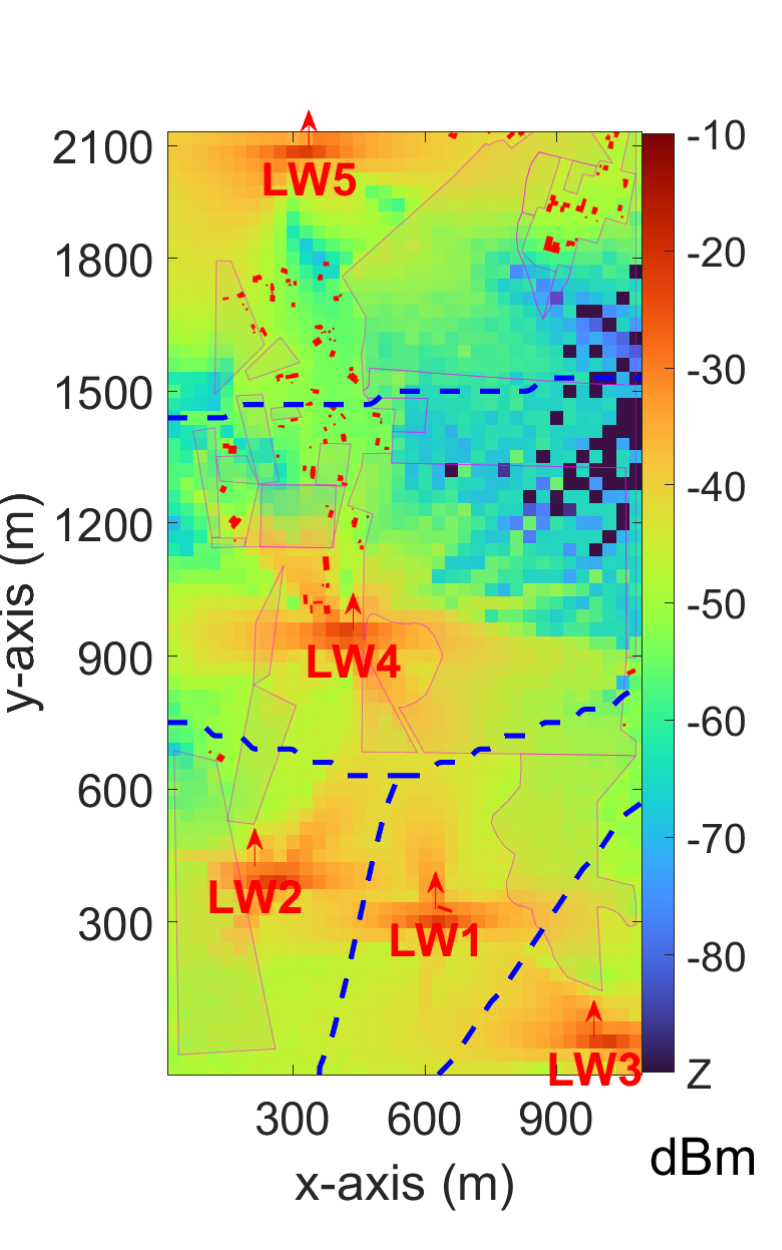}
    \label{fig:RSSI_MIMO_30_joint}
    }
    \subfigure[SISO $30$~m]{
    \includegraphics[trim={0.1cm 0.8cm 0.5cm 1.7cm},clip,width=0.45\columnwidth]{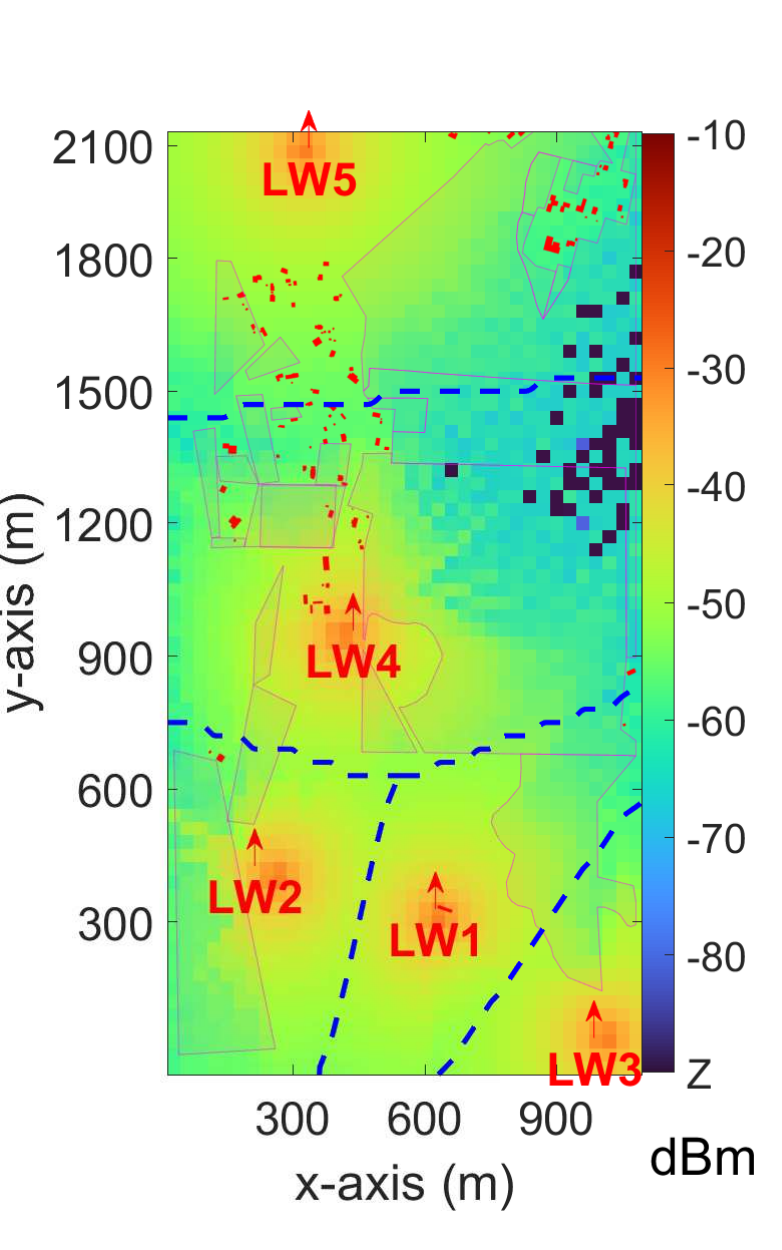}
    \label{fig:RSSI_SISO_30_joint}
    }    
    \subfigure[MIMO $70$~m]{
    \includegraphics[trim={0.1cm 0.8cm 0.5cm 1.7cm},clip,width=0.45\columnwidth]{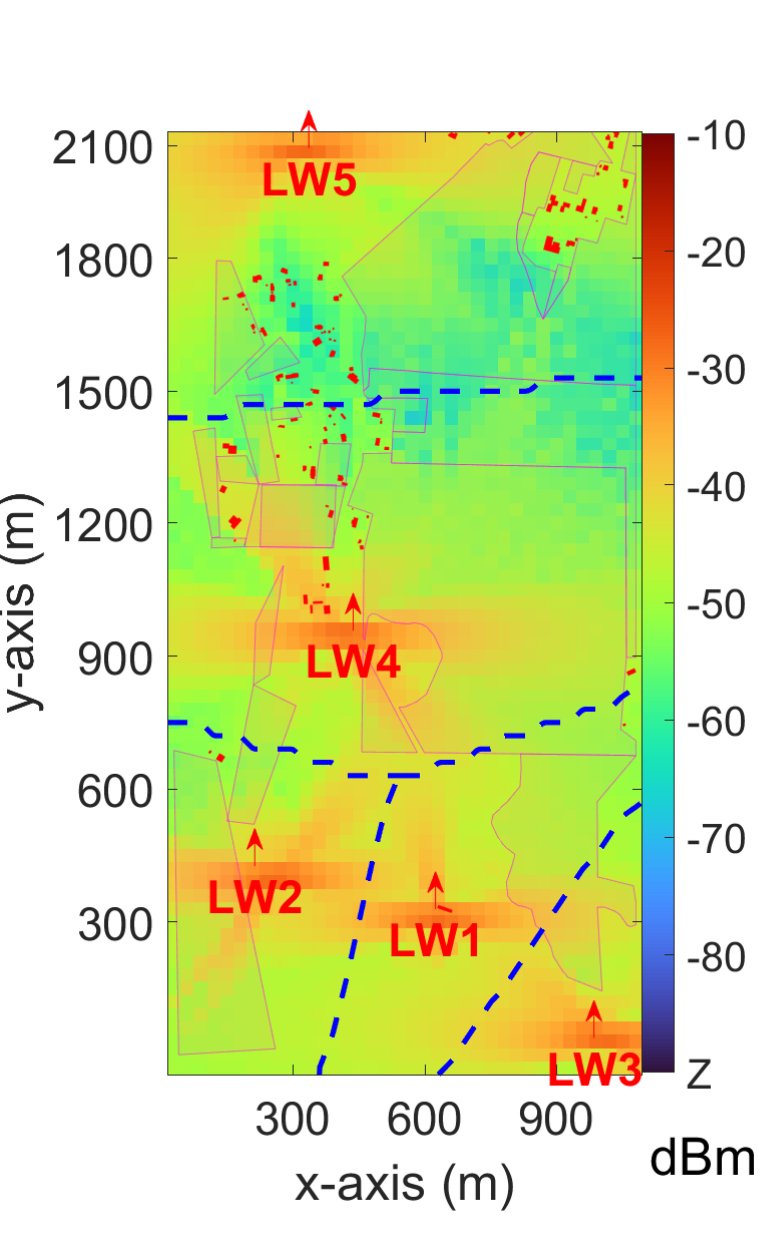}
    \label{fig:RSSI_MIMO_70_joint}
    }
    \subfigure[SISO $70$~m]{
    \includegraphics[trim={0.1cm 0.8cm 0.5cm 1.7cm},clip,width=0.45\columnwidth]{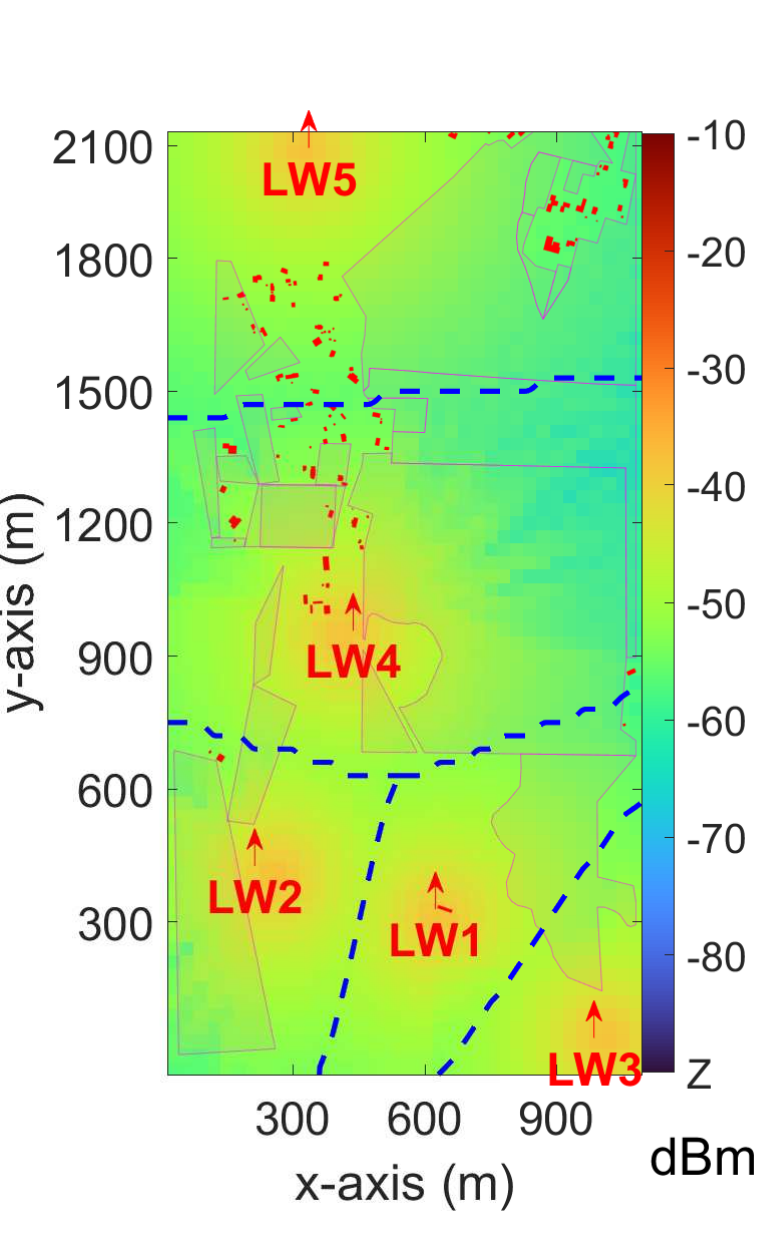}
    \label{fig:RSSI_SISO_70_joint}
    }
    \subfigure[MIMO $110$~m]{
    \includegraphics[trim={0.1cm 0.8cm 0.5cm 1.7cm},clip,width=0.45\columnwidth]{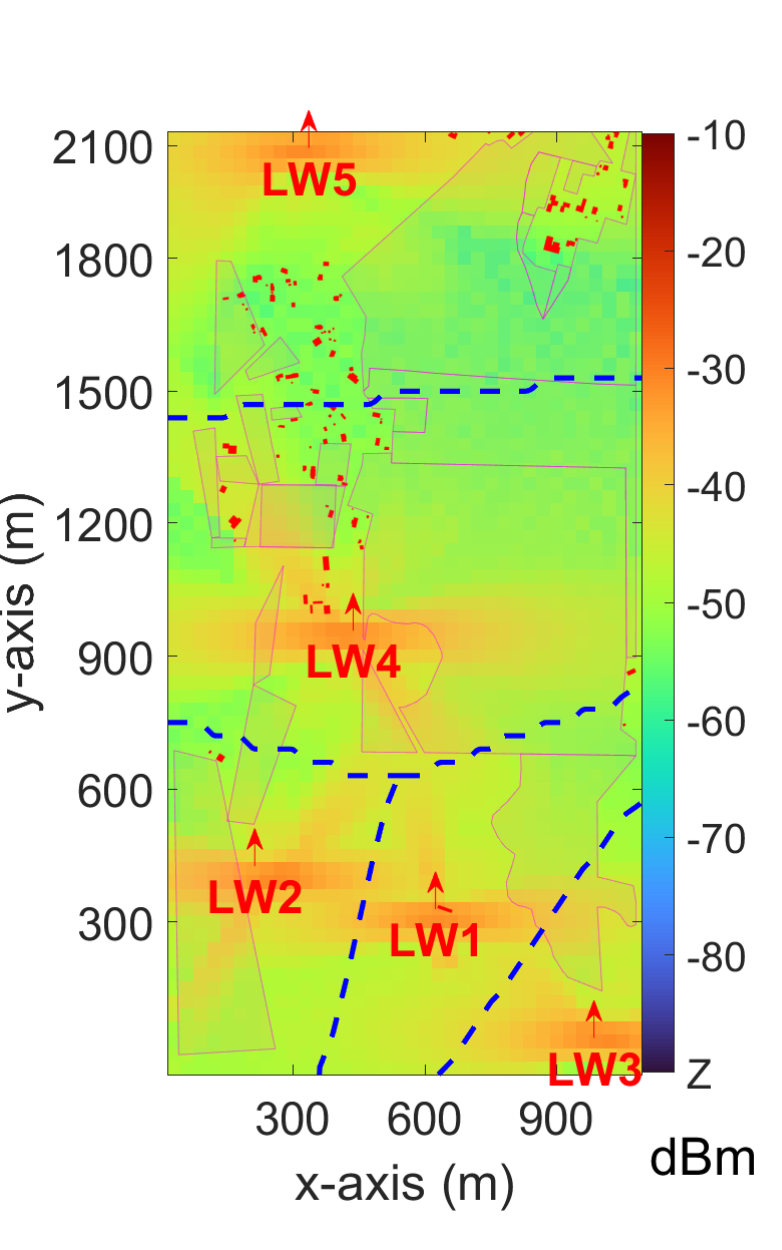}
    \label{fig:RSSI_MIMO_110_joint}
    }
    \subfigure[SISO $110$~m]{
    \includegraphics[trim={0.1cm 0.8cm 0.5cm 1.7cm},clip,width=0.45\columnwidth]{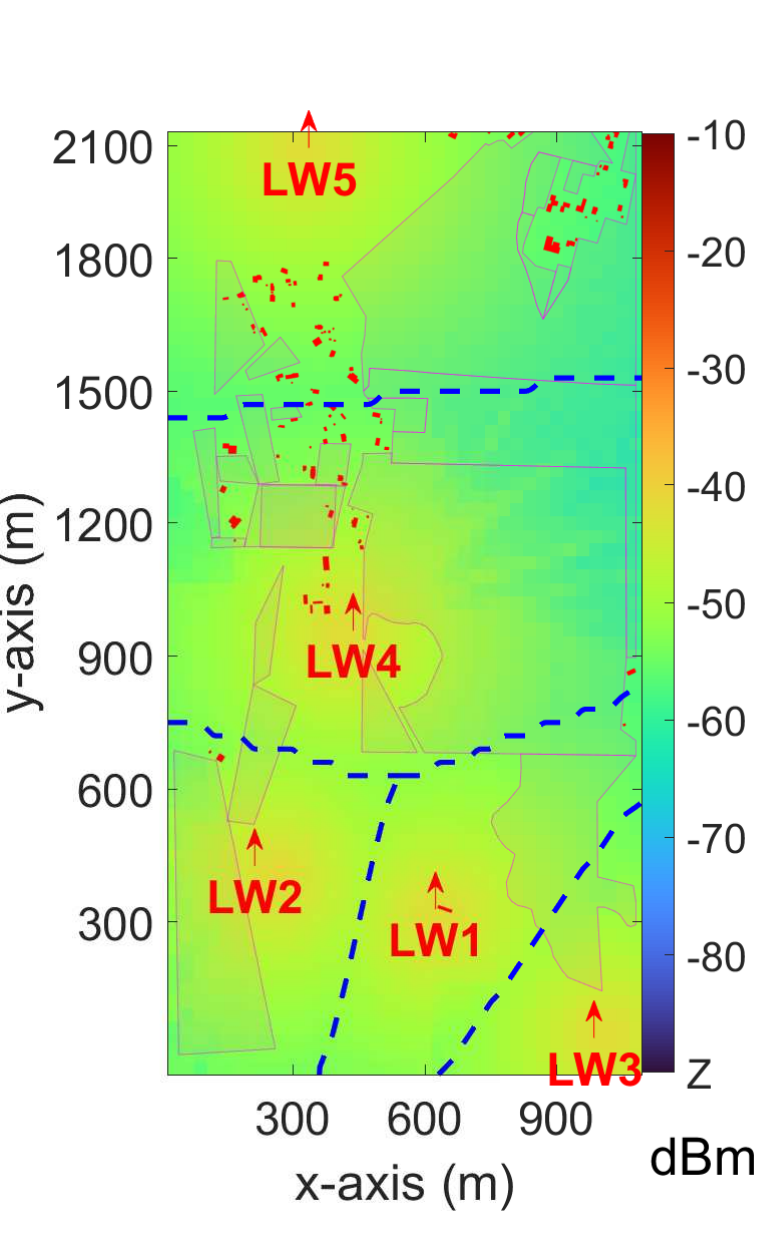}
    \label{fig:RSSI_SISO_110_joint}
    }
    \caption{RSS in the Lake Wheeler Field Labs area with the joint coverage and $30, 70,$ and $110$~m altitude configurations.}
    \label{fig:RSSI_SISO_MIMO_all_joint}
\end{figure}

 \begin{figure*}[t!]
    \centering    
    \subfigure[$30$~m MIMO]{
    \includegraphics[trim={0.4cm 0cm 0.8cm 0.6cm},clip,width=0.63\columnwidth]{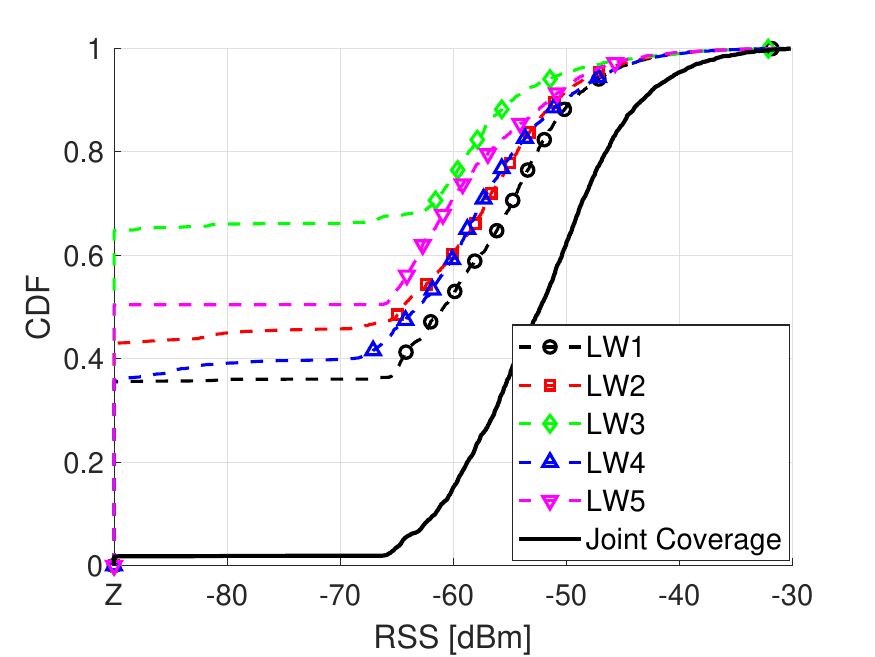}
    \label{fig:RSSI_CDF_30m_MIMO}
    }
    \subfigure[$70$~m MIMO]{
    \includegraphics[trim={0.4cm 0cm 0.8cm 0.6cm},clip,width=0.63\columnwidth]{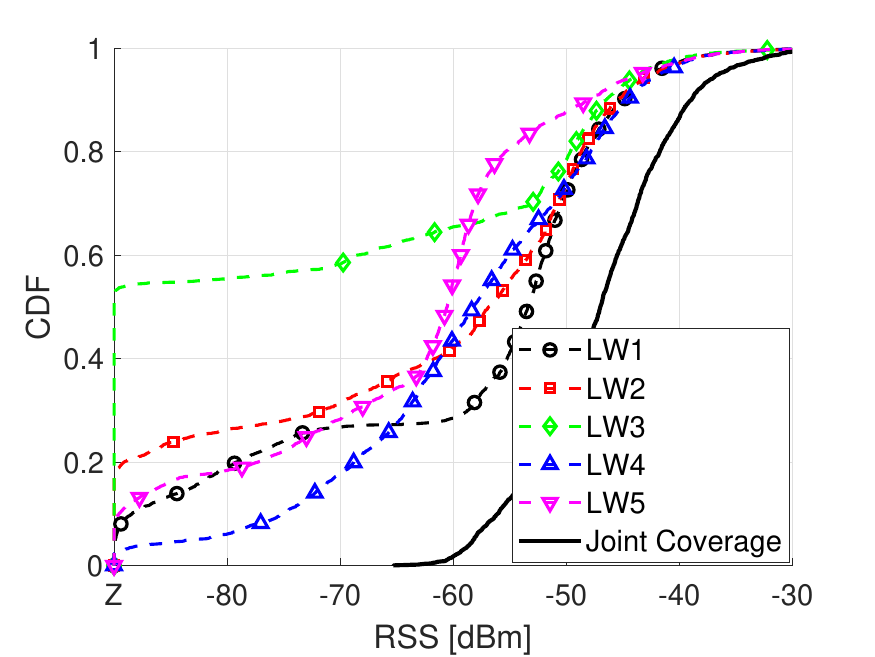}
    \label{fig:RSSI_CDF_70m_MIMO}
    }
    \subfigure[$110$~m MIMO]{
    \includegraphics[trim={0.4cm 0cm 0.8cm 0.6cm},clip,width=0.63\columnwidth]{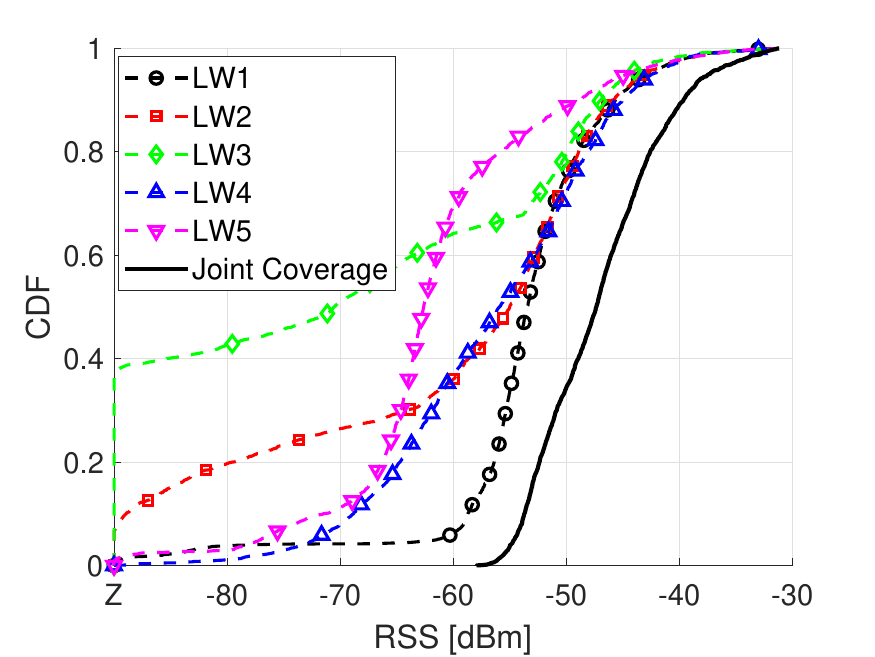}
    \label{fig:RSSI_CDF_110m_MIMO}
    }
    \subfigure[$30$~m SISO]{
    \includegraphics[trim={0.4cm 0cm 0.8cm 0.6cm},clip,width=0.63\columnwidth]{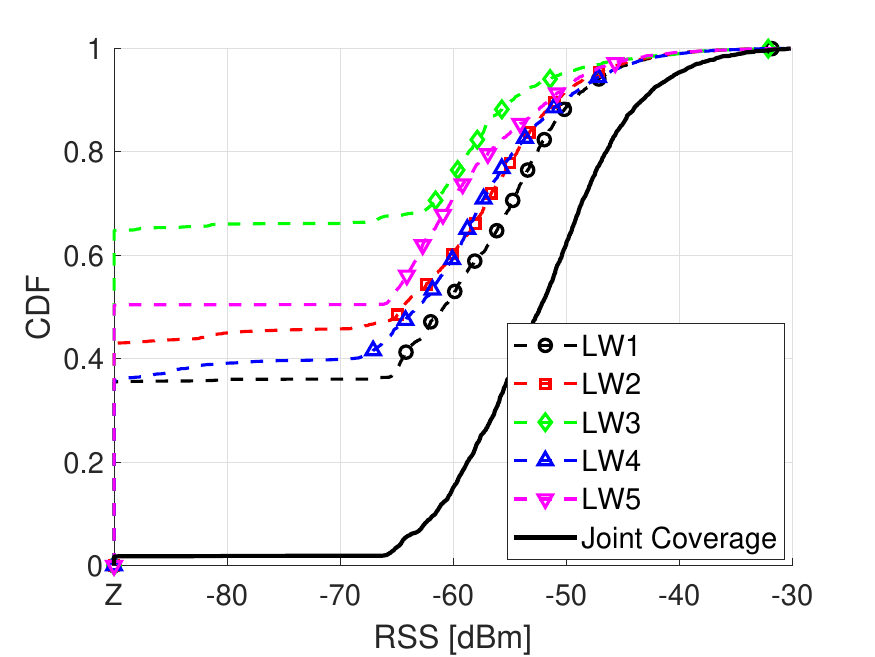}
    \label{fig:RSSI_CDF_30m_SISO}
    }
    \subfigure[$70$~m SISO]{
    \includegraphics[trim={0.4cm 0cm 0.8cm 0.6cm},clip,width=0.63\columnwidth]{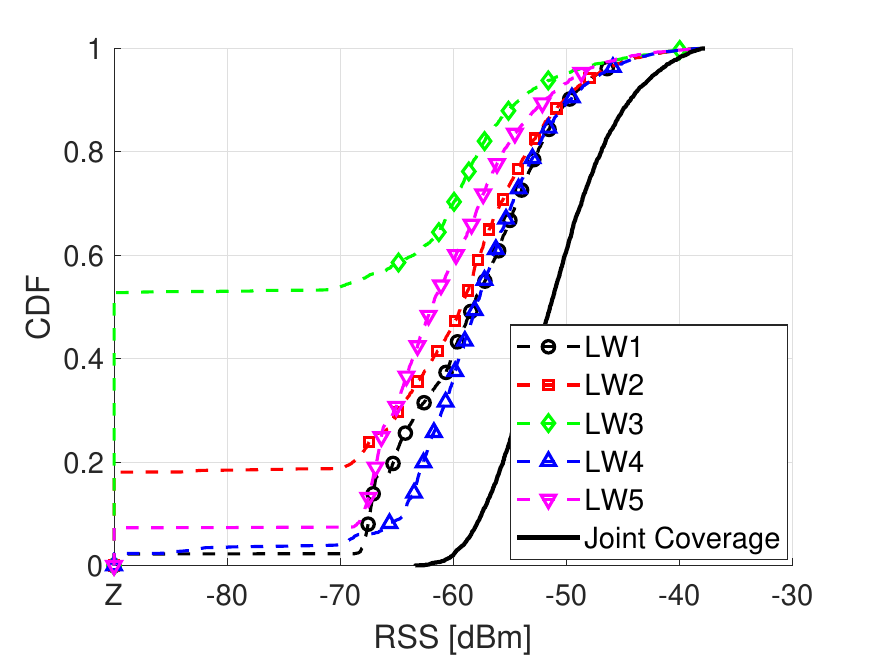}
    \label{fig:RSSI_CDF_70m_SISO}
    }
    \subfigure[$110$~m SISO]{
    \includegraphics[trim={0.4cm 0cm 0.8cm 0.6cm},clip,width=0.63\columnwidth]{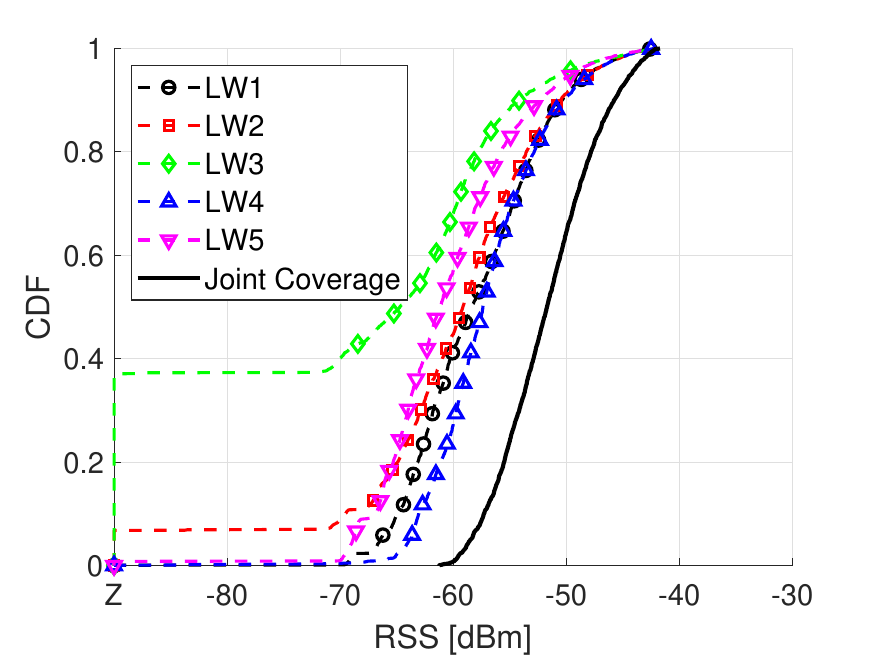}
    \label{fig:RSSI_CDF_110m_SISO}
    }
   \caption{CDFs of RSS in the Lake Wheeler Field Labs area with MIMO and SISO configurations.}
    \label{fig:RSSI_CDFs}
\end{figure*}

 In this section, we analyze the RSS for UAV channels using the RT simulation tool. We consider both single-input single-output (SISO) and MIMO antenna configuration ($4\times 1$ linear array at the tower) with three different UAV altitudes: $30$~m, $70$~m, and $110$~m  from the ground. For MIMO, we consider a $4$-element linear array antenna placed along the y-axis at each tower. We study the coverage of each tower individually and when they operate simultaneously with the other towers. For the latter scenario, each UAV receiver is connected to the closest tower. The Voronoi boundaries of the coverages under the joint coverage assumption are marked in blue dashed lines in the figures. 

 Figures \ref{fig:RSSI_MIMO_all_altitudes} and \ref{fig:RSSI_SISO_all_altitudes} show the simulation results of RSS distribution over the Lake Wheeler Field Labs for three different altitudes with SISO and MIMO configurations, respectively. Moreover, the simulation results of RSS distribution with the joint coverage assumption are shown in Figure~\ref{fig:RSSI_SISO_MIMO_all_joint}. Here, towers are marked in red arrows and buildings are marked in red polygons. Moreover, the forest areas are marked in magenta polygons. The letter $\mathsf{Z}$ represents the out-of-coverage area with the given tower and UAV, where the signal strength is below a threshold value (e.g. due to blockage). In Figure \ref{fig:RSSI_MIMO_all_altitudes}, narrow distribution patterns through the horizontal axis and side lobes in the diagonal direction are observed due to the beam pattern of the MIMO linear array. It is observed that the beam pattern leads to the narrow coverage, which can be seen around the LW2, LW3, and LW5 towers. Meanwhile, due to the omnidirectional beam pattern of SISO configuration, the RSS distribution in Figure \ref{fig:RSSI_SISO_all_altitudes} tends to have a circular pattern near the base stations. 
 The blockage effects from the trees are observed in $30$~m in both SISO and MIMO cases, which is gradually mitigated at the higher UAV altitudes. 
 
 The RSS simulation results of  Figures~\ref{fig:RSSI_MIMO_all_altitudes} -~\ref{fig:RSSI_SISO_MIMO_all_joint} are summarized with CDFs in Figure \ref{fig:RSSI_CDFs}. The portion of the blockage area with the LW3 site is the highest of all cases due to LW3 being at the edge of the flight area and being most blocked by the trees. It is observed that the LW1 tower has the least blockage due to its location being in an open area close to the center of the field. The blockage is reduced as the altitude of the UAV increases, while the higher values of the RSS are slightly degraded due to larger path loss at higher altitudes. The blockage with the joint coverage is significantly reduced even with the $30$~m altitude, where there is no blockage with the higher altitudes.  Moreover, it can be observed that the range of RSS tends to be wider in the MIMO cases that have multiple directional beam patterns than the omnidirectional SISO cases.  

 \subsection{Channel Rank Analysis}\label{ch:channel_rank}

  \begin{figure*}[t!]
    \centering    
    \subfigure[$K_1$, $3$~m]{
    \includegraphics[trim={0.15cm 1.0cm 1.3cm 1.7cm},clip,width=0.47\columnwidth]{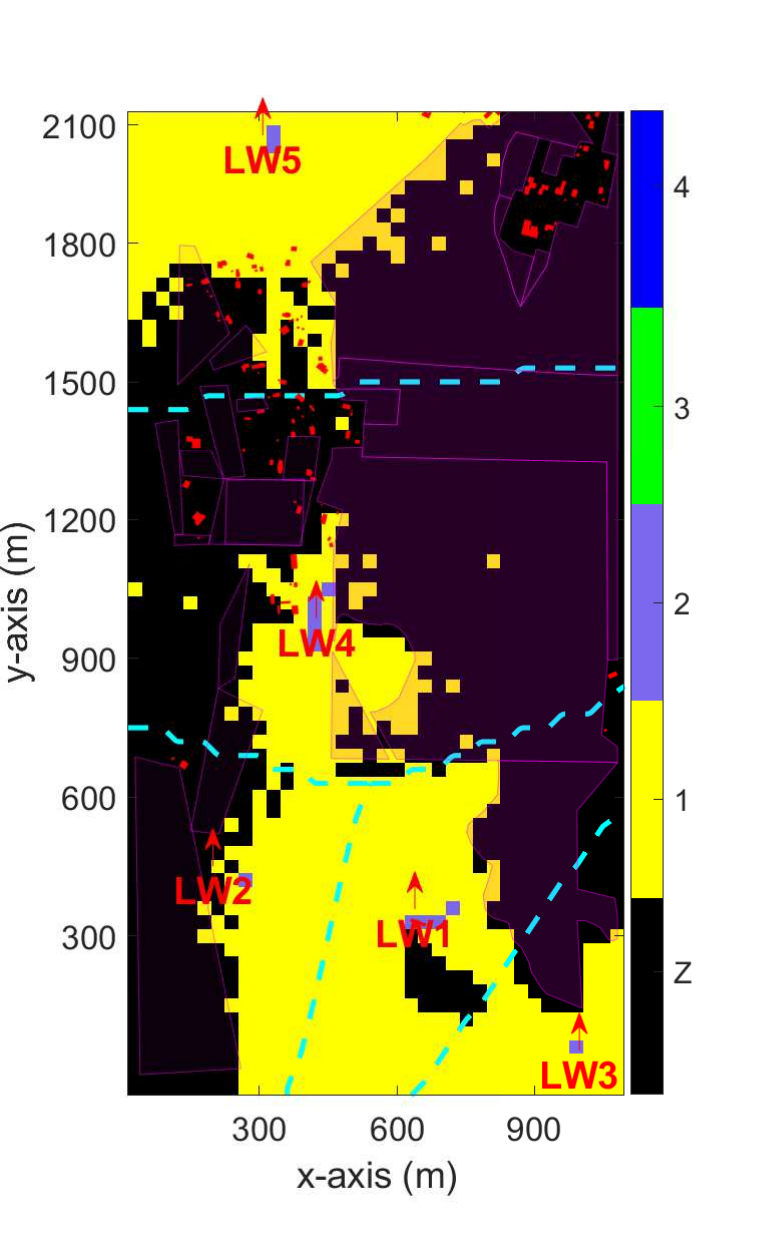}
    \label{fig:Rank_MIMO_3_10}
    }
    \subfigure[$K_1$, $30$~m]{
    \includegraphics[trim={0.15cm 1.0cm 1.3cm 1.7cm},clip,width=0.47\columnwidth]{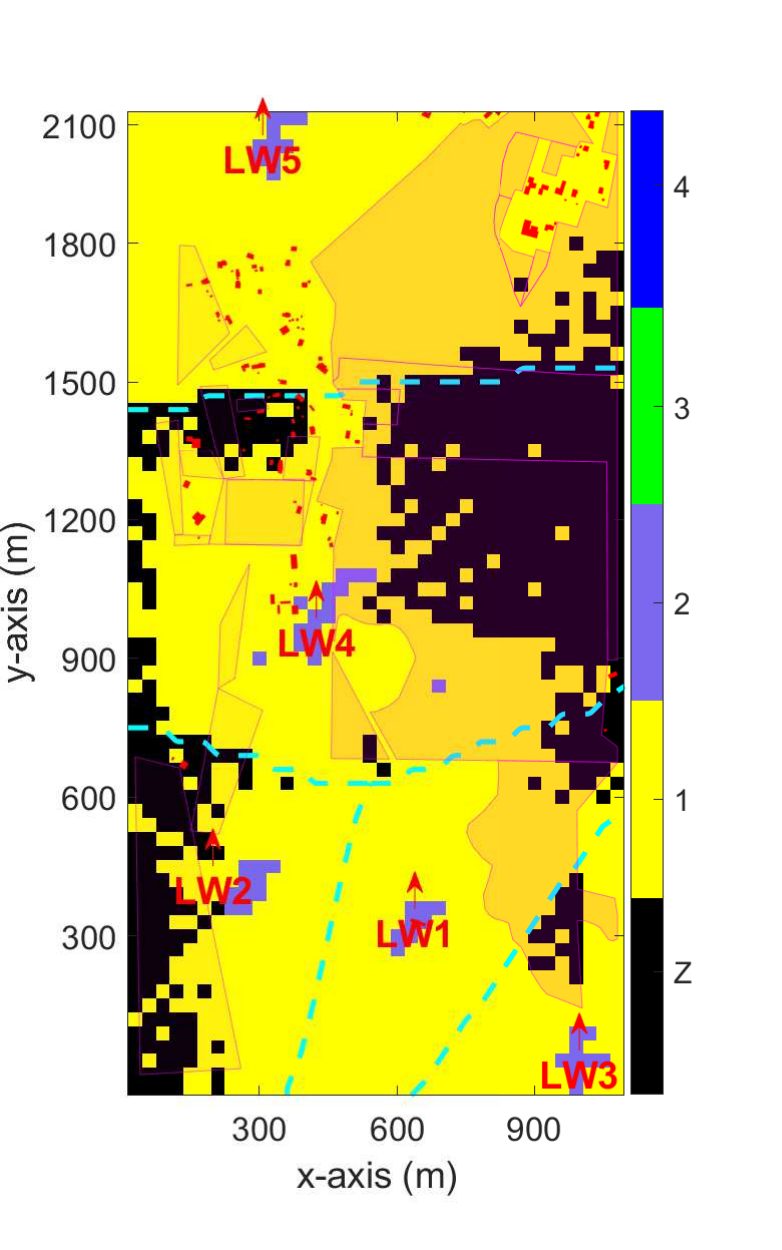}
    \label{fig:Rank_MIMO_30_10}
    }
    \subfigure[$K_1$, $70$~m]{
    \includegraphics[trim={0.15cm 1.0cm 1.3cm 1.7cm},clip,width=0.47\columnwidth]{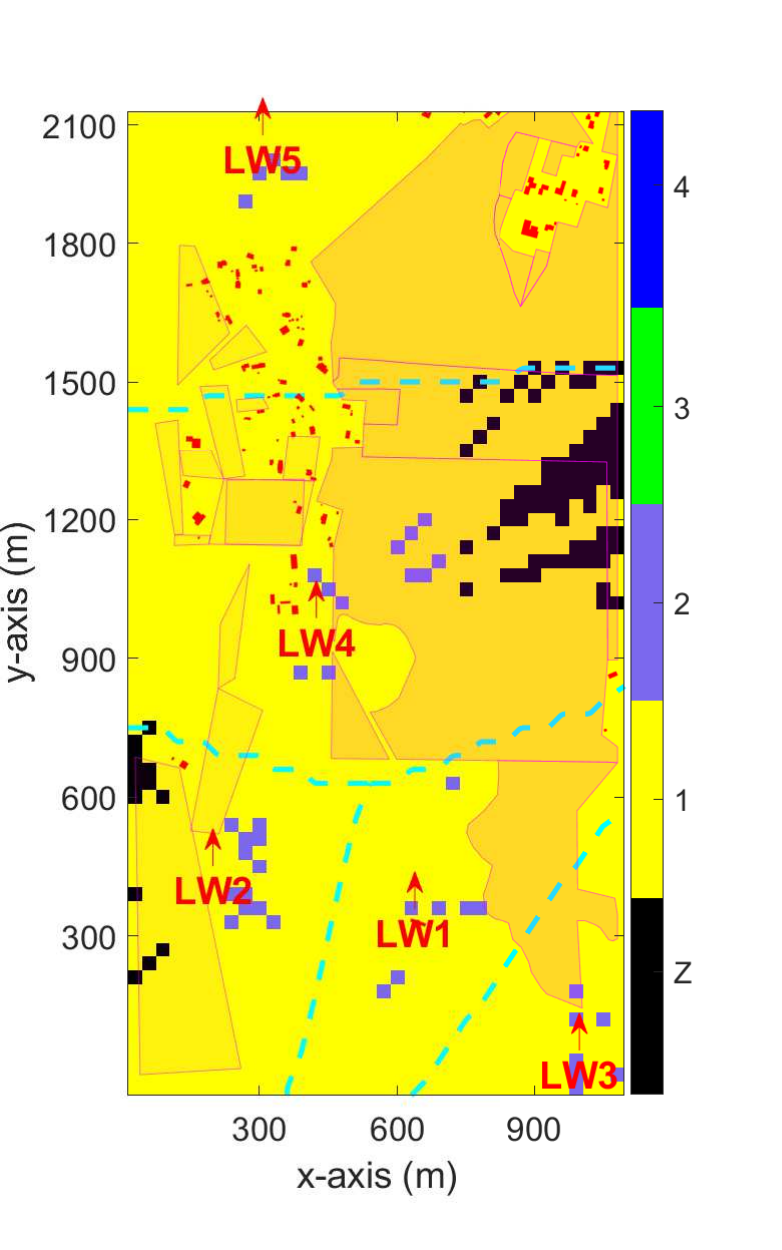}
    \label{fig:Rank_MIMO_70_10}
    }
    \subfigure[$K_1$, $110$~m]{
    \includegraphics[trim={0.15cm 1.0cm 1.3cm 1.7cm},clip,width=0.47\columnwidth]{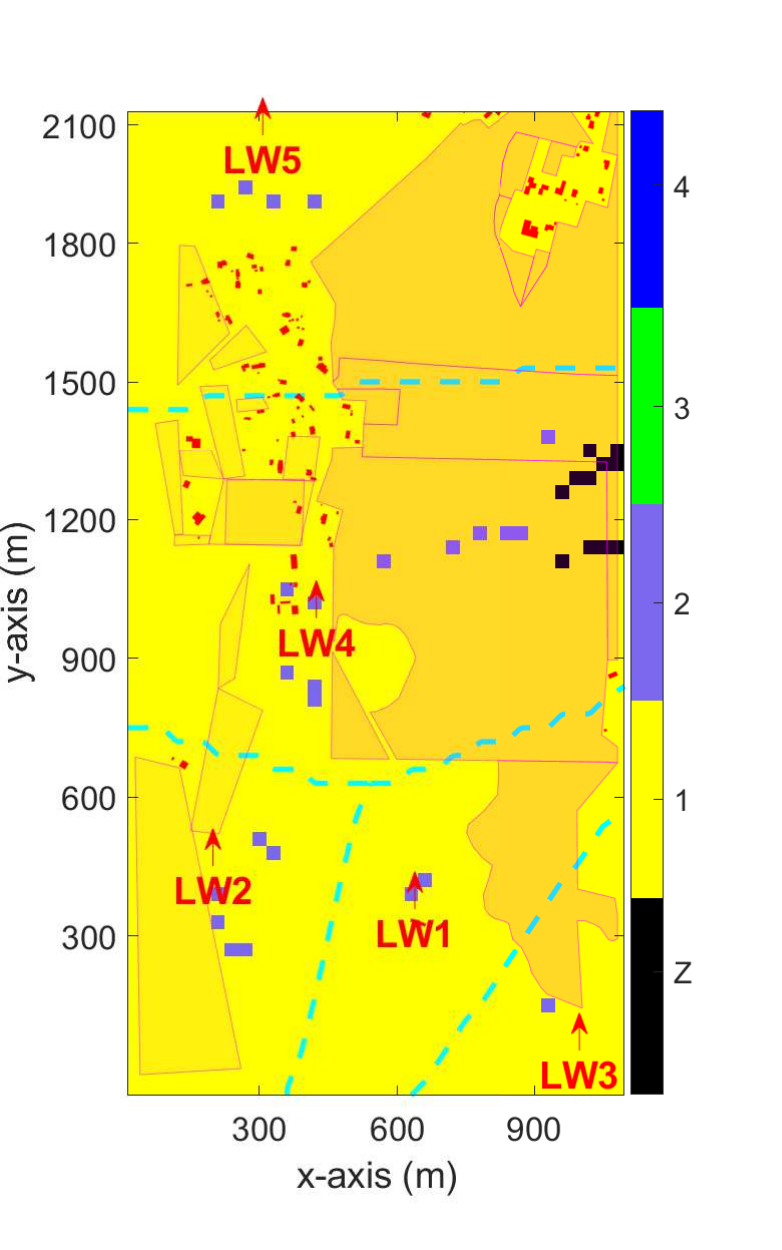}
    \label{fig:Rank_MIMO_110_10}
    }   
    \subfigure[$K_2$, $3$~m]{
    \includegraphics[trim={0.15cm 1.0cm 1.3cm 1.7cm},clip,width=0.47\columnwidth]{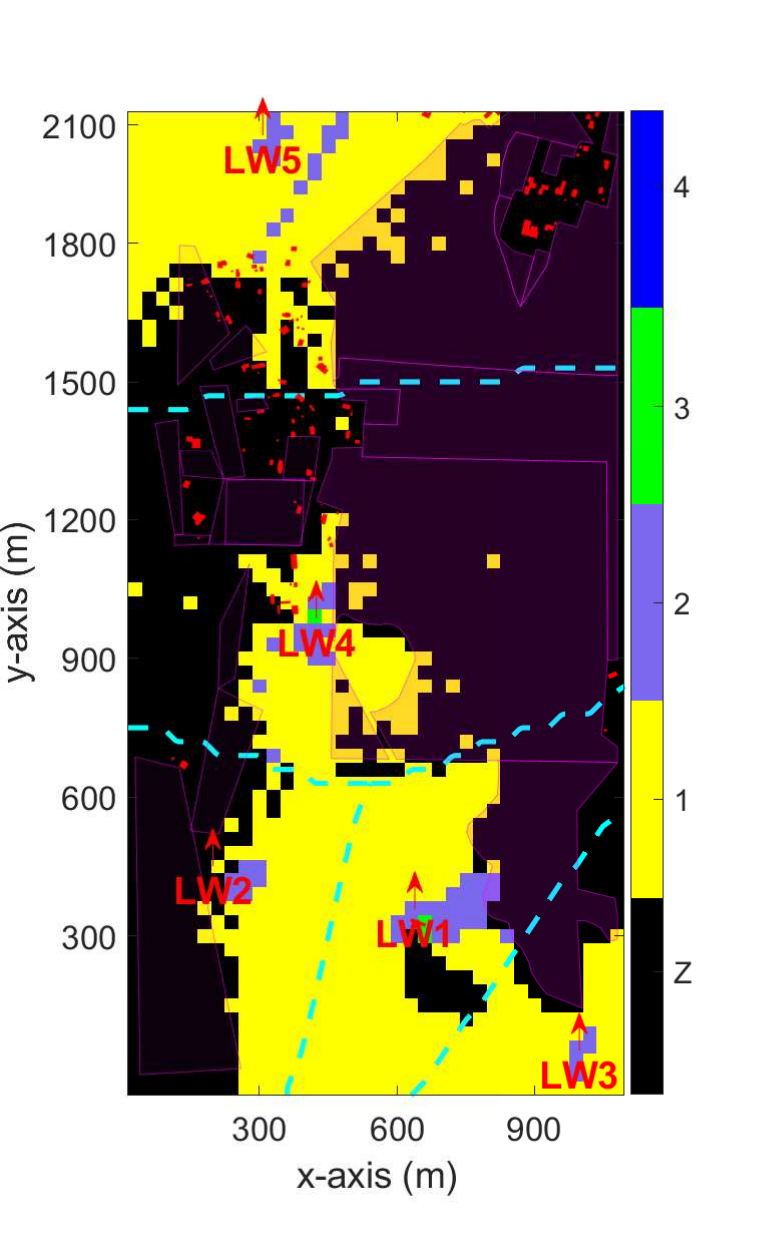}
    \label{fig:Rank_MIMO_3_100}
    }
    \subfigure[$K_2$, $30$~m]{
    \includegraphics[trim={0.15cm 1.0cm 1.3cm 1.7cm},clip,width=0.47\columnwidth]{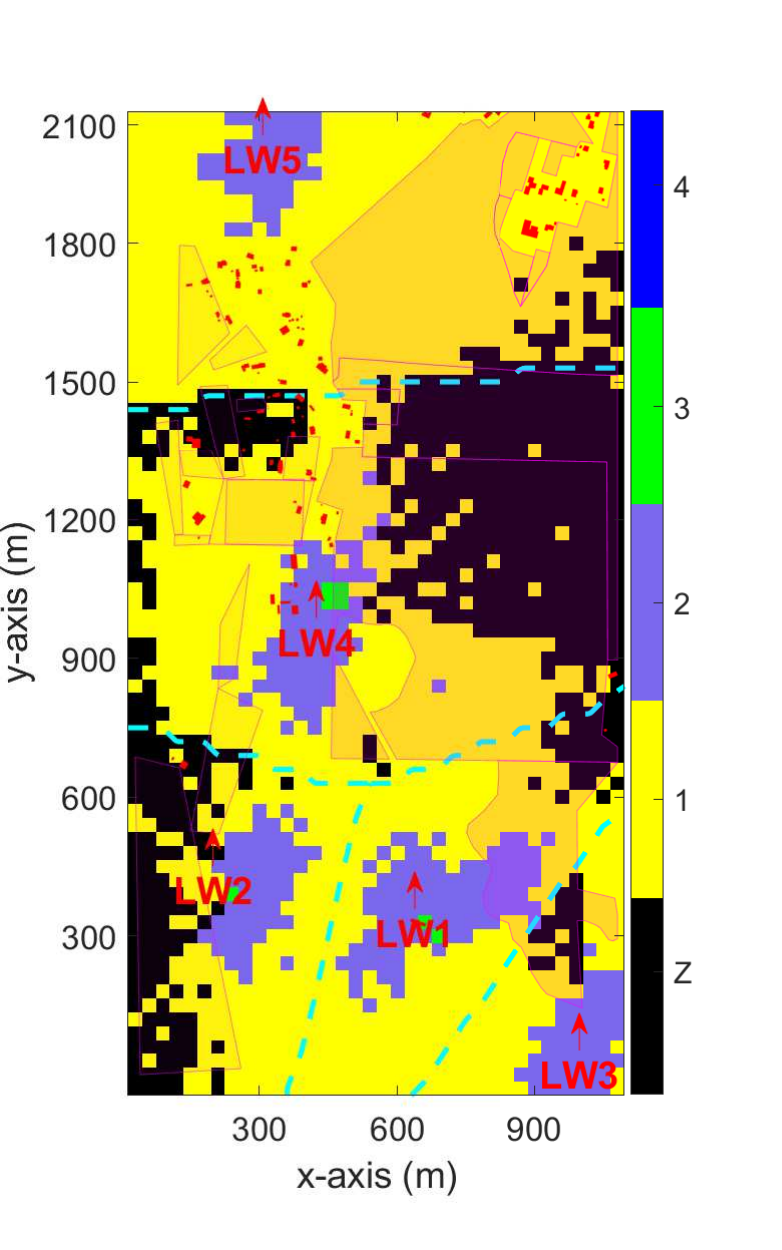}
    \label{fig:Rank_MIMO_30_100}
    }
    \subfigure[$K_2$, $70$~m]{
    \includegraphics[trim={0.15cm 1.0cm 1.3cm 1.7cm},clip,width=0.47\columnwidth]{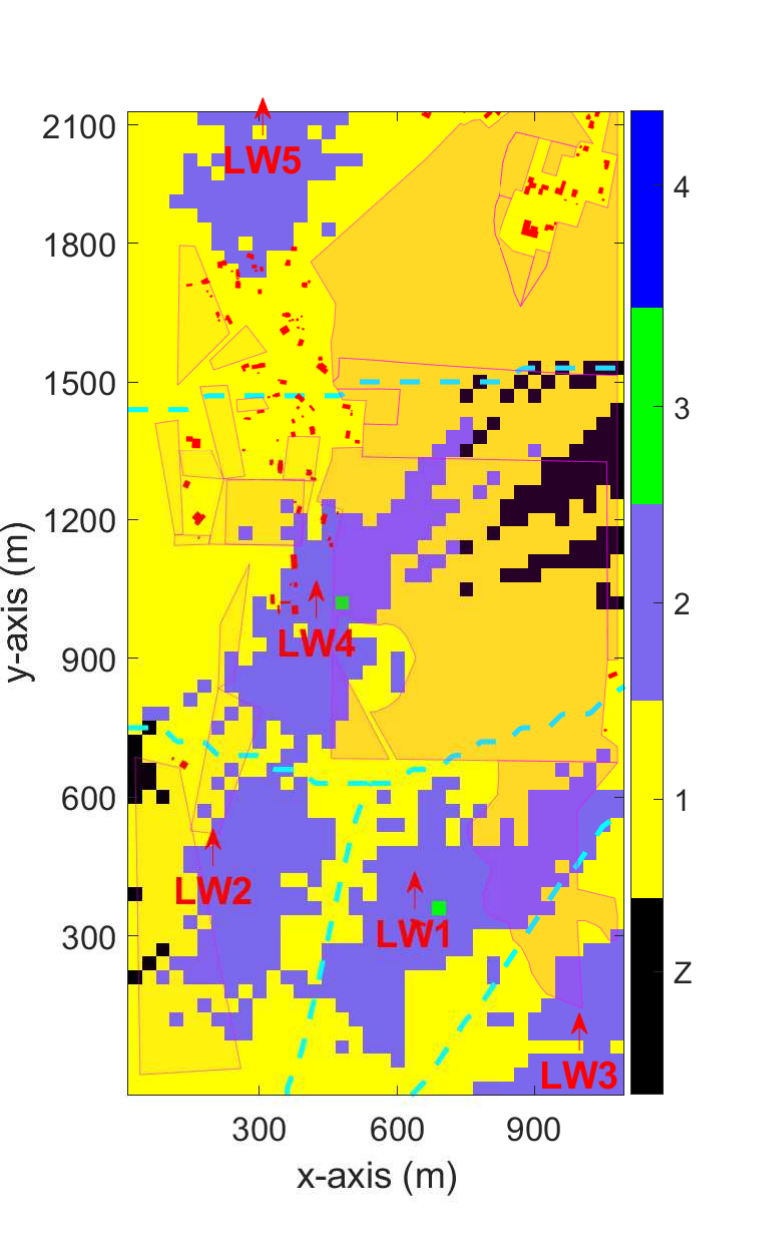}
    \label{fig:Rank_MIMO_70_100}
    }
    \subfigure[$K_2$, $110$~m]{
    \includegraphics[trim={0.15cm 1.0cm 1.3cm 1.7cm},clip,width=0.47\columnwidth]{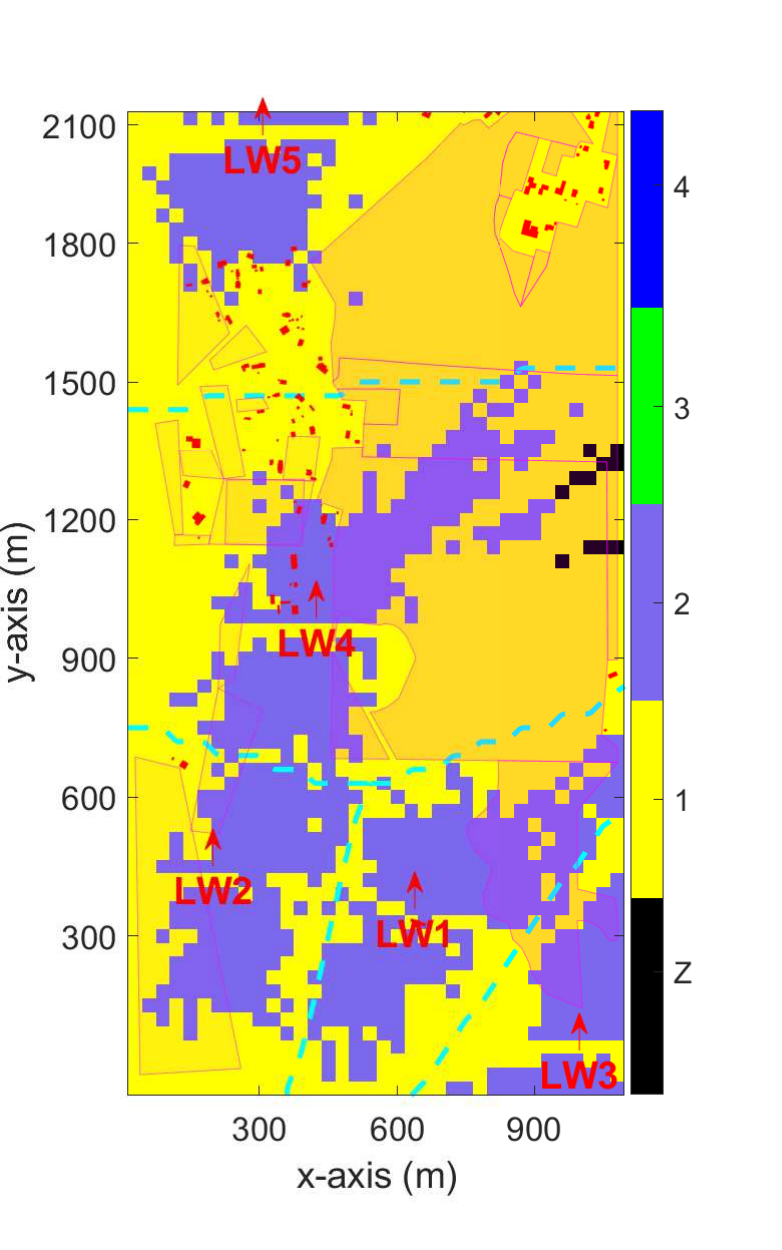}
    \label{fig:Rank_MIMO_110_100}
    }
    \subfigure[$K_3$, $3$~m]{
    \includegraphics[trim={0.15cm 1.0cm 1.3cm 1.7cm},clip,width=0.47\columnwidth]{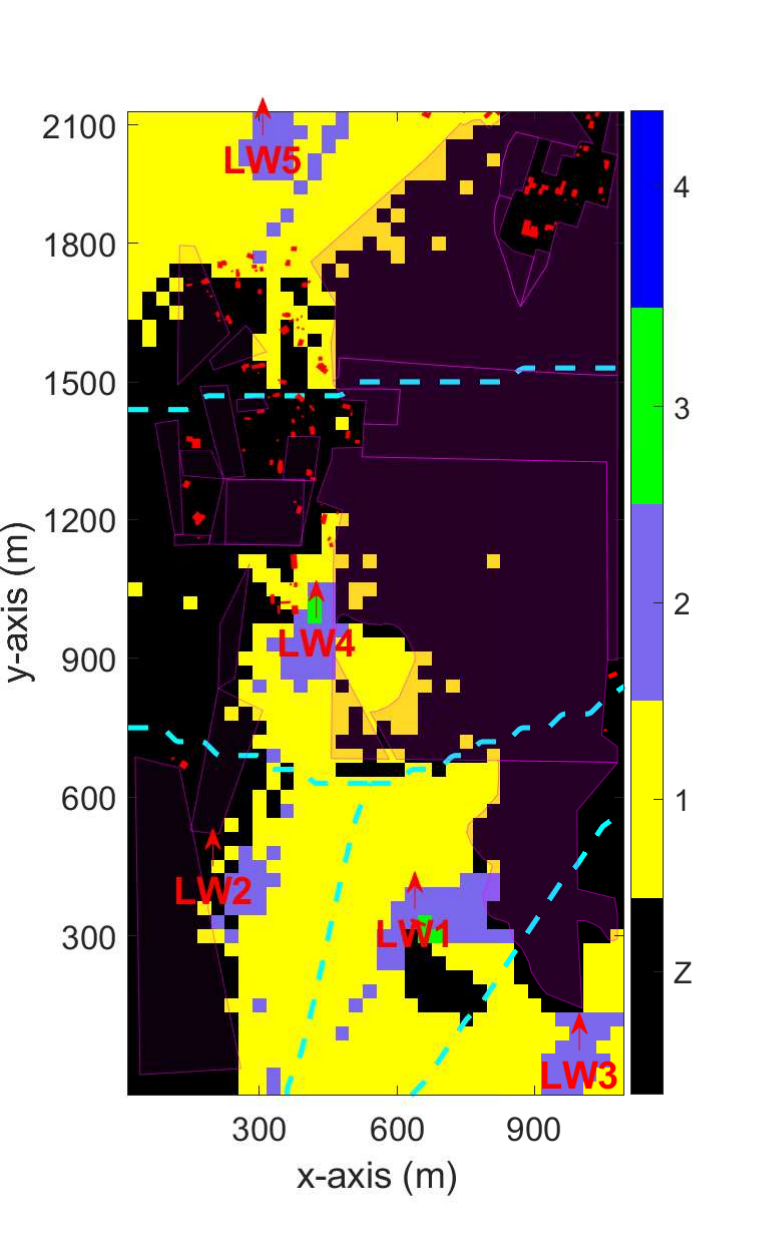}
    \label{fig:Rank_MIMO_3_1000}
    }
    \subfigure[$K_3$, $30$~m]{
    \includegraphics[trim={0.15cm 1.0cm 1.3cm 1.7cm},clip,width=0.47\columnwidth]{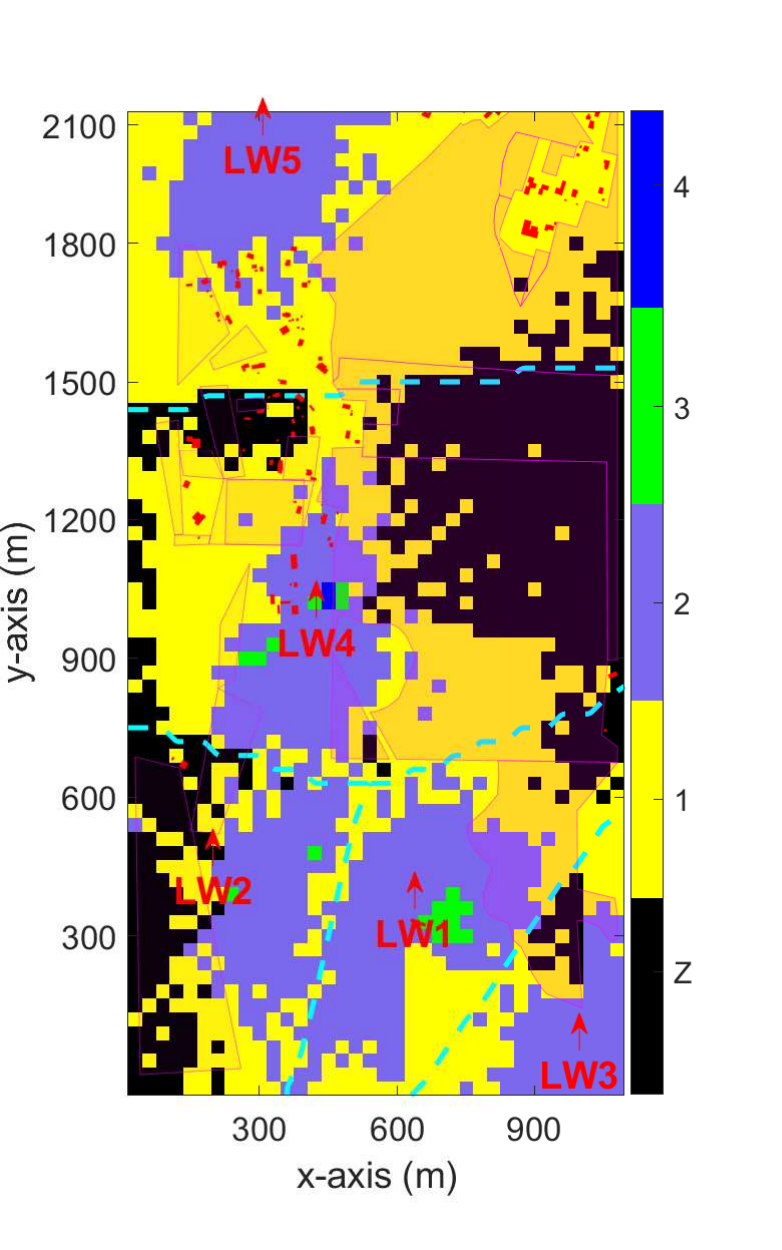}
    \label{fig:Rank_MIMO_30_1000}
    }
    \subfigure[$K_3$, $70$~m]{
    \includegraphics[trim={0.15cm 1.0cm 1.3cm 1.7cm},clip,width=0.47\columnwidth]{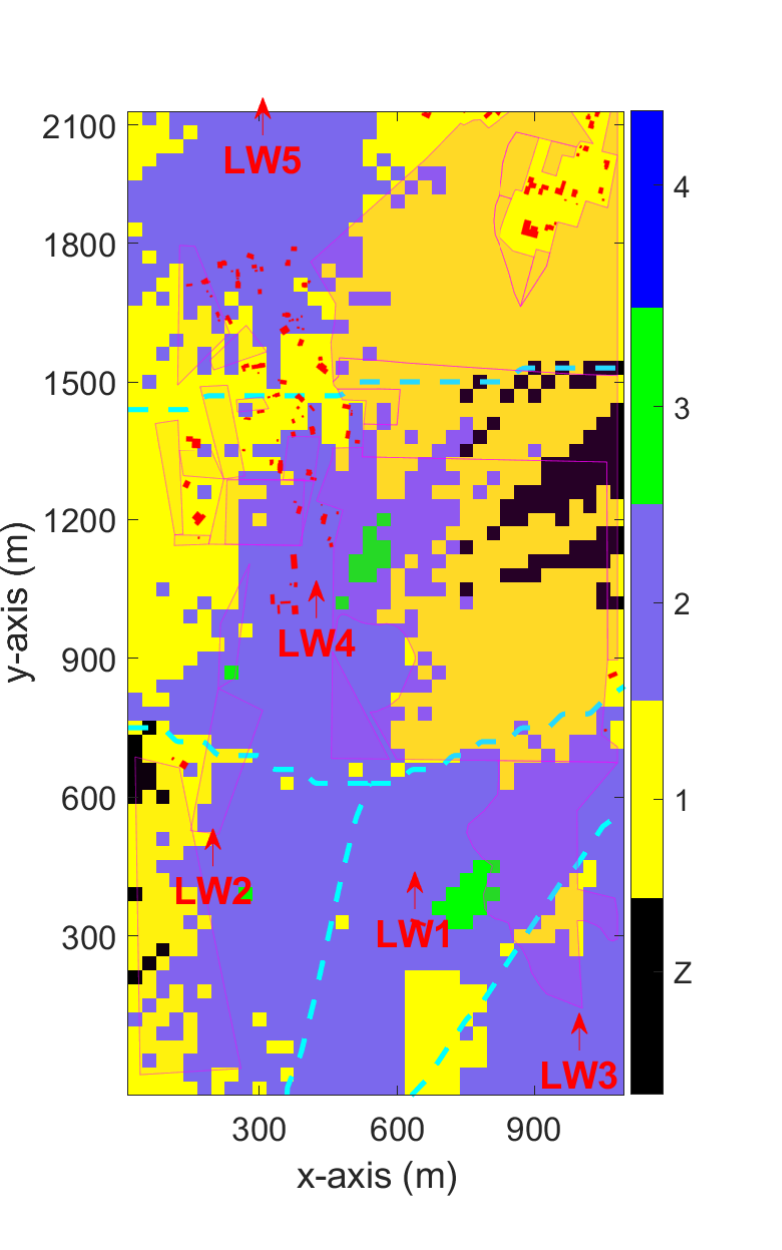}
    \label{fig:Rank_MIMO_70_1000}
    }
    \subfigure[$K_3$, $110$~m]{
    \includegraphics[trim={0.15cm 1.0cm 1.3cm 1.7cm},clip,width=0.47\columnwidth]{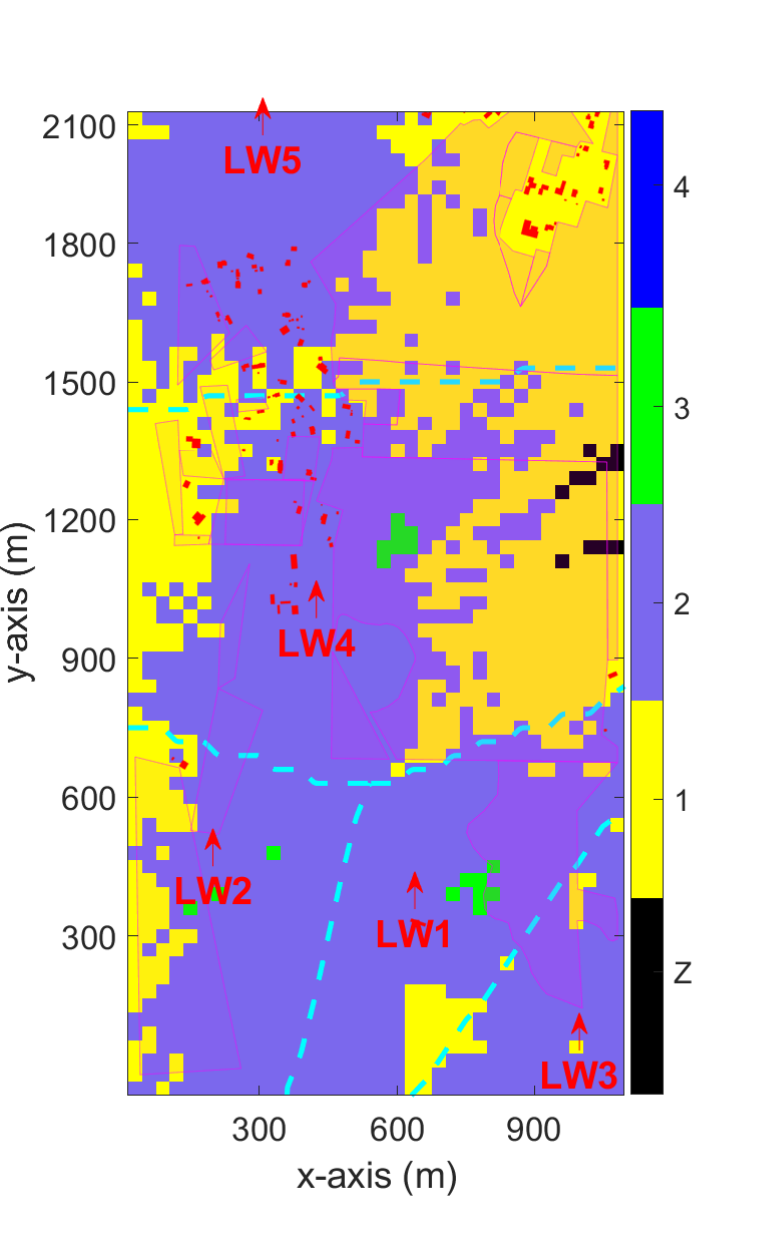}
    \label{fig:Rank_MIMO_110_1000}
    }
   \caption{Channel rank in the Lake Wheeler Field Labs with various rank thresholds $K_1$, $K_2$, $K_3$ and receiver altitudes of  $3$~m, $30$~m, $70$~m, and $110$~m.}
    \label{fig:Rank_distribution}
\end{figure*}

\begin{figure*}[t!]
    \centering    
    \subfigure[Distribution with $K_1$]{
    \includegraphics[trim={0.4cm 0.1cm 1.3cm 0.7cm},clip, width=0.64\columnwidth]{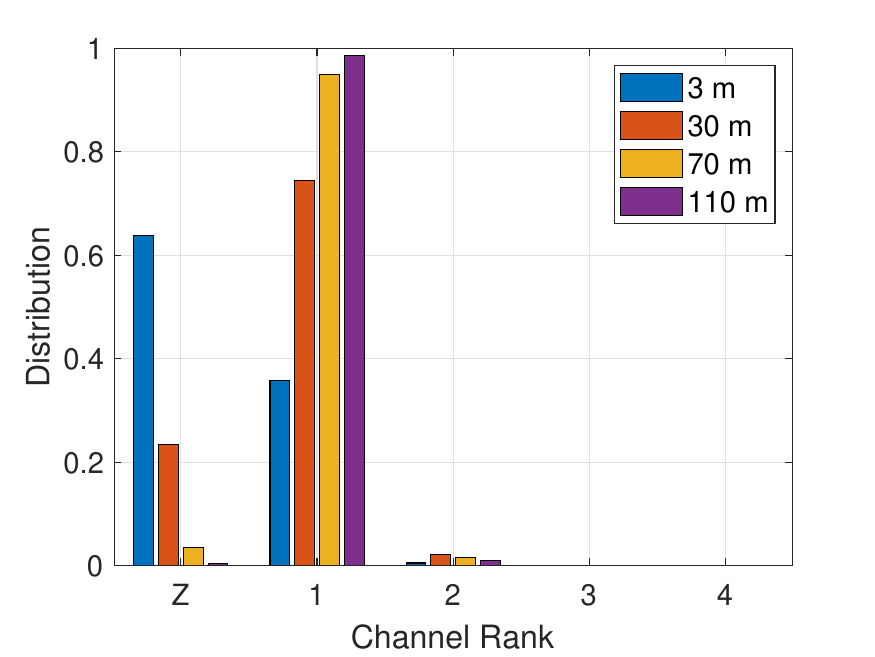}
    \label{fig:Rank_PDF_MIMO_10}
    }
    \subfigure[Distribution with $K_2$]{
    \includegraphics[trim={0.4cm 0.1cm 1.3cm 0.7cm},clip, width=0.64\columnwidth]{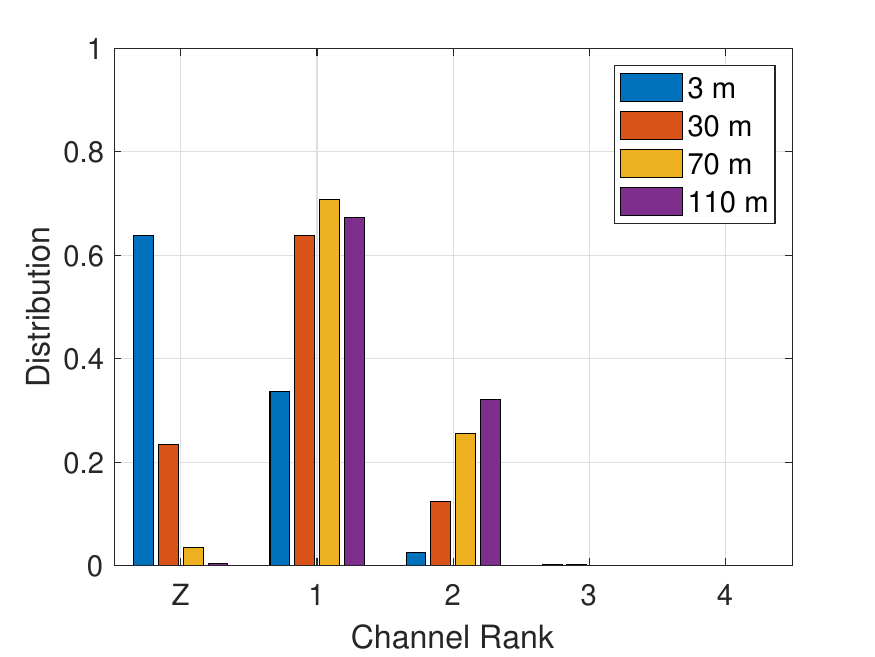}
    \label{fig:Rank_PDF_MIMO_100}
    }
    \subfigure[Distribution with $K_3$]{
    \includegraphics[trim={0.4cm 0.1cm 1.3cm 0.7cm},clip, width=0.64\columnwidth]{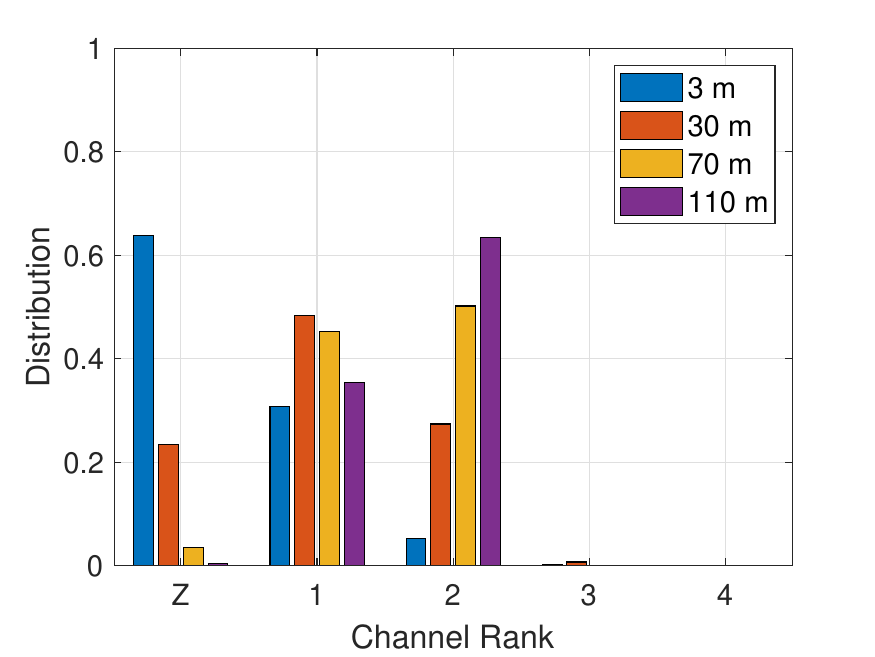}
    \label{fig:Rank_PDF_MIMO_1000}
    }
    \caption{Distribution of the channel rank in the Lake Wheeler Field Labs area with different threshold constants and different UAV altitudes.}
    \label{fig:Rank_PDF_LWs}
\end{figure*}

The channel rank distribution in the Lake Wheeler area based on RT simulations, with different rank thresholds at $3$~m, $30$~m, $70$~m, and $110$~m UAV altitudes are shown in Figure \ref{fig:Rank_distribution}, where $\mathsf{Z}$ indicates the receiver sites that have no connectivity to the corresponding tower. Here, the locations of the buildings and the tower are marked in red polygons and red arrows. The forest areas with trees are marked in magenta polygons in the figures. The UAV is assumed to be connected to the closest tower when the channel rank is evaluated. The boundaries of each coverage are marked in cyan dashed curves in the figures.

As seen in the cases with $3$~m and $30$~m altitudes, the blockage from the trees is observed in the forest area for all tower cases. The blockage area in the forest tends to be mitigated as the altitude of the UAV increases. We have to note the difference in the blockage areas of Figures \ref{fig:RSSI_MIMO_30_joint} and \ref{fig:Rank_MIMO_30_10}. This can be due to the fact that the number of samples in calculating the candidates of the rays using the Fibonacci unit sphere is set to $10^3$ for channel rank and $10^6$ for the RSS simulations to set a reasonable balance in heavy computational load in channel rank simulations. 

It is also observed that UAV locations near the tower have a channel rank of $2$, which can be expected to have two multipath components with LoS and ground-reflected rays. The area of channel rank $2$ near the tower gets wider as the threshold constant is relaxed from $K_1$ to $K_3$. Meanwhile, the channel rank tends to be $1$ as the distance from the tower increases. It can be interpreted that the link is LoS-dependent as the singular values from the ground-reflected ray are not strong enough compared to the strongest singular value. Moreover, some receiver locations that are close to the tower and other objects (e.g., trees or buildings) have a channel rank of $3$ due to richer reflections in those locations. 

The distribution of the channel rank in the Lake Wheeler area with different threshold constant $K$ are shown in Figure~\ref{fig:Rank_PDF_LWs}. Here, the $3$~m altitude cases show more than $60$ percent of the out-of-coverage area due to significant blockage by the trees. The likelihood of observing channel rank 1 reduces as the threshold constant $K$ increases. Most receiver sites have channel rank 1 or 2, which can be interpreted as most of the links having dominant LoS and ground-reflected ray in the given rural scenario.

 \subsection{Kriging Based 3D Channel Rank Interpolation}\label{ch:Kriging_interpolation}

  \begin{figure}
     \centering
     \includegraphics[trim={0.6cm 0 1.1cm 0.5cm},clip,width=0.98\columnwidth]{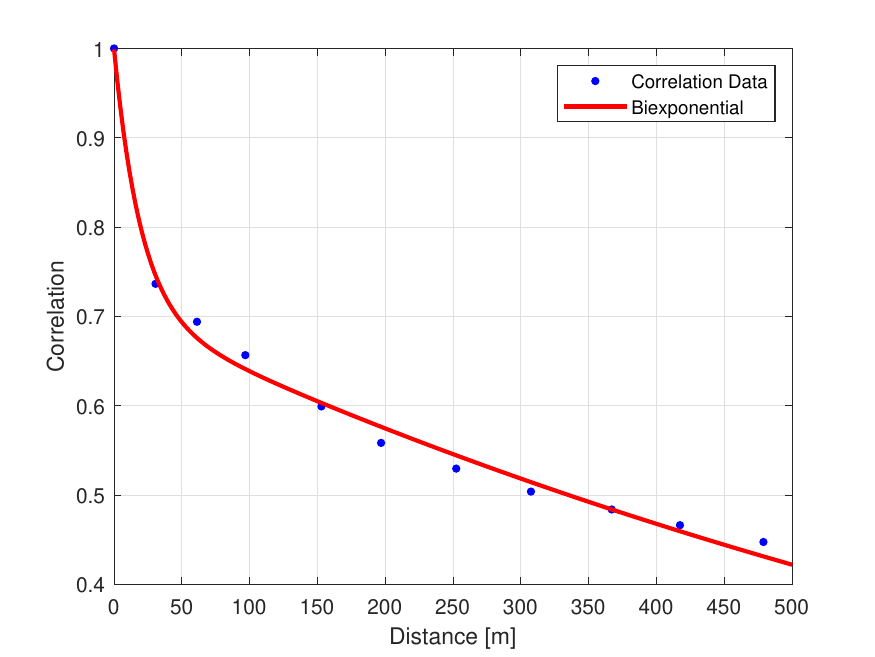}
     \caption{Simulation result of 3D correlation vs. distance and curve fitting model in (\ref{eq:bi_exponential}).}
     \label{fig:correlation_distance_curvefitting}
 \end{figure}

 In this section, we present results on the Kriging-based 3D interpolation scheme for channel rank prediction. First, the semi-variogram has to be estimated based on RT simulations. The simulation results of the correlation of channel rank and 3D distance over every UAV position are shown in Figure \ref{fig:correlation_distance_curvefitting}, where blue dots indicate the correlation data calculated by the procedure in Section \ref{ch:Kriging_3D_Channel_interpolation}-\ref{ch:spatial_correlation} and the red line represents the bi-exponential model derived by the curve fitting expression in (\ref{eq:bi_exponential}). We limit the maximum horizontal distance between two locations of interest for spatial correlation analysis, $\Delta_{\mathrm{2D}}(\boldsymbol{p}_{i}, \boldsymbol{p}_{j})$ as in~(\ref{eq:horizontal_distance}) and (\ref{eq:bi_exponential}), as $500$~m to prevent taking into account less correlated UAV positions. The coefficients $c_{1}$ - $c_{4}$ for equation (\ref{eq:bi_exponential}) are derived by the Matlab curve fitting tool \cite{matlab_curve_fitting}, which are $c_{1} = 0.2932, c_{2} = -0.0508, c_{3} = 0.7057,$ and $c_{4} = -0.001$, respectively. The RMSE between actual correlation data and the bi-exponential model is $0.0153$. 

\begin{figure*}
     \centering
     \subfigure[Kriging Interpolation]{\includegraphics[trim={0.4cm 0.1cm 1.3cm 0.6cm},clip,width=0.66\columnwidth]{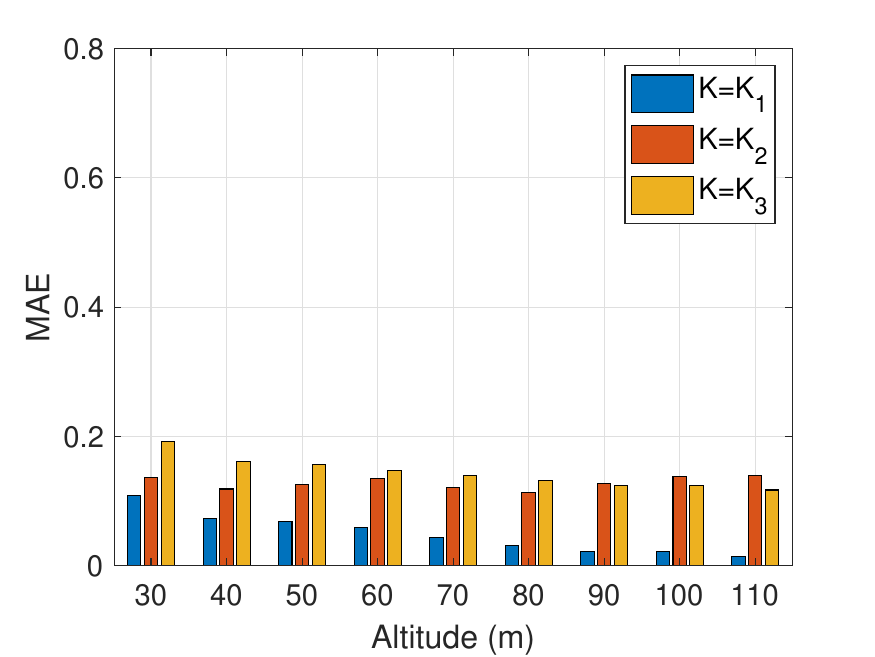}
     \label{fig:MAE_kriging}
     }
     \subfigure[Baseline Interpolation: Makima Method]{\includegraphics[trim={0.4cm 0.1cm 1.3cm 0.6cm},clip,width=0.66\columnwidth]{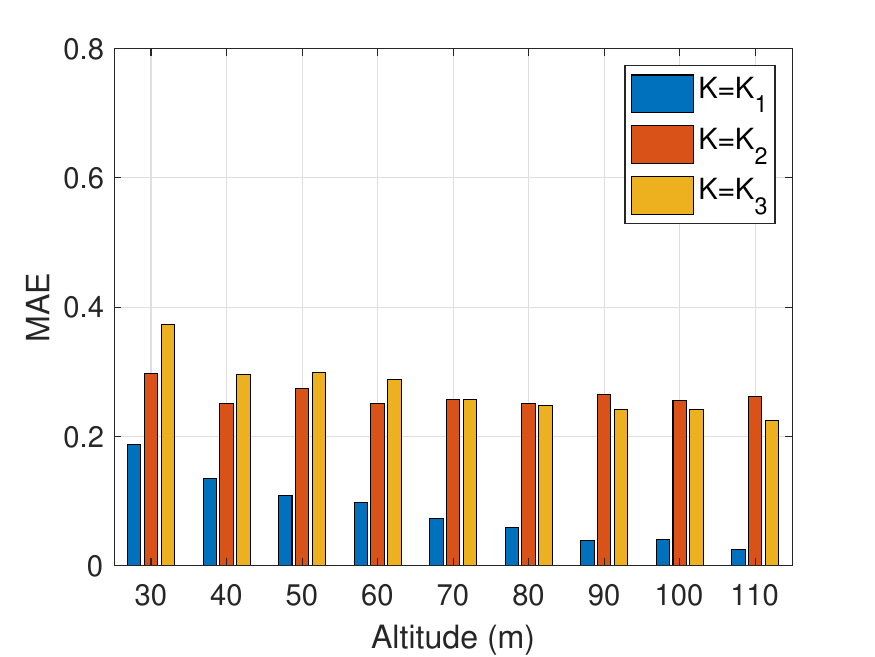}
     \label{fig:MAE_makima}
     }
     \subfigure[Baseline Interpolation: Spline Method]{\includegraphics[trim={0.4cm 0.1cm 1.3cm 0.6cm},clip,width=0.66\columnwidth]{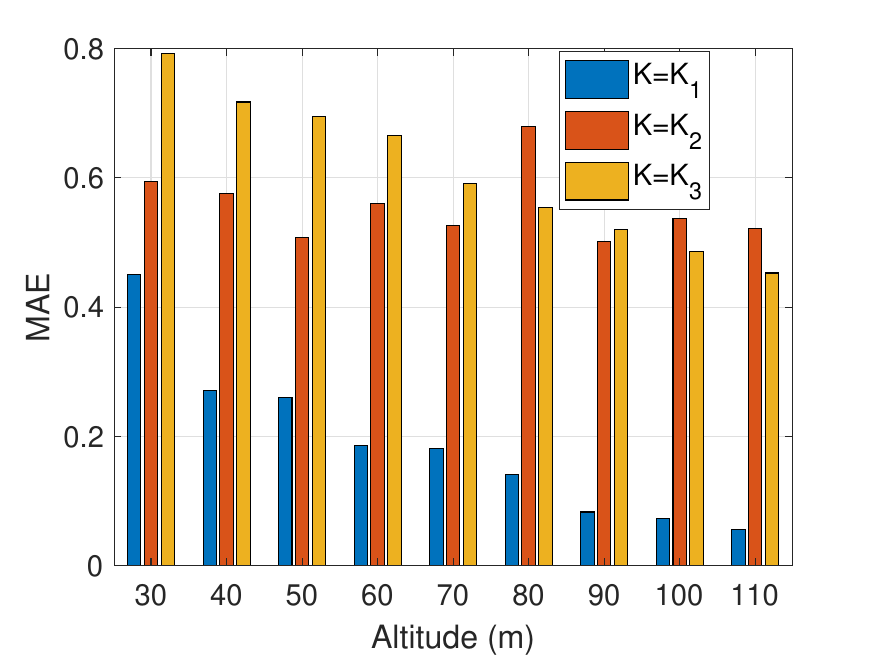}
     \label{fig:MAE_spline}
     }
     \caption{MAE performance comparison of Kriging interpolation-based 3D channel rank interpolation scheme and baseline interpolation-based approaches.}
     \label{fig:MAE_interpolations}
 \end{figure*}

Using the results in Figure~\ref{fig:correlation_distance_curvefitting}, Kriging-based channel rank interpolation is carried out. The interpolated channel rank for all positions of the UAV with $9$ attitudes and $3$ thresholds for the channel rank calculation is evaluated using the MAE error metric, which can be expressed as follows:
\begin{align} \label{eq:MAE}
    E_{\mathrm{MAE}, \delta_{K_j}, h}  = \frac{1}{N_{\mathrm{loc}}}  \sum_{i=1}^{N_{\mathrm{loc}}}\Big|R_{\delta_{K_j}}(\boldsymbol{p}_i) - \hat{R}_{\delta_{K_j}}(\boldsymbol{p}_i)\Big|~, 
\end{align}
for all $j=1, ..., N_K$ and
$h \in (h_1, h_2, ..., h_{N_h})$. Here, $N_{\mathrm{loc}}$ is the total number of possible UAV locations to be interpolated in the uniformly discretized 2D target area as described right after ~(\ref{eq:channel_rank_vector}) and Section~\ref{ch:numerical_results}. Moreover, the number of samples for interpolation $M$ as in (\ref{eq:Kriging_estimation}) is set to $20$ and the radius for sampling from the $\boldsymbol{p}_{0}$, $r_{0}$, as in Figure~\ref{fig:Kriging_description} is $150$~m. The MAE performance evaluations of the Kriging interpolation-based 3D channel rank interpolation scheme and baseline interpolation-based approaches are shown in Figure \ref{fig:MAE_interpolations}. 

Using the previously defined MAE in (\ref{eq:MAE}) and the procedure illustrated in Figure~\ref{fig:Kriging_description}, the MAE of the channel rank is computed for each altitude and threshold setting. The target location $\boldsymbol{p}_0$ is sequentially processed across all UAV location index $i=1,..., N_{\mathrm{loc}}$ as described in Section~\ref{ch:other_baseline}. For each $\boldsymbol{p}_0$, interpolation is performed using $M$ samples within the sampling radius $r_0$ under the assumption that the channel rank at $\boldsymbol{p}_0$ is unknown. This interpolation process is repeated for all possible UAV locations, and the MAE is computed for a given altitude and threshold setting. By iterating this procedure across different altitude and threshold combinations, the MAE is evaluated for all possible configurations.
  
 Results in Figure~\ref{fig:MAE_interpolations} show that the Kriging interpolation-based 3D channel rank interpolation scheme outperforms the MAE of the two baseline interpolation-based approaches. It can be interpreted that the Kriging interpolation yields an accurate interpolation of channel rank using spatial correlation in the given area, while the baseline approaches employ channel rank data and the UAV location index. For all approaches, $K_1$ has the lowest MAE because most channel ranks at UAV locations have a channel rank of $1$ as seen in Figure \ref{fig:Rank_PDF_MIMO_10}. Moreover, it has been observed that the MAE with the altitude of $30$ m yields the highest MAE performance for all thresholds and approaches due to the blockage from the trees. The MAE performance in the lower altitudes, e.g., $30$ - $70$ m, tends to get worse as the threshold constant increases, which is from the different portion of channel rank distribution with the relaxed threshold. Meanwhile, an irregular pattern can be observed at higher altitudes, e.g., $80$ - $110$ m cases. This can be interpreted as being due to relatively low spatial correlation in the case of $K_2$ and $K_3$, where the ratio of channel ranks 1 and 2 is low.

 \subsection{Measurement Comparison}\label{ch:measurement}
In this section, we present the comparison of RT simulations and real-world measurements of RSS and channel rank. In addition to the RT simulation and measurement campaign setups described in Section~\ref{ch:RT_setup} and Section~\ref{ch:measurement_setup}, the following assumptions are used for the channel rank RT simulation for comparison purposes: 1) $2 \times 2$ equally spaced in vertical and horizontal axis by $0.5 \lambda$ is used for multi-antenna receiver setup with four antennas; 2) non-uniformly spaced linear array antenna is used for transmitter multi-antenna configuration, which has antenna element spacing of $60$~mm, $1.68$~m, and $60$~mm to simulate two sector antenna configuration used in our measurements; 3) $90^\circ$ of roll angle is applied for the receiver at the UAV to implement bottom mounted antenna module;  and 4) the number of samples for Fibonacci lattice unit sphere of Sionna RT is set to $10^6$ to get accurate ray information.
 
\subsubsection{Coverage Measurements} 

\begin{figure*}
     \centering
     \subfigure[LW1]{\includegraphics[trim={0.3cm 0.1cm 1.35cm 0.6cm},clip,width=0.64\columnwidth]{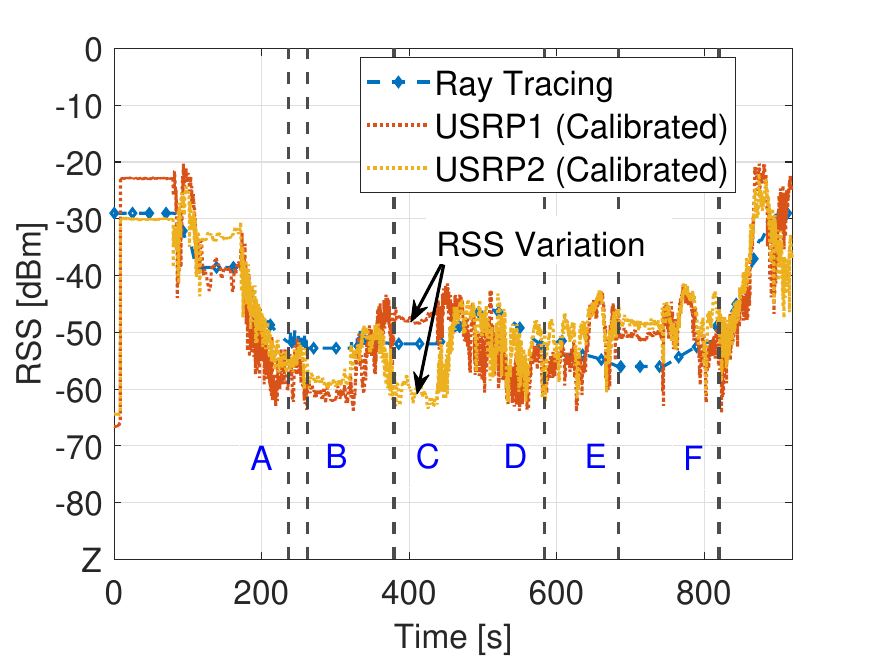}
     \label{fig:measurement_vs_sionna_LW1}
     }
     \subfigure[LW2]{\includegraphics[trim={0.3cm 0.1cm 1.35cm 0.6cm},clip,width=0.64\columnwidth]{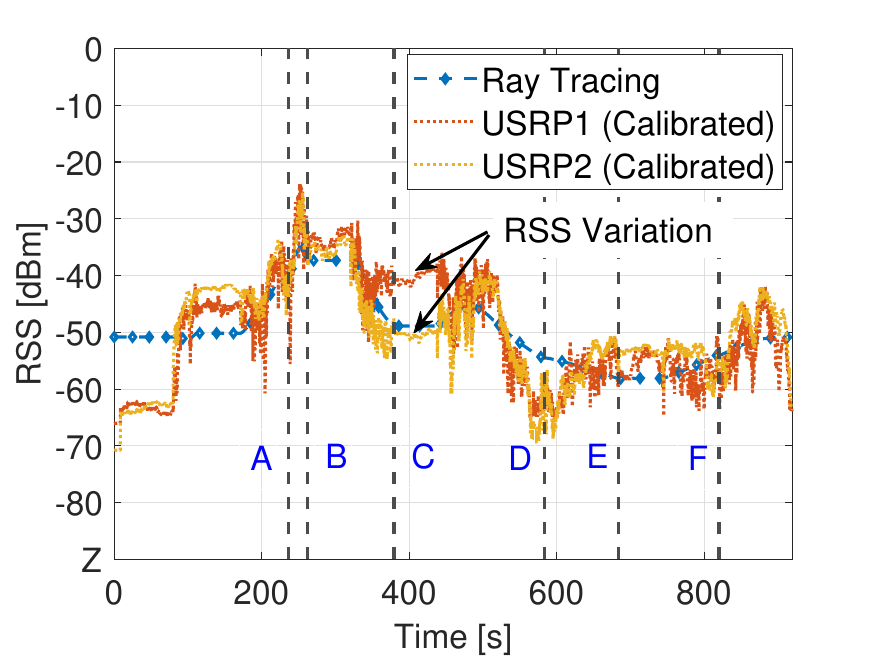}
     \label{fig:measurement_vs_sionna_LW2}
     }
     \subfigure[LW3]{\includegraphics[trim={0.3cm 0.1cm 1.35cm 0.6cm},clip,width=0.64\columnwidth]{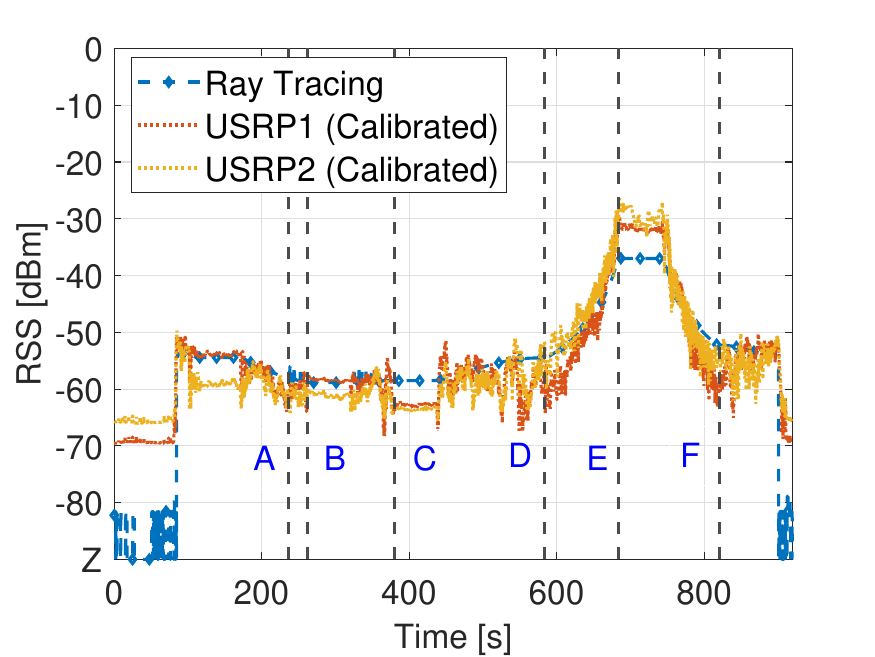}
     \label{fig:measurement_vs_sionna_LW3}
     }
     \subfigure[LW4]{\includegraphics[trim={0.3cm 0.1cm 1.35cm 0.6cm},clip,width=0.64\columnwidth]{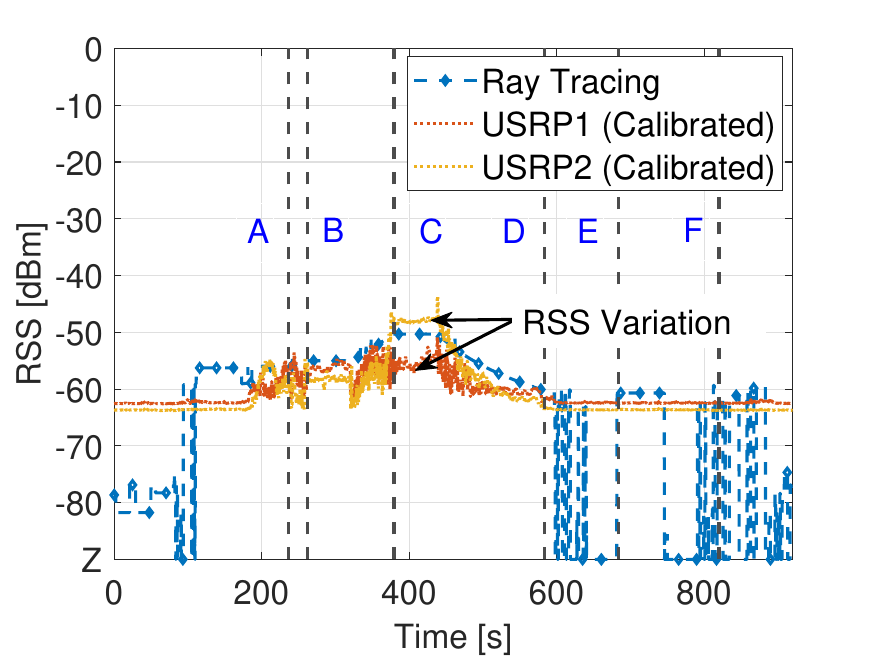}
     \label{fig:measurement_vs_sionna_LW4}
     }
     \subfigure[LW5]{\includegraphics[trim={0.3cm 0.1cm 1.35cm 0.6cm},clip,width=0.64\columnwidth]{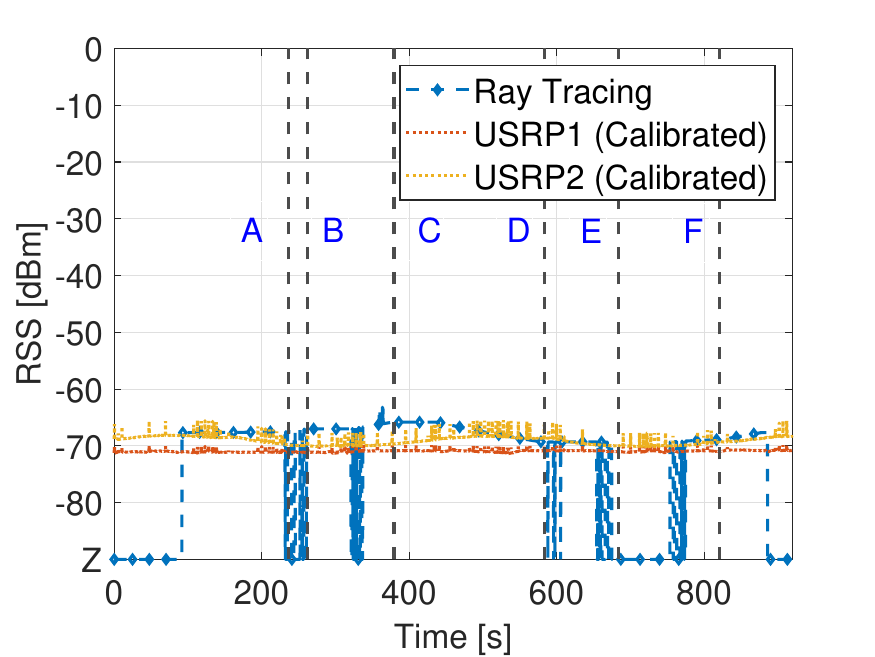}
     \label{fig:measurement_vs_sionna_LW5}
     }
     \caption{RSS measurement and RT simulation results with the predefined trajectory for the specific UAV locations {A}-{F} in Figure \ref{fig:trajectory_altitude_measurement_RSS}. }     \label{fig:measurement_vs_sionna_LW1_LW5}
 \end{figure*}

The RSS simulation results of RT and the measurements from each tower, over the trajectory of the UAV, as illustrated in Figure~\ref{fig:trajectory_measurement_RSS}, are shown in Figure \ref{fig:measurement_vs_sionna_LW1_LW5}. Measurement methodology for coverage measurements is presented earlier in Section~\ref{ch:measurement_setup}. The labels USRP1 and USRP2 indicate measurements captured simultaneously at two different antennas connected to a dual channel USRP at each tower, and $\mathsf{Z}$ is the out-of-coverage area. The RT approach shows a similar pattern with the measurements over LW1 to LW5 tower cases. It is observed that the measurements show fluctuations throughout the trajectory due to fading effects, changing roll/yaw/pitch of the UAV with respect to each tower, and Doppler effects, among other factors. 

An interesting behavior is observed at the RSS values observed by USRP1 and USRP2 antennas. While at some locations the measurements at USRP1 and USRP are very similar, at other locations, there may be over 10~dB of difference between them, see e.g. the interval right after waypoint C, for LW1, LW2, and LW4 cases, as highlighted on the figures. The RSS variation can be from the geometry-dependent LoS blockage and channel conditions due to different relative orientations of the USRP1 and USRP2 antennas with respect to the UAV. 

We can notice the out-of-coverage and low RSS conditions in RT simulation for LW3, LW4, and LW5 at lower altitudes, i.e., time intervals before $100$~s and after $850$~s when the drone is taking off and landing, respectively. These are due to altitude-dependent blockage at those towers. Moreover, blockage is also observed at LW4 and LW5 (further towers to the UAV's flight area) with RT for other UAV locations. This can be interpreted to be due to the RT ray calculation not being able to find a proper ray with the simulation settings of the number of samples for the Fibonacci lattice unit sphere and assumptions for the height of the trees. The RSS measured at LW5 and some intervals at LW4 show to be constant over time when it is possible to detect it, where the RSS is very close to the background noise due to the far distance with each of these towers.

 \subsubsection{Channel Rank Measurements}
 
\begin{figure*}
     \centering       
      \subfigure[5G NR RI measurement]{\includegraphics[trim={0.2cm 0.1cm 0.9cm 0.65cm},clip,width=0.64\columnwidth]{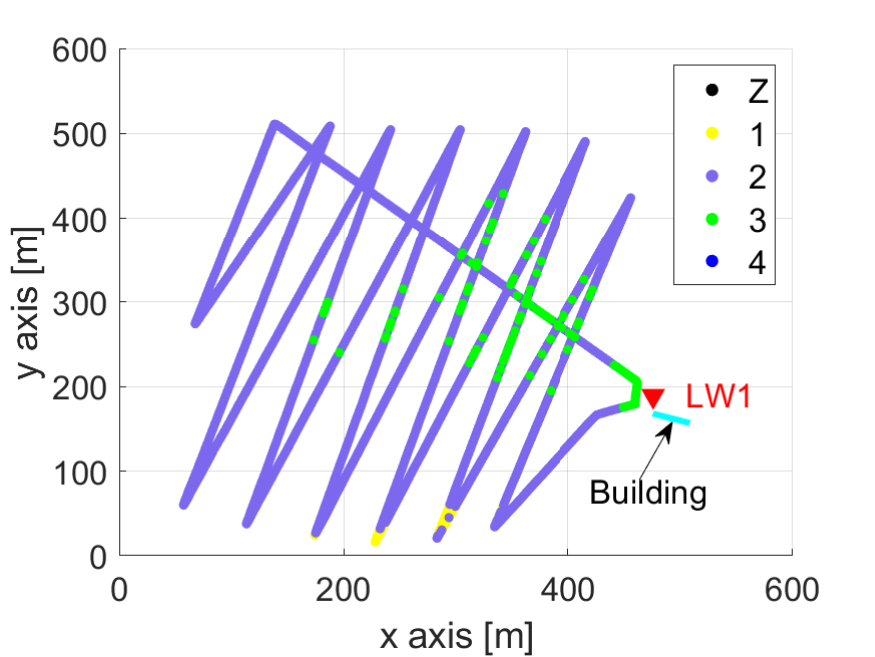}
     \label{fig:rank_zigzag_measurement}
     }     
     \subfigure[RT simulation with $K_3$]{\includegraphics[trim={0.2cm 0.1cm 0.9cm 0.65cm},clip,width=0.64\columnwidth]{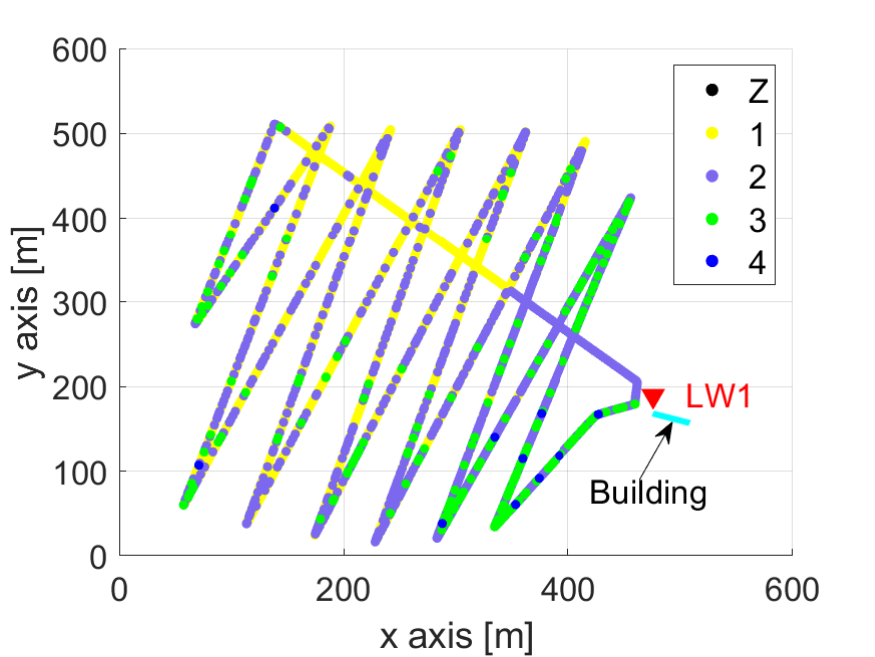}
     \label{fig:rank_zigzag_RT}
     }
     \subfigure[Channel rank distribution with different $K$ of RT and measurements]{\includegraphics[trim={0.2cm 0.15cm 1.1cm 0.7cm},clip,width=0.6\columnwidth]{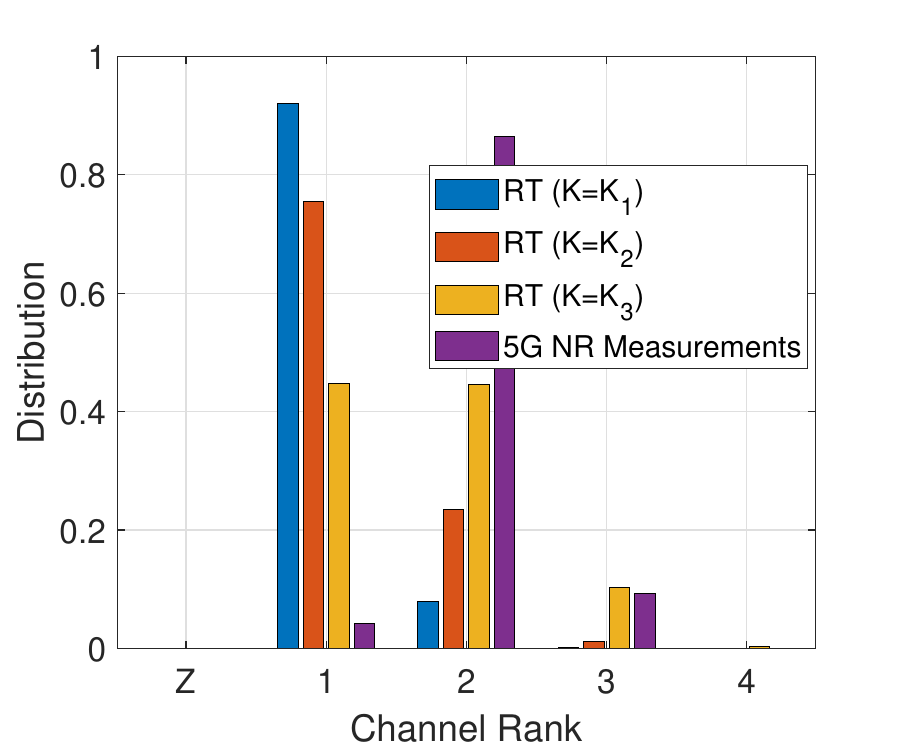}
     \label{fig:distribution_rank_RT_measurement}
     }     
     \caption{5G NR RI measurement and RT simulation results with the predefined trajectory in Figure \ref{fig:trajectory_altitude_measurement_rank}.}
     \label{fig:measurement_vs_sionna_rank}
 \end{figure*}

 RT simulations and real-world measurements to study the channel rank are done based on the trajectory shown earlier in Figure~\ref{fig:trajectory_measurement_rank}. Measurement methodology is presented earlier in Section~\ref{ch:measurement_setup}.   
 The channel rank simulation results of RT and the corresponding measurements are shown in Figure~\ref{fig:measurement_vs_sionna_rank}, where the LW1 tower is marked as a red triangle and the building in the target area is marked as a cyan polygon. As seen in the 5G NR RI measurements in Figure~\ref{fig:rank_zigzag_measurement}, the channel rank at most locations is 2, where one of the dual ports of each sector antenna supports spatial multiplexing. A channel rank of 3 has been observed at the locations on the northwest side of the LW1 tower, which can be interpreted as an additional spatial stream from a reflective path through the building near the tower. Moreover, a channel rank of 1 can also be observed on the south side of the target area implying a lack of angular separation over the given MIMO channel. 
 
 On the other hand, the RT simulation results are shown in Figure \ref{fig:rank_zigzag_RT}. Channel rank in RT simulation tends to be more variant compared to the measurements. A channel rank of 1 or 2 is observed in most UAV locations on the northwest side of the LW1 tower due to the orientation of the LW1 tower and the receiver of the UAV. On the other hand, a channel rank of 3 can be observed on the edge of each waypoint because of the reflections. It is worthwhile to note that it is inevitable to have a mismatch between RT and real-world measurements at specific locations due to the following reasons.
 \begin{itemize}
     \item \textbf{Angular mismatch.} For the LW1 tower antenna configuration, the dual port AW3232 sector antenna supports $+/- 45^\circ$ slant linear polarization. Moreover, the receiver antenna module mounted at the UAV also has a dual slant orientation structure. However, the element-wise rotation of the linear array configuration is not supported in the Sionna RT tool.
     \item \textbf{Dynamic vs. static environments.} There are environmental gaps between measurement and RT scenarios such as foliage patterns, temporal fadings, antenna behavior in real scenarios, and among others.
     \item \textbf{Rank calculation.} A threshold-based channel rank calculation is used for the RT scenarios. However, in the 5G NR RI measurements, a mobile device mounted on the UAV determines the RI to report to the LW1 tower. It is known that RI calculation is implement-dependent, which leads to the differences.
 \end{itemize}

The distribution of channel rank of RT with different $K$ can be a more meaningful comparison between RT simulations and measurements in the same environment. The histograms of the channel rank with RT and measurements are shown in Figure~\ref{fig:distribution_rank_RT_measurement}. The likelihood of having a channel rank of 2 or 3 is increased as $K$  increases from $K=K_1$ to $K=K_3$, which results in a higher similarity with the measurements. The mismatch between measurements and RT occurs mainly when the channel rank is 1. This suggests that, at most UAV locations, the second singular value for RT simulations does not fall within the range required for $K_3$ ($30$~dB) from the strongest singular value. Even though there are gaps in the channel rank between RT and measurement, RT can be a reasonable benchmark as a baseline in the controlled environment.

 \section{Conclusions}
In this paper, we investigated RF coverage and channel rank in rural environments using the NVIDIA Sionna RT tool, extending our previous work \cite{previous_work} with a broader analysis of the Lake Wheeler Field Labs. Our study incorporated realistic foliage modeling, multiple base stations, and the Kriging interpolation-based 3D channel rank prediction scheme. We applied a constant threshold ratio for singular value selection and analyzed its distribution across different UAV altitudes and threshold settings. The observed spatial correlation of channel rank motivated the application of Kriging interpolation, where we derived the semi-variogram using the correlation-distance relationship at UAV positions. The proposed Kriging interpolation scheme was evaluated using the MAE metric and compared with two baseline interpolation methods. Results demonstrated that the Kriging-based approach outperformed baseline interpolation techniques by leveraging spatial correlation. Finally, we compared our RT-based RSS and channel rank simulation results with the real-world measurements collected from the NSF AERPAW testbed. The reasonable consistency between the RT results and the measurements was demonstrated in the foliage-dense rural scenarios.

For future work, unique challenges can be considered in UAV-based 6G and V2X communications including managing time-varying channel characteristics, ensuring ultra-reliable low-latency communication (URLLC), and optimizing spatial multiplexing under dynamic network conditions. Addressing the time-varying nature of signal sources, obstacles, and UAVs in dynamic environments requires advanced RT techniques to capture the rapid fluctuation of channel conditions. It enables analyzing adaptive spatial multiplexing strategies that leverage real-time channel conditions, geographical information, and mobility of UAVs. Moreover, improving the Kriging interpolation framework by incorporating data-driven and ML-aided techniques is another potential direction. By integrating ML with statistics of spatial correlation into Kriging interpolation techniques, it is expected that accuracy and computational complexity can be improved over various environment settings.

\bibliographystyle{IEEEtran}




\end{document}